\documentclass[11pt, oneside]{article}
\usepackage[paper=a4paper, total={16cm,23cm}]{geometry}
\usepackage[T1]{fontenc}
\usepackage{setspace} 
\usepackage[hang]{footmisc} 
\usepackage{titling} 
\usepackage{tocloft} 
\usepackage{parskip} 
\usepackage{comment}
\usepackage{booktabs}

\usepackage{amsfonts}
\usepackage{amsmath}
\usepackage{amssymb}
\usepackage{mathrsfs, dsfont, amsthm}
\usepackage{physics}
\usepackage{mathtools}
\usepackage{upgreek}
\numberwithin{equation}{section} 
\usepackage{wasysym}
\usepackage{marvosym}

\newcommand{\e}{\mathrm{e}}

\newcommand{\Order}{\mathcal O}

\usepackage{xcolor}
\usepackage{colortbl}
\definecolor{dark-red}{rgb}{0.50,0.12,0.12} 
\definecolor{mblue}{rgb}{0.30, 0.45, 0.70}
\definecolor{mred}{rgb}{0.70, 0.20, 0.20}
\definecolor{mgray}{rgb}{0.63, 0.63, 0.63}

\usepackage{hyperref} 
\hypersetup{colorlinks=true, allcolors=dark-red, linktoc=all}

\usepackage{cite}

\usepackage{tikz}
\usepackage{tikzlings}
\usetikzlibrary{calc, patterns, decorations, decorations.markings, shapes, positioning, intersections, quotes, angles}
\usepackage[subrefformat=parens]{subcaption}
\usepackage{pgfplots}
\pgfplotsset{compat=newest}
\newcommand{\mathdefault}[1][]{}

\newcommand{\beq}{\begin{equation}}
\newcommand{\eeq}{\end{equation}} 

\newcommand{\lc}{\left(}
\newcommand{\rc}{\right)}

\newcommand{\ic}{\mathrm{i}}
\newcommand{\diff}{\mathrm{d}}

\usetikzlibrary{decorations.pathmorphing}

\begin{document}
\begin{titlingpage}
    \vspace*{3em}
    \onehalfspacing
    \begin{center}

   {\LARGE A signature of the timelike singularity in holography}

    \end{center}
    \singlespacing
    \vspace*{2em}
      \begin{center}
        \textbf{
      Joydeep Chakravarty$^{1,2}$ \ and \ Yoav Zigdon$^3$
        }
    \end{center}
    \vspace*{1em}
    \begin{center}
        \textsl{
       $^1$ Department of Physics, Harvard University, Cambridge, MA 02138, USA \\
       \vspace*{1em}
       $^2$ Department of Physics, McGill University,
        Montr\'eal, QC H3A 2T8, Canada \\
        \vspace*{1em}
        $^3$ School of Physics and Astronomy, Tel Aviv University, Ramat Aviv, 69978, Israel
        \\[\baselineskip]
         }
    \href{mailto:joydeepchakravarty@fas.harvard.edu}{\small joydeepchakravarty@fas.harvard.edu}, $\quad$\href{mailto:yoavzi@tauex.tau.ac.il}{\small yoavzi@tauex.tau.ac.il}
    \end{center}
    \vspace*{3em}
    \begin{abstract}
In the context of the AdS/CFT correspondence, we provide a simple observable, the imaginary part of the retarded Green's function, which captures the timelike singularity of a charged black brane. Solving wave equations of bulk fields with different spins outside the black brane, we calculate exponentially small contributions to retarded Green's functions in the dual CFT at large real frequency.  These contributions admit an interpretation in terms of paths in an analytically continued classical black brane interior via a WKB approximation: they provide the complex time elapsed along null geodesics that reflect off the singularity, and incorporate the corresponding reflection coefficients.
    \end{abstract}
\end{titlingpage}
\tableofcontents


\section{Introduction}
\label{sec:introduction}

Improving our understanding of spacetime singularities is an important goal in quantum gravity. The gauge/gravity correspondence provides a setting in which questions about singularities can be formulated in terms of well-defined observables in a dual quantum field theory. In the large-$N$ and large-'t~Hooft coupling limits, a thermal system in holographic conformal field theories (CFTs) is described by  a classical, asymptotically anti-de Sitter (AdS) black hole geometry (with a temperature above the Hawking-Page transition). One can then
ask which boundary observables capture physics associated with bulk regions behind the event horizon and, in particular, about the black hole curvature singularity. 

A number of works have established that holographic correlation functions have signatures of the black hole interior and the singularity. 
Reference \cite{Fidkowski:2003nf} evaluated correlation functions of operators with large scaling dimensions in the thermofield double state of two CFTs in the context of AdS/CFT \cite{Maldacena:2001kr}. The authors studied the singularities of the correlator, which appear in a secondary sheet in the complex time plane, and tied them to spacelike geodesics probing the black hole interior. The spacelike geodesics become almost null, approaching the singularity and turn into another asymptotic region~\footnote{Earlier discussions of how correlators can encode information behind an AdS black hole horizon include references~\cite{Louko:2000tp,Kraus:2002iv}.}. A complementary line of work was initiated by \cite{Festuccia:2005pi}, which showed
that the spacelike singularity of AdS black holes is encoded in thermal correlation functions that exhibit exponentially suppressed large-imaginary frequency behavior. A subsequent Wentzel, Kramers, and Brillouin (WKB) analysis related asymptotic quasinormal-mode spectra and boundary Green's functions to bulk geodesics in large-mass and large-momentum regimes~\cite{Festuccia:2008zx} (see also \cite{Amado:2008hw} for null bouncing geodesics in different dimensions). 

The case of a charged black hole has several distinctive features. Reference \cite{Brecher:2004gn} considered two-sided correlation functions on the CFT side in a state dual to the charged AdS black hole solution. The spacelike geodesics relevant for their heavy-operator correlation functions probe the region between the outer and inner horizons, rather than the immediate vicinity of the timelike singularity. This analysis has also emphasized the characteristic global structure of charged black holes in asymptotically AdS spacetimes, in which the timelike singularities ``bow out'' when the AdS boundaries are represented by straight lines.

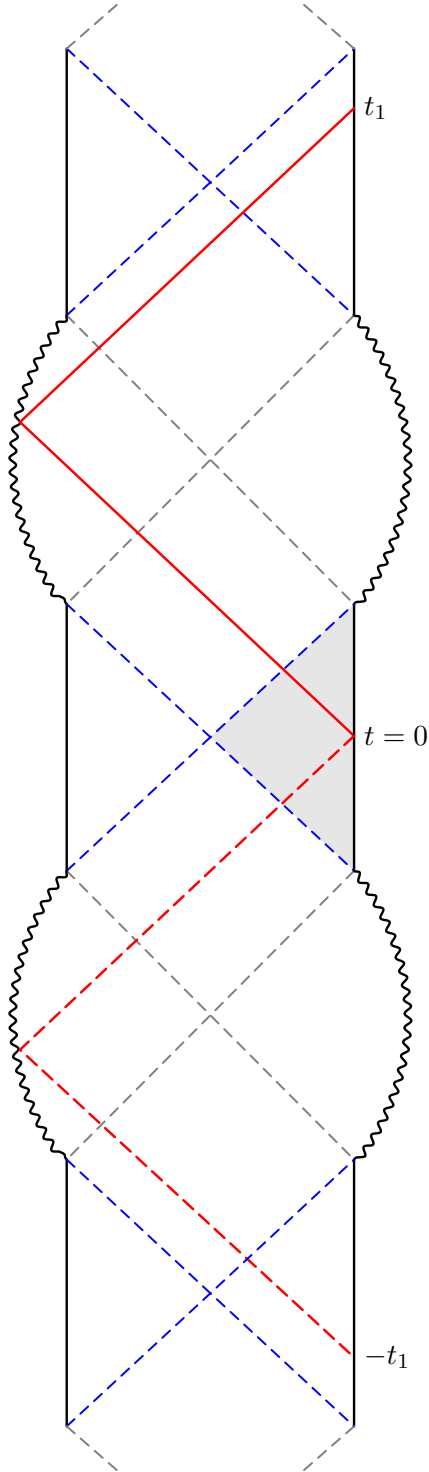
\begin{figure}[tbp]
\centering

\begin{tikzpicture}[
    x=0.95cm,
    y=0.83cm,
    line cap=round,
    line join=round
]

\tikzset{
    horizon/.style={
        blue,
        dashed,
        line width=0.75pt,
        dash pattern=on 4pt off 3pt
    },
    extension/.style={
        gray,
        dashed,
        line width=0.75pt,
        dash pattern=on 4pt off 3pt
    },
    boundary/.style={
        black,
        line width=0.9pt
    },
    singularity/.style={
        black,
        line width=0.9pt,
        decorate,
        decoration={
            snake,
            amplitude=1.25pt,
            segment length=5.4pt
        }
    },
    futuregeodesic/.style={
        red,
        line width=0.95pt
    },
    pastgeodesic/.style={
        red,
        dashed,
        line width=0.95pt,
        dash pattern=on 5pt off 3pt
    },
    exterior/.style={
        fill=gray!20,
        draw=none
    }
}

\def\xside{2.00}


\fill[exterior]
    (\xside,-2.30)
    -- (0,-4.42)
    -- (\xside,-6.55)
    -- cycle;


\draw[boundary]
    (-\xside,6.55) -- (-\xside,2.30);

\draw[boundary]
    (\xside,6.55) -- (\xside,2.30);

\draw[horizon]
    (-\xside,6.55) -- (0,4.42) -- (\xside,6.55);

\draw[horizon]
    (-\xside,2.30) -- (0,4.42) -- (\xside,2.30);

\draw[extension]
    (-\xside,6.55) -- (-1.30,7.25);

\draw[extension]
    (\xside,6.55) -- (1.30,7.25);


\draw[extension]
    (-\xside,2.30) -- (\xside,-2.30);

\draw[extension]
    (\xside,2.30) -- (-\xside,-2.30);

\draw[singularity]
    (-\xside,2.30)
    .. controls (-2.48,1.62) and (-2.72,0.72) .. (-2.72,0)
    .. controls (-2.72,-0.72) and (-2.48,-1.62) .. (-\xside,-2.30);

\draw[singularity]
    (\xside,2.30)
    .. controls (2.48,1.62) and (2.72,0.72) .. (2.72,0)
    .. controls (2.72,-0.72) and (2.48,-1.62) .. (\xside,-2.30);

\draw[futuregeodesic]
    (\xside,5.6)
    -- (-2.65,0.6)
    -- (\xside,-4.4);

\draw[fill]
    (\xside,5.6) node[right] {$t_1$};

\draw[fill]
    (\xside,-4.4) node[right] {$t=0$};


\draw[boundary]
    (-\xside,-2.30) -- (-\xside,-6.55);

\draw[boundary]
    (\xside,-2.30) -- (\xside,-6.55);

\draw[horizon]
    (-\xside,-2.30) -- (0,-4.42) -- (\xside,-2.30);

\draw[horizon]
    (-\xside,-6.55) -- (0,-4.42) -- (\xside,-6.55);


\draw[extension]
    (-\xside,-6.55) -- (\xside,-11.15);

\draw[extension]
    (\xside,-6.55) -- (-\xside,-11.15);

\draw[singularity]
    (-\xside,-6.55)
    .. controls (-2.48,-7.23) and (-2.72,-8.13) .. (-2.72,-8.85)
    .. controls (-2.72,-9.57) and (-2.48,-10.47) .. (-\xside,-11.15);

\draw[singularity]
    (\xside,-6.55)
    .. controls (2.48,-7.23) and (2.72,-8.13) .. (2.72,-8.85)
    .. controls (2.72,-9.57) and (2.48,-10.47) .. (\xside,-11.15);

\draw[pastgeodesic]
    (\xside,-4.4)
    -- (-2.65,-9.4)
    -- (\xside,-14.3);

\draw[fill]
    (\xside,-14.3) node[right] {$-t_1$};


\draw[boundary]
    (-\xside,-11.15) -- (-\xside,-15.40);

\draw[boundary]
    (\xside,-11.15) -- (\xside,-15.40);

\draw[horizon]
    (-\xside,-11.15) -- (0,-13.28) -- (\xside,-11.15);

\draw[horizon]
    (-\xside,-15.40) -- (0,-13.28) -- (\xside,-15.40);

\draw[extension]
    (-\xside,-15.40) -- (-1.30,-16.10);

\draw[extension]
    (\xside,-15.40) -- (1.30,-16.10);

\end{tikzpicture}

\caption{
A part of a Penrose diagram of the charged black brane. The solid red curve describes a geodesic that reflects off the future timelike singularity and continues to another region. The dashed red curve represents the time reversal of that bouncing geodesic, reflecting off the past timelike singularity. These geodesics have an imprint on the leading non-perturbative contribution to the spectral density. The imaginary part of $G_{\text{ret}}(\omega,\vec{k})$ evaluated in the exterior region (in gray), with the vacuum contribution subtracted off, encodes the time elapsed along bouncing geodesics in the extended geometry. Higher bounces correspond to multiple reflections off the singularities in the (further) extended geometry and contribute too.}

\label{fig:penrose}

\end{figure}

Reference~\cite{Ceplak:2024bja} expressed the thermal two-point function on the CFT side using the operator product expansion (OPE), and showed that a contribution coming from exchanges of the identity, stress-tensor, and multi-stress-tensor operators leads to a singularity in a complex time associated with null geodesics that bounce off a spacelike singularity. This perspective was extended to charged black branes in reference~\cite{Ceplak:2025dds}, where the relevant analytic continuation probes trajectories bouncing off the timelike singularity. Related approaches have used the thermal product formula and asymptotic quasinormal-mode data~\cite{Dodelson:2023vrw} to suggest that stringy effects split and shift bouncing geodesic singularities~\cite{Dodelson:2025jff}, and studied WKB continuations past inner horizons in charged and rotating backgrounds~\cite{AliAhmad:2026wem}. Related papers about the SYK model and holography have investigated complex time singularities and the temperature dependence of quasinormal modes~\cite{Buric:2026qsp,Dodelson:2026gak}.

In the context of the AdS Schwarzschild geometry, it was recently pointed out that the spectral density, i.e., twice the imaginary part of retarded Green's function at finite temperature, admits an asymptotic large-frequency expansion, with exponentially small contributions that can be calculated by a steepest-descent prescription in the bulk \cite{Afkhami-Jeddi:2025wra}. The relevant complex trajectory can be interpreted as a null geodesic bouncing off the spacelike singularity. A WKB analysis along the lines of \cite{Caron-Huot:2025hmk, Caron-Huot:2025she} (see also \cite{Caron-Huot:2022lff, Chakravarty:2025ncy}) determines the exponential damping and oscillatory phase, while a connection problem near the singularity determines a reflection coefficient.  Analyses of thermal spectral density asymptotics at both large frequency and momenta and their relation to black brane singularities have provided further insight as to the interior interpretation \cite{Jia:2025jbi, Jia:2026ryl}.  Similar exponentially suppressed contributions have also been noticed in defect correlators, where contributions from the retarded two-point function of displacement operators on a Wilson line were obtained in \cite{Giombi:2026kdz}.  

An active field of research focuses on mechanisms that resolve bouncing singularities. Upper bounds were placed on the magnitude of the thermal two-point correlation at finite $N$, implying that the bouncing singularity is absent \cite{Simon_unpub,Komargodski:2026oob}. In the cases studied, this shows that the bouncing singularity is an artifact of the strict $N\to \infty$ limit. In models of two-dimensional dilaton gravity, the gravitational backreaction on the geodesic worldline smoothens the bouncing singularity~\cite{Batra:2026gbq} (see also \cite{Batra:2026otu}).  In the context of the $\text{SL}(2,\mathbb{R})_k/\text{U}(1)$ Euclidean cigar, reference~\cite{Buric:2026sit} showed that the reflection coefficient of the exact CFT is an analytical function of the energy, and interpreted this as a resolution of the bouncing singularity in the two-sided Lorentzian black hole.

\subsection{Summary and results}
In this paper, we study four-dimensional $\text{SU}(N)$, $\mathcal N = 4$ super Yang-Mills on $\mathbb{R}^3 \times \mathbb{R}_t$ in the large number of colors $N\to \infty$ and large 't Hooft coupling $\lambda\to \infty$ limit, in a grand canonical ensemble with fixed inverse temperature $\beta$ and chemical potential $\mu$. The bulk dual of this ensemble is the planar Reissner-Nordstr\"om-AdS$_5$ black brane, whose classical extension contains an inner horizon and a timelike singularity. 

Our observable is a difference of spectral densities
\begin{equation}
\Delta \rho(\omega,\vec{k})=2\,\lc {\rm Im}G_{\rm ret}(\omega,\vec{k})-\,{\rm Im}G_{\rm ret,vacuum}(\omega,\vec{k}) \rc~,
\end{equation}
where $G_{\rm ret}(\omega,\vec{k})$ is the retarded Green's function of the CFT, and the difference above is between the spectral density of the grand canonical ensemble and the spectral density of the vacuum state.
We will compute the spectral density using a conventional prescription that relies on the physics of waves in the black brane exterior and the AdS/CFT correspondence. We will show that the large real-frequency behavior of $\Delta \rho$ contains exponentially small non-perturbative corrections that admit a very interesting interior interpretation. These contributions to $\Delta \rho(\omega,\vec{k}=0)$ can be explained by radial null geodesics that bounce off the timelike singularity in the extended charged geometry (see Fig. \ref{fig:penrose}). The coordinate time along them, $\Delta t_Q=\tau_Q+\mathrm{i}\lambda_Q$, features in the result for $\Delta \rho$:
\begin{equation}
  \Delta \rho\propto   \exp\!\left(\mathrm{i}\omega\Delta t_Q\right)
    =
    \exp(-\lambda_Q\omega)\exp(\mathrm{i}\tau_Q\omega)~.
    \label{Deltarho}
\end{equation}
The parameter $\lambda_Q$ fixes the exponential suppression of the vacuum-subtracted spectral density, and $\tau_Q$ fixes its oscillation frequency (explicit expressions of these parameters in terms of the charge of the solution will be provided). The proportionality constant in Eq.~(\ref{Deltarho}) is determined by connecting solutions to bulk perturbation equations near the singularity. 

For the charged case, we utilize tools from \cite{Afkhami-Jeddi:2025wra} to analyze the spectral density associated with a dual charged planar black brane solution that gives interesting, novel features not present in the uncharged AdS-Schwarzschild case: The nonzero chemical potential modifies both the time along bouncing geodesics and the reflection coefficients from the singularity. At a finite bulk charge parameter and sufficiently large frequency, the WKB approximation breaks down near the black brane singularity. To proceed, we solve the wave equations approximately in this region and find Hankel functions solutions. The subleading terms  lead to characteristic fractional-power expansions in the vacuum-subtracted current and stress-tensor spectral densities. We connect the solution near the singularity to solutions in the asymptotically AdS region. 

By contrast, at small chemical potential and for an appropriate high-frequency window, the scale at which the WKB approximation breaks down is outside the inner-horizon scale. We provide evidence that the reflection coefficients are continuously connected to their AdS-Schwarzschild values, even though the time along the radially null geodesics receives charge-dependent corrections. In this scenario, to venture beyond the inner horizon and to probe the black brane singularity, one should tune the frequency to sufficiently high values. 

Our analysis distinguishes two ingredients in the leading non-perturbative spectral density correction. The first is the complex WKB propagation factor, fixed by $\Delta t_Q$. The second is a local near-singularity connection coefficient, which determines the reflection amplitude and its fractional-power corrections. We test these predictions
using high-precision numerical solutions of wave equations. For a finite nonzero charge parameter, the extracted leading coefficients agree with the Hankel connection coefficients predicted by the charged near-singularity analysis. For a small charge parameter, the effective reflection coefficients remain close to their uncharged AdS-Schwarzschild values.

An advantage of the spectral density is that it is a directly Lorentzian, single-sided observable. It is defined in terms of a retarded correlation function, which is computed using bulk equations outside the outer horizon, with standard infalling boundary conditions there. The two-sided geometry and the bouncing geodesic picture arise only after analytic continuation and large-frequency asymptotics. The results can be interpreted as saying that the observable \textit{creates} interior paths associated with the extended black brane geometry, without requiring a physical two-sided measurement or an analytic continuation of the measured frequency.

A separate set of questions concerns the status of the inner horizon itself. The classical Reissner-Nordstr\"om geometry admits an analytic extension across its Cauchy horizon, but this extension is expected to be sensitive to perturbations. The high blueshift of field modes near the inner horizon motivates a scenario in which the Cauchy horizon becomes singular~\cite{Simpson:1973ua,Poisson:1989zz,Poisson:1990eh,
Ori:1991zz}. Further developments have clarified that the strength of the resulting singularity, and the possibility of extending the metric beyond it at low regularity, depend on the matter model and on the precise regularity class imposed on the evolution~\cite{Dafermos:2003wr,Dafermos:2003yw}.
Quantum effects add further subtleties. In particular, \cite{Papadodimas:2019ery} formulated a necessary smoothness criterion in terms of the entanglement of near-horizon modes and provided numerical evidence that the Hartle-Hawking state of asymptotically AdS Reissner-Nordstr\"om black holes is singular at the inner horizon. In the context of the AdS/CFT correspondence, it was proposed that CFT two-point functions in a grand canonical ensemble satisfy Kubo–Martin–Schwinger (KMS) relations both with respect to the outer horizon temperature and the inner horizon temperature, which conflict with each other for generic temperatures~\cite{Balasubramanian:2019qwk}. 

Our use of the maximally extended classical geometry is strictly an analytic continuation that organizes the single-sided exterior calculation in the probe limit $G_N \propto N^{-2} \to 0$. The results of the exterior computation of the retarded Green's function are independent of statements about the actual Cauchy horizon or the spacetime beyond it. This distinction is important here: The spectral density is extracted by an exterior computational prescription, whereas an interpretation is given in terms of bouncing geodesics that emerge upon analytic continuation of the bulk exterior geometry.

The remainder of the paper is organized as follows.
Section~\ref{sec:cft} reviews two-point functions in the grand canonical ensemble and their relation to spectral densities. Section~\ref{sec:retarded-green-function} introduces the charged black brane background, the holographic prescription for evaluating the retarded Green's function, and the numerical spectral density calculation. In Section~\ref{sec:two-sided-geometry}, we
relate the leading non-perturbative contribution to complex WKB trajectories and bouncing null geodesics in the extended geometry. Section~\ref{sec:reflection-coefficient} presents the solution to the connection problem between the near-singularity region and the asymptotically AdS region at finite and small charges.
Section~\ref{sec:discussion} concludes with a discussion of the results and
open questions. Several appendices are devoted to recording technical details about analytical and numerical computations.

\section{Two-point functions on the CFT side}
\label{sec:cft}
We begin by introducing the various correlation functions for a grand canonical ensemble of fixed temperature and chemical potential, and study their analytic properties. We then discuss the OPE and enumerate the non-perturbative contributions to two-point functions.

\subsection{Correlators and analyticity}
The purpose of this subsection is to set notations and write observables in a CFT that will be important for the rest of the paper. We also extend the arguments of \cite{Iliesiu:2018fao}~\footnote{The thermal OPE and its holographic realization were studied further in references~\cite{Alday:2020vdk,Karlsson:2022mzg,Esper:2023fiv}.} to the case of the grand canonical ensemble.

Starting with a CFT on $\mathbb{R}_t \times \mathbb{R}^{d-1}$ with a set of R-charges, we consider the grand canonical ensemble where the inverse temperature $\beta$ and a chemical potential $\mu$, conjugate to one of the $R$-charges, are both fixed.   The Wightman two-point functions of an uncharged operator $O$ with scaling dimension $\Delta_{O}$ that we are interested in are given by:
\begin{align}
    G^{>} (t,\vec{x} ) &= \frac{1}{\mathcal{Z}(\beta,\mu)} \text{Tr} \Big( e^{-\beta (H -  \mu Q)}O(t,\vec{x})O(0,0) \Big) ~~,\nonumber\\
    G^{<} (t,\vec{x} ) &= \frac{1}{\mathcal{Z}(\beta,\mu)} \text{Tr} \Big( e^{-\beta (H -  \mu Q)}O(0,0)O(t,\vec{x}) \Big) ~,
\end{align}
where $H$ is the Hamiltonian of the CFT, $Q$ is an R-charge, and $\mathcal{Z}$ is the grand-canonical partition function. 

The retarded Green's function of the uncharged operator $O$ is defined using a commutation relation of $O(t,\vec{x})$ with $O(0,0)$
\begin{equation}
     G_{\text{ret}} (t,\vec{x}) =\ic \theta(t) \frac{1}{\mathcal{Z}(\beta,\mu)} \text{Tr} \Big( e^{-\beta (H -  \mu Q)}[O(t,\vec{x}),O(0,0)] \Big) ~.
     \label{RetadedGreen} 
\end{equation}
In Fourier space, labeled by a frequency $\omega$ and a wave-vector $\vec{k}$, the retarded Green's function is related to $ G_{\text{ret}}(t,\vec{x})$ through
\begin{equation}
     G_{\text{ret}}(\omega,\vec{k} ) = \int \diff^{d-1} x \int_{0} ^{\infty} \diff t ~e^{\ic\omega t -\ic \vec{k}\cdot \vec{x}}  G_{\text{ret}} (t,\vec{x})~.  
     \label{eq:FT}
\end{equation}
The spectral density $\rho(\omega,\vec{k})$ is defined as twice the imaginary part of the retarded Green's function:
\begin{equation}
      \rho(\omega,\vec{k})\equiv 2 \text{Im} \big(G_{\text{ret}} (\omega,\vec{k})\big)~.
      \label{eq:spectral-density-definition}
\end{equation}
The spectral density is related to the Wightman two-point function via a KMS relation:
\begin{equation}
   G^{>} (\omega, \vec{k}) = \frac{\rho(\omega,\vec{k})}{1-e^{-\beta \omega}}~.
\end{equation}
A generalization for a two-point function of charged operators with charges $q$ and $-q$, of the form $O_q(t,\vec{x}) \, O_{-q}(0,0)$, follows from a charged version of the KMS relation: For an operator satisfying $[Q,O_q]=qO_q$, cyclicity of the trace gives rise to following relation:
\begin{equation} 
G^{>}(\omega,\vec{k}) = e^{\beta(\omega-\mu q)}G^{<}(\omega,\vec{k}) \quad \implies \quad
G^{>} (\omega, \vec{k}) = \frac{\rho(\omega,\vec{k})}{1-e^{-\beta (\omega-\mu q)}}~. 
\end{equation}
\subsubsection{Convergence} 
Next, we discuss whether the Wightman two-point function analytically continued to imaginary time $t\to -\ic \tau$ is well defined. The fixed-charge version of the Wightman two-point correlation function with uncharged operators $O$ is
\begin{align}
G^{>}(-\ic\tau,\vec{x}=0)|_Q = \frac{1}{{Z}(\beta,Q)}\text{Tr}_{Q}\Big(e^{-(\beta-\tau) H } O(0,0) e^{-\tau H} O(0,0)  \Big)~,
\label{eq:fixed-charge-two-point}
\end{align} 
where ${Z}(\beta,Q)$ is the partition function of a superselection sector with a fixed charge $Q$. For suitably smeared operators, Eq.~\eqref{eq:fixed-charge-two-point} is well defined in the range $0<\tau<\beta$, provided that the fixed-charge thermal ensemble exists. For local operators, the correlation function should instead be understood as a distribution, with the usual ultraviolet singularities at coincident points. 

Now consider the grand canonical ensemble. The Wightman function analytically continued to imaginary time is 
\begin{align} 
G^{>}(-\ic\tau,\vec{x}=0)|_{\mu} = \frac{1}{\mathcal{Z}(\beta,\mu)} \text{Tr}\Big( e^{-\beta(H-\mu Q)} e^{\tau H}O(0,0)e^{-\tau H}O(0,0) \Big)~. 
\end{align} 
Since the operator $O$ is uncharged, i.e. $[Q,O]=0$, this can be written as 
\begin{align} 
G^{>}(-\ic\tau,\vec{x}=0)|_{\mu} = \frac{1}{\mathcal{Z}(\beta,\mu)} \text{Tr}\Big( e^{-(\beta-\tau)(H-\mu Q)} O(0,0) e^{-\tau(H-\mu Q)} O(0,0) \Big)~. 
\label{eq:grand-canonical-strip} 
\end{align} 
It follows that the two-point function is analytic in the strip $0 < \tau < \beta$. 
\subsection{Comments on the OPE}
For uncharged operators $O$, the Wightman two-point function at imaginary time admits an expansion that follows in part from the isometries of $S^1 _{\beta} \times \mathbb{R}^{d-1}$ \cite{Iliesiu:2018fao}:
\begin{align}
G^{>}(\tau,\vec{x})
=
\sum_{\Delta',\ell}
C_{\ell}^{\left(\frac{d}{2}-1\right)}
\Bigg(
\frac{\tau}{\sqrt{|\vec{x}|^2+\tau^2}}
\Bigg)
\frac{
c_{\Delta',\ell}(\mu)\,
\left\langle \mathcal{O}_{\Delta',\ell}\right\rangle_{\beta,\mu}
}{
\left(|\vec{x}|^2+\tau^2\right)^{\Delta_O-\frac{\Delta'}{2}}
}~,
\label{Green}
\end{align}
where $\Delta'$ and $\ell$ label the scaling dimensions and spins of the exchanged operators $O_{\Delta',\ell}$, $C_{\ell} ^{\left(\frac{d}{2}-1
\right)}(x)$ are Gegenbauer's polynomials, and $c_{\Delta',\ell}$ are constants. 

In the large-$N$ limit, the generalized-free double-trace operators constructed from the external operator $O$ have dimensions 
\begin{equation} 
\Delta_{n,\ell}=2\Delta_O+2n+\ell+O(N^{-2})~. 
\end{equation} 
At large $N$, their contributions to the Euclidean OPE are polynomial after the overall short-distance prefactor is extracted. They therefore have no discontinuity under continuation to Lorentzian signature, and cancel in the commutator of the retarded Green's function. 

Non-vanishing contributions to the retarded OPE come from the identity, the stress tensor, combinations involving the current, and products thereof: $I, J^{\mu}J_{\mu}, J_{\mu}J_{\nu}, T_{\mu\nu}, T_{\mu\nu}J^{\mu}J^{\nu}$ etc. We specialize to zero momentum vector $\vec{k}=0$ 
\begin{align} 
G_{\text{ret}}(t,\vec{k}=0) = \frac{1}{t^{2\Delta_O-d+1}} \sum_{n=0}^{\infty} \sum_{m=0}^{\infty} c_{n,m}\, t^{nd+2m(d-1)}~. 
\label{GreenSeries} 
\end{align}
In the R.H.S. of Eq.~(\ref{GreenSeries}), $n$ counts the powers associated with stress-tensor exchanges (each of which contributes a scaling dimension $d$), whereas $2m$ counts the current insertions (each of which has scaling dimension $d-1$). The coefficients $c_{n,m}$ are generally real functions of temperature and chemical potential.

The Fourier transform of Eq.~(\ref{GreenSeries}) to frequency space has the form
\begin{align} 
G_{\text{ret}}(\omega,\vec{k}=0) \propto \sum_{n=0}^{\infty} \sum_{m=0}^{\infty} c_{n,m} \frac{ \Gamma\big(d(n+2m+1)-2m-2\Delta_O\big) }{ (-\ic\omega)^{d(n+2m+1)-2m-2\Delta_O} }~. \label{GreenSeries2} 
\end{align}
Here and below, the term-by-term Fourier transform is understood up to contact terms and logarithmic terms that arise at special integer powers. The series on the R.H.S. of Eq.~(\ref{GreenSeries2}) is asymptotic due to the Gamma function in the numerator. In principle, one can perform a Borel resummation of the series. We note that for an integer $\Delta_O$ and even $d$, the powers $d(n+2m+1)-2m-2\Delta_O$ are even integers. Apart from logarithmic and contact terms, the corresponding powers of $(-\ic\omega)^{-1}$ are therefore real for real $\omega$. 

The retarded Green's function in general possesses non-analytical log-terms and other ``non-perturbative'' terms that are exponential in $\omega$.  The vacuum pieces of the retarded Green's functions of the current operator and the stress tensor are given by \cite{Afkhami-Jeddi:2025wra}:
\begin{equation}
G^J_{\rm vac}(\omega)\sim {\cal N}_J \,\omega^2
\log\!\left[-\frac{(\omega+\ic 0)^2}{\Lambda^2}\right]~,
\qquad
G^T_{\rm vac}(\omega)\sim {\cal N}_T \,\omega^4
\log\!\left[-\frac{(\omega+\ic 0)^2}{\Lambda^2}\right]~,
\label{eq:cft-vacuum-logarithms}
\end{equation}
where ${\cal N}_J$, ${\cal N}_T$ are overall constants and $\Lambda$ is a UV cutoff.  

The imaginary parts coming from the discontinuities of the vacuum logarithms in Eq.~(\ref{eq:cft-vacuum-logarithms}) are nonzero. They give the vacuum spectral densities, 
\begin{equation}
\rho_{J,\rm vac}(\omega)=\pi\omega^2\quad,\quad
\rho_{T,\rm vac}(\omega)=\pi\omega^4~ ,
\label{eq:vacuum-spectral-densities}
\end{equation}
where we choose the normalizations $\mathcal N$ as written in equation \eqref{eq:vacuum-spectral-densities}. 

 \subsubsection{Non-perturbative contributions}
\begin{figure}[tbp]
\centering
\begin{tikzpicture}[scale=0.7]
  \draw[ ->] (0, -2.35) -- (0, 4.8) (-4.8,0)--(4.8,0);
      \node at (0, 0) [mred, cross out, draw, thick, inner sep=2pt, minimum size=2pt] {};
      \node at (2,2) [mred, cross out, draw, thick, inner sep=2pt, minimum size=2pt] {};
      \node at (-2,2) [mred, cross out, draw, thick, inner sep=2pt, minimum size=2pt] {};
    \node at (2,-2) [mred, cross out, draw, thick, inner sep=2pt, minimum size=2pt] {};
      \node at (-2,-2) [mred, cross out, draw, thick, inner sep=2pt, minimum size=2pt] {};
          \node at (4,-4) [mred, cross out, draw, thick, inner sep=2pt, minimum size=2pt] {};
      \node at (-4,-4) [mred, cross out, draw, thick, inner sep=2pt, minimum size=2pt] {};
               \node at (4,4) [mred, cross out, draw, thick, inner sep=2pt, minimum size=2pt] {};
      \node at (-4,4) [mred, cross out, draw, thick, inner sep=2pt, minimum size=2pt] {};
 \draw[thick, mblue] (2 -0.2,2) arc[start angle=-180, end angle=0, radius=0.2];
   \draw[thick, mblue] (2 -0.2,4.8) --  (2 -0.2,2) ;
      \draw[thick, mblue, ->] (2 +0.2,2) --  (2 +0.2,4.8) ;
      \draw[thick, mblue, ->] (0 +0.2,0) --  ( +0.2,4.8) ;
\draw[thick, mblue] (0, -0.2) arc[start angle=-90, end angle=0, radius=0.2];
 \draw[thick, mblue] (4 -0.2,4) arc[start angle=-180, end angle=0, radius=0.2];
   \draw[thick, mblue] (4 -0.2,4.8) --  (4 -0.2,4) ;
      \draw[thick, mblue, ->] (4 +0.2,4) --  (4 +0.2,4.8) ;
  \draw[thick, mblue, ->] (0.25,0) -- (4.8,0) ;
  \draw[thick, mblue] (0, -0.25) arc[start angle=-90, end angle=0, radius=0.25];
   \draw[decorate,decoration=zigzag, mred, thick] (0,0) --(0,4.8);
 \draw[decorate,decoration=zigzag, mred, thick] (2,2) --(2,4.8);
 \draw[decorate,decoration=zigzag, mred, thick] (-2,-2) --(-2,-4);
  \draw[decorate,decoration=zigzag, mred, thick] (2,-2) --(2,-4);
  \draw[decorate,decoration=zigzag, mred, thick] (-2,2) --(-2,4.8);
\draw[decorate,decoration=zigzag, mred, thick] (4,4) --(4,4.8);
  \draw[decorate,decoration=zigzag, mred, thick] (-4,4) --(-4,4.8);
\node at (6,5) {$t$};
 \draw[] (6.2, 4.8) --  (5.8, 4.8) --(5.8, 5.2);
\end{tikzpicture}
\caption{Instead of performing the Fourier transform of the retarded Green's function in \eqref{eq:FT} for real time, one can deform the contour of integration to the horizontal blue contour in the complex time plane. The location of the non-analyticities in the upper-half-plane,  denoted by $t_\star$, correspond to the time elapsed along the bouncing geodesic, which give the exponentially suppressed weights $e^{\ic \omega t_\star}$, see \eqref{eq:complex-bounce-factor}.  
}
\label{fig:contour}
\end{figure}
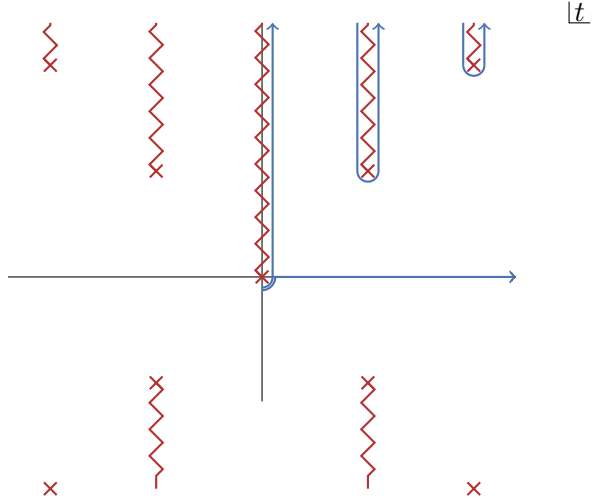

We now present a perspective that non-perturbative contributions to the retarded Green's function are associated with the transseries completion of the perturbative large-frequency expansion encountered above. For large positive frequency, the Fourier transform of the time-domain correlator in Eq.~\eqref{eq:FT} can be deformed into the upper half of the complex boundary-time plane, where the factor $\exp(\mathrm{i}\omega t)$ is exponentially damped (see Fig.~\ref{fig:contour}). When the retarded Green's function has singularities in the upper half of the complex time plane, the deformed contour captures them. As a result, the following contributions to the frequency-domain Green's function arise
\begin{equation}
    \exp\!\left(\mathrm{i}\omega \Delta t_Q\right)
    =
    \exp\!\left(-\lambda_Q\omega\right)
    \exp\!\left(\mathrm{i}\tau_Q\omega\right)~,
    \qquad
    \Delta t_Q\equiv \tau_Q+\mathrm{i}\lambda_Q~,
    \qquad
    \lambda_Q>0~.
    \label{eq:complex-bounce-factor}
\end{equation}
These contributions are non-perturbatively suppressed in $\omega$. See Fig.~\ref{fig:contour} for an illustration of the singularities in complex time plane that give rise to these exponentially-small corrections. The singularities in Fig.~\ref{fig:contour} are precisely the same singularities that one encounters upon a Borel resummation of the asymptotic series.
Non-perturbative contributions to the retarded Green's functions can be parametrized as follows
\begin{equation}
G^s_{\rm np}(\omega,Q)
\sim
\omega^{2s}
e^{-\lambda_Q\omega+\ic\tau_Q\omega}
\left[
C_{s,0}
+C_{s,1}\omega^{-\gamma_s}
+C_{s,2}\omega^{-2\gamma_s}
+\cdots
\right]~,
\label{eq:Gnp-general-structure}
\end{equation}
where $s=1$ is the value relevant for the current and $s=2$ for the transverse stress tensor.  In subsequent sections, we will exploit the gravity side of the AdS/CFT correspondence and use analytical methods to predict parameters in equation~(\ref{eq:Gnp-general-structure}). The exponentially small correction therefore is contained in the vacuum-subtracted spectral density:
\begin{equation}
 \Delta\rho_J=\rho_J-\pi\omega^2~,
 \qquad
 \Delta\rho_T=\rho_T-\pi\omega^4~.
 \label{eq:vacuum-subtracted-rho}
\end{equation}
In particular, using Eq.~(\ref{GreenSeries2}) for an integer $\Delta_O$ and even $d$, we conclude that in the large-$N$ limit, the spectral densities of the current and stress tensor do not receive perturbative contributions in $1/\omega$. 
\section{Retarded Green's functions at large frequency} 
\label{sec:retarded-green-function}

In this section, we review the relevant charged black brane solution in asymptotically AdS, write wave equations for bulk perturbations, and solve them in the exterior region using numerics. We also review a Lorentzian prescription for calculating the retarded Green's function from the bulk solutions, and use it to evaluate spectral densities on the dual CFT  at large real frequency.

\subsection{Charged black brane background}
Our starting point is the Einstein-Maxwell theory in five spacetime dimensions, described by the following action:
\begin{equation} 
S_{\mathrm{background}} = \frac{1}{16\pi G_N} \int \diff ^5x\,\sqrt{-g} \left( R-2\Lambda -\frac{1}{4g_5^2} \mathcal F^{MN}\mathcal F_{MN} \right)~,
\label{action}
\end{equation}
where the background field strength $\mathcal F$ is related to the background gauge field $\mathcal A$ by $\mathcal F=\diff \mathcal A$, and the cosmological constant is $\Lambda= -\frac{6}{\ell_{\rm AdS}^2}$.  Below, we mainly work in units with AdS radius $\ell_{\rm AdS}=1$, and choose the normalization $g_5^2=\frac{1}{12}$. We consider the electrically charged, planar Reissner-Nordstr\"om (RN) AdS$_5$ black brane solution
\begin{equation}
\label{metric}
\diff s^2 = -r^2 f(r)\,\diff t^2 + \frac{\diff r^2}{r^2 f(r)} + r^2 \diff \vec{x}_3^2~,
\qquad
\mathcal A= \Big(\mu - \frac{Q}{2r^2} \Big) \diff t~.
\end{equation}
The blackening function of the planar RN-AdS$_5$ black brane is
\begin{equation}\label{fdef}
f(r) \;=\; 1 - \frac{r_0^4}{r^4} + \frac{Q^2}{r^6}~,
\end{equation}
where $r_0$ and $Q$ are related to the mass and charge density, respectively. 

We next describe the outer and inner horizons of the black brane solution. The following is a summary of a longer text appearing in Appendix~\ref{app:Time}. The horizon equation $f(r)=0$ can be written as a cubic equation in $r^2$, 
\begin{equation} (r^2)^3-r_0^4r^2+Q^2=0~. 
\end{equation} The number of real roots of this cubic depends on the sign of its discriminant. Defining 
\begin{equation} 
q\equiv\frac{3\sqrt{3}Q^2}{2r_0^6}~, 
\end{equation}
the discriminant of the cubic equation is $4-4q^2$. If $q>1$, the discriminant is negative, which implies that $f(r^2)$ has a single real root and two complex conjugate roots.  Physically, a naked singularity appears in the bulk in this regime. We will focus on the opposite regime where $q<1$, in which case the cubic equation in $r^2$ has three real roots: two positive roots corresponding to the outer and inner horizons, and one negative root. Therefore, $f(r^2)$ can be written in terms of real parameters $r_+$ and $r_-$ that correspond to the outer and inner horizon, respectively,
\begin{equation}
  f(r) =\frac{  (r^2 - r_+ ^2) (r^2 - r_-^2) (r^2+r_q ^2)}{r^6} ~,
\end{equation}
where  $r_q ^2$ is defined as follows:
\begin{equation}
 r_q ^2 = r_+ ^2 + r_-^2~.
\end{equation}
In terms of $q$, the scales of the outer and inner horizons are given by
     \begin{equation}
     	r_+ = r_0 \sqrt{\frac{2}{\sqrt{3}} \cos\Big(\frac{\pi - \arccos(q)}{3}\Big)}~~,~~
     	r_- = r_0 \sqrt{\frac{2}{\sqrt{3}} \cos\Big(\frac{\pi + \arccos(q)}{3}\Big)}~. 
     \end{equation}
The horizons merge at extremality when $q=1$. In the uncharged case $q=0$, the inner horizon $r_-(q=0)=0$ disappears, and the outer horizon is at $r_+(q=0)=r_0$, as expected.
The inverse temperature $\beta_+$ can be computed by requiring that the Euclidean continuation of the metric in Eq.~(\ref{metric}) is smooth at $r=r_+$: 
	\begin{equation}
    \label{betap}
		\beta_+ = \frac{4\pi}{r_+^2 f'(r_+)}=\frac{2\pi r_+ ^3}{ (r_+ ^2 - r_-^2) (2r_+ ^2 + r_- ^2)}~.
	\end{equation}
	Similarly, one can associate an inverse temperature to the inner horizon
		\begin{equation}
        \label{betam}
		\beta_- =\frac{4\pi}{r_-^2 f'(r_-)}=  -\frac{2\pi r_- ^3}{ (r_+ ^2 - r_-^2) (2r_- ^2 + r_+ ^2)}~.
	\end{equation}
    Note that in general $\beta_-<0$.
Imposing regularity of the Euclidean continuation of the gauge field $\mathcal A$ in Eq.~(\ref{metric}) at the outer horizon determines the chemical potential: $\mu = \frac{Q}{2r_+^2}.$ The Penrose diagram of the solution consists of infinitely many copies, a subset of which are depicted in Fig~\ref{fig:penrose}. 

\subsection{Wave equations for bulk fields}
\label{sec:son-starinets}
We first review the wave equations of bulk perturbations in the black brane background. 
We consider a scalar field $\phi$, a gauge field $A_M$, and a graviton perturbation $h_{MN}$. These fields are dual to a scalar operator $\mathcal O$, a conserved current $J_\mu$, and the stress-energy tensor $T_{\mu\nu}$, respectively. 

We consider a gauge field perturbation along the $x$ direction, $A_x$, and a transverse spin-two metric perturbation along the $x-y$ directions of $\vec{x}_3$, $h_{xy}$. In general, the perturbation equations couple to each other; however, we explain below that in the special case where the perturbations do not carry momentum along the $\vec{x}_3$ directions, and by picking a gauge choice that includes $g_{rx}=0$, one can write a decoupled equation for $A_x$.

Writing the fields as plane waves with time dependence $e^{-\ic\omega t}$ with vanishing momenta vectors along the $\vec{x}_3$ dimensions, the radial parts of the wave equations that govern the perturbations are given by
\begin{align}
 {}&\phi''+
 \left(\frac{5}{r}+\frac{f'}{f}\right)\phi'
 +\left(\frac{\omega^2}{r^4f^2}
 -\frac{m^2}{r^2f}\right)\phi=0~, 
 \label{eq:scalar-r}\\
 {}&A_x''+
 \left(\frac{3}{r}+\frac{f'}{f}\right)A_x'
 +\left(\frac{\omega^2}{r^4f^2}-
\frac{12\,Q^2}{r^8f} \right)A_x=0~, 
 \label{eq:vector-r}\\
 {}&h_{xy}''+
 \left(\frac{5}{r}+\frac{f'}{f}\right)h_{xy}'
 +\frac{\omega^2}{r^4f^2}
 h_{xy}=0~. 
 \label{eq:tensor-r}
\end{align}
Here, primes denote derivatives with respect to $r$.  The scalar field equation follows from the conventional Klein-Gordon equation in the curved geometry~(\ref{metric}). The tensor field equation is identical to the massless scalar field equation. An explanation for the gauge field equations is as follows.  Linearising the Einstein--Maxwell equations that follow from Eq.~(\ref{action}), the $r$-$x$
component of the Einstein's equations and the $x$ component of the gauge field equations are:
\begin{align}
  &\frac{\ic\omega}{2r^5f}\Big(r^5\,{h_t^{\;x}}'+12Q\,A_x\Big)=0~, \nonumber \\
  &\big(r^3fA_x'\big)'+\frac{\omega^2}{rf}A_x+Q\,{h_t^{\;x}}'=0~. 
  \label{gaugemix}
\end{align}
For $\omega\neq0$, the first line is a first-order constraint,
\begin{equation}
  {h_t^{\;x}}'=-\frac{12Q}{r^5}A_x~ ,
  \label{constraint}
\end{equation}
and Eq.~(\ref{eq:vector-r}) then follows from combining Eqs.~(\ref{constraint}) and the second line of (\ref{gaugemix}) into a single equation for the gauge field. 

We impose a purely ingoing boundary condition at the outer horizon, written as follows
\begin{equation}
 \Phi(r;\omega,\vec{k})\sim
 (r-r_+)^{-\ic\omega\beta_+/(4\pi)},
 \qquad r\rightarrow r_+~.
 \label{eq:retarded-infalling-condition}
\end{equation}
Defining a tortoise coordinate via 
\begin{equation}
 \diff r_*=\frac{\diff r}{r^2f(r)}~, 
 \label{tortoise}
\end{equation}
the wave behaves as $e^{-\ic\omega(t+r_*)}$ near the outer horizon, corresponding to the purely incoming boundary condition.

Several subsequent calculations are more convenient using the variable
\begin{equation}
    u = \frac{r_+^2}{r^2}~,
\end{equation}
which ranges from $0$ to $1$ in the black brane exterior. We introduce a dimensionless frequency: 
\begin{equation} 
\widehat\omega = \frac{\omega_{}}{2r_+}~. 
\label{eq:dimensionless-frequency-momentum} 
\end{equation} 
The $r$-coordinate equations above are written in terms of the physical frequency. The $u$-coordinate equations and the numerical analysis below are written in terms of $\widehat\omega$.  From this point onward, in the $u-$coordinate, all frequencies are expressed in units of $2r_+$, and we suppress the hats. The quantities $\tau_Q$ and $\lambda_Q$ are rescaled to the corresponding conjugate units. Likewise, the spectral densities will be normalized using these dimensionless frequency units.

In terms of $u$, the equations (\ref{eq:scalar-r})-(\ref{eq:tensor-r}) take the form
\begin{equation}
    \phi''(u) + \left( \frac{f'(u)}{f(u)} - \frac{1}{u} \right) \phi'(u) + \frac{\omega^2 }{u f(u)^2} \, \phi(u) - \frac{m^2}{4 u^2 f(u)} \, \phi(u) = 0 ~,
\end{equation}
\begin{equation}
 A_x''(u)+\frac{f'(u)}{f(u)}A_x'(u)
 +\Big(\frac{\omega^2}{u f(u)^2}-\frac{3Q^2u}{f(u)}\Big)A_x(u)=0~,
 \label{eq:current-u-equation}
\end{equation}
\begin{equation}
 h_{xy}''(u)+\left(\frac{f'(u)}{f(u)}-\frac{1}{u}\right)h_{xy}'(u)
 +\frac{\omega^2}{u f(u)^2}h_{xy}(u)=0~.
 \label{eq:tensor-u-equation}
\end{equation}
Near the AdS boundary, the gauge field and graviton in transverse directions can be expanded as follows 
\begin{equation}
 A_x(u)=A_x^{(0)}
 +u\left(A_x^{(2)}+\widetilde A_x^{(2)}\log u\right)+\cdots~,
 \label{eq:current-boundary-expansion}
\end{equation}
\begin{equation}
 h_{xy}(u)=h_{xy}^{(0)}+u h_{xy}^{(2)}
 +u^2\left(h_{xy}^{(4)}+\widetilde h_{xy}^{(4)}\log u\right)+\cdots
 \label{eq:tensor-boundary-expansion}
\end{equation}
The leading coefficients $A_x^{(0)}$ and $h_{xy}^{(0)}$ are the corresponding boundary sources.
For any of these fields, denoted collectively by $\Phi$, we separate the boundary source $\Phi_{(0)}$ from a normalized radial solution that we refer to as $F(u;\omega,\vec{k})$,
\begin{equation}
 \Phi(u;\omega,\vec{k})
 =\Phi_{(0)}(\omega,\vec{k})F(u;\omega,\vec{k})~,~
 \qquad
 \lim_{u\rightarrow0}F(u;\omega,\vec{k})=1~.
 \label{eq:bulk-to-boundary-propagator}
\end{equation}
\subsection{AdS/CFT prescription for the retarded Green's function}
This subsection reviews a prescription for evaluating the retarded Green's function from the solutions to the bulk perturbation equations~\cite{Son:2002sd,Policastro:2001yc,Policastro:2002se}.
We define canonical radial momenta in terms of the bulk action $S_{\mathrm{bulk}}$ for perturbations:
\begin{equation}
 \Pi_\Phi(u;\omega,\vec{k})
 =\frac{\delta S_{\mathrm{bulk}}}{\delta(\partial_u\Phi)}~.
\end{equation}
The radial momentum is used to calculate the response per unit boundary source $\Phi_{(0)}$, which is given by:
\begin{equation}
 \mathcal{G}(u;\omega,\vec{k})
 =\frac{\Pi_\Phi(u;\omega,\vec{k})}{\Phi_{(0)}(\omega,\vec{k})}~.
 \label{eq:radial-response}
\end{equation}
We refer to the function $\mathcal{G}$ as the response kernel. The imaginary parts of the responses are obtained from radial, conserved fluxes.
Up to positive normalization factors inherited from the quadratic actions, the conserved Wronskians in the outward-increasing $r$ coordinate are 
\begin{equation} 
\mathcal F_J^{(r)} \propto r^3f(r)\, {\rm Im}\!\left[ A_x^*(r)\partial_rA_x(r) \right]~, \qquad \mathcal F_T^{(r)} \propto r^5f(r)\, {\rm Im}\!\left[ h_{xy}^*(r)\partial_rh_{xy}(r) \right]~.
\label{Wronsk}
\end{equation} 
In the variable $u=r_+^2/r^2$, Eq.~(\ref{Wronsk}) becomes 
\begin{equation} 
\mathcal F_J^{(r)} \propto -f(u)\, {\rm Im}\!\left[ A_x^*(u)\partial_uA_x(u) \right], \qquad \mathcal F_T^{(r)} \propto -\frac{f(u)}{u}\, {\rm Im}\!\left[ h_{xy}^*(u)\partial_uh_{xy}(u) \right]~, 
\label{eq:tensor-flux} 
\end{equation} 
where powers of $r_+$ and the overall action normalizations have been suppressed.  The two-point functions are related to the responses above in the limit $u\to0$.
To leading order, the on-shell action of the perturbed solution is expressible in terms of the boundary sources and the kernel, 
\begin{equation} 
S_{\mathrm{bulk}}^{(2)}(u_c) = \frac{1}{2} \int\frac{\diff\omega\,\diff^3k}{(2\pi)^4}\, \Phi_{(0)}(-\omega,-\vec{k}) \mathcal{G}(u_c;\omega,\vec{k}) \Phi_{(0)}(\omega,\vec{k})~, 
\label{eq:on-shell-boundary-action} 
\end{equation}
where $u_c$ is a radial cutoff near the asymptotic boundary, with $u_c\rightarrow0$. The retarded Green's function is then the renormalized boundary value of the response kernel:
\begin{equation} 
G_{\text{ret}}(\omega,\vec{k}) = \lim_{u_c\rightarrow0} \left[ \mathcal{G}(u_c;\omega,\vec{k}) +G_{\mathrm{ct}}(u_c;\omega,\vec{k}) \right]~. 
\end{equation}
Here, $G_{\mathrm{ct}}$ contains the local holographic counterterms that remove ultraviolet divergences.

\subsection{Evaluation of the spectral density}
\label{sec:numerical-spectral-density}
This subsection presents a summary of a numerical computation of the spectral density of two types of operators: (a) the transverse current and (b) the transverse stress tensor, both evaluated at zero spatial momentum $\vec{k}=0$.  The calculation of the Green's function is presented first; the spectral density follows from it.

The practical prescription is to solve the appropriate wave equation with infalling horizon data, normalize the solution by its boundary source, and evaluate the imaginary part of its canonical response at a small-$u_c$ cutoff surface near the asymptotic boundary.

Since the conserved radial flux is independent of $u$, the spectral density can be evaluated directly from the horizon flux divided by the squared magnitude of the boundary source. This form is useful numerically: 
\begin{equation} 
\rho_s(\omega,Q) \propto \frac{\mathcal F_s} {|\Phi_{s,(0)}(\omega,Q)|^2}~,~ \qquad s=J,T~.
\end{equation} 
Here, the proportionality constants are fixed by the quadratic bulk action and by the vacuum normalization.

Physically, the infalling boundary condition selects the causal response and measures absorption by the outer horizon. At large frequency, the wave propagates almost geometrically and the spectral density approaches its vacuum value. The charged geometry nevertheless produces an exponentially small  contribution, which interferes with the leading infalling solution thereby producing the damped oscillations visible only after subtracting the much larger vacuum contribution.

The numerical analysis is as follows. Firstly, we subtract the vacuum spectral density, which isolates the exponentially small tail. Then, we multiply by $e^{\lambda_Q\omega}$, which removes the exponential envelope. The remaining oscillation tests the real propagation time $\tau_Q$. Finally, subtracting successive terms in the fractional inverse-frequency expansion tests Eq.~(\ref{eq:Gnp-general-structure}), which we copy below for convenience: 
\begin{equation}
G^s_{\rm np}(\omega,Q)
\sim
\omega^{2s}
e^{-\lambda_Q\omega+\ic\tau_Q\omega}
\left[
C_{s,0}
+C_{s,1}\omega^{-\gamma_s}
+C_{s,2}\omega^{-2\gamma_s}
+\cdots
\right]~.
\label{eq:Gnp-general-structureV2}
\end{equation}
In the next two sections, the parameters in this equation - the damping exponent, oscillation period, and slowly varying amplitude, will be accounted for using a semiclassical picture.

Setting $r_+=1$ and using the $u$-coordinate, the blackening factor is
\begin{equation}
 f(u)=1-(1+Q^2)u^2+Q^2u^3~.
\end{equation}
We now define a variable $\eta$ through
\begin{equation}
    \eta=1-u~,
\end{equation}
which we use to write the infalling solution is as a series of ``$N_h$'' terms near the outer horizon 
\begin{equation}
 \Phi(\eta)=\eta^\alpha\sum_{n=0}^{N_h}a_n \eta^n~,
 \qquad \alpha=-\frac{\ic\omega}{2-Q^2}~.
 \label{eq:numerical-horizon-series}
\end{equation}
The near-horizon series is evaluated at a small positive value $\eta=\epsilon\ll1$ and supplies the initial value of the field and its first derivative for radial integration towards the AdS boundary at $\eta=1$. The overall normalization of the ingoing solution is arbitrary, so we set $a_0=1$. The remaining coefficients $a_n$ are fixed recursively by substituting Eq.~\eqref{eq:numerical-horizon-series} into the corresponding radial equation.  
Near $u=0$, the numerical solution is matched to the boundary expansions in Eqs.~\eqref{eq:current-boundary-expansion} and \eqref{eq:tensor-boundary-expansion}, which determine the source and response coefficients.

The extraction becomes increasingly delicate at large frequency because the exponentially small signal is obtained by subtracting the vacuum contribution from two quantities that agree to many significant digits. We therefore check stability under changes of $N_h$, the starting point $\epsilon$, the radial integration precision, the boundary fitting window, and the frequency interval used in the asymptotic fit. Further details are given in Appendix~\ref{app:fixed-charge-numerics}.

\paragraph{Large-frequency fitting ansatz:} As discussed in Section~\ref{sec:cft}, the perturbative inverse-frequency corrections in the $k=0$ channels considered here are real on the positive-frequency axis, apart from the imaginary factor appearing in the discontinuity of the vacuum logarithmic term. We analyze the vacuum-subtracted quantities:
\begin{equation} 
\Delta\rho_J(\omega,Q) = \rho_J(\omega,Q)-\pi\omega^2~, \qquad \Delta\rho_T(\omega,Q) = \rho_T(\omega,Q)-\pi\omega^4~. \label{eq:numerical-vacuum-subtracted-rho}
\end{equation} Their leading large-frequency behavior is parametrized as 
\begin{equation} 
\Delta\rho_s(\omega,Q) \sim 2\rho_{s,\mathrm{vac}}(\omega) e^{-\lambda_Q\omega} {\rm Re}\!\left[ \mathcal C_s(\omega,Q)e^{\ic\tau_Q\omega} \right], \qquad s=J,T~. 
\label{eq:observed-wkb-form}
\end{equation} 
Here, $\Delta t_Q=\tau_Q+\ic\lambda_Q$. The real part $\tau_Q$ determines the oscillation scale, while $\lambda_Q>0$ determines the exponential damping scale.

\paragraph{Leading order:} The first non-vacuum contribution to the spectral density is non-perturbative in large real frequencies.  Its leading term takes the form in \eqref{eq:observed-wkb-form}.

\begin{figure}[tbp]
    \centering
    \includegraphics[width=0.75\textwidth]{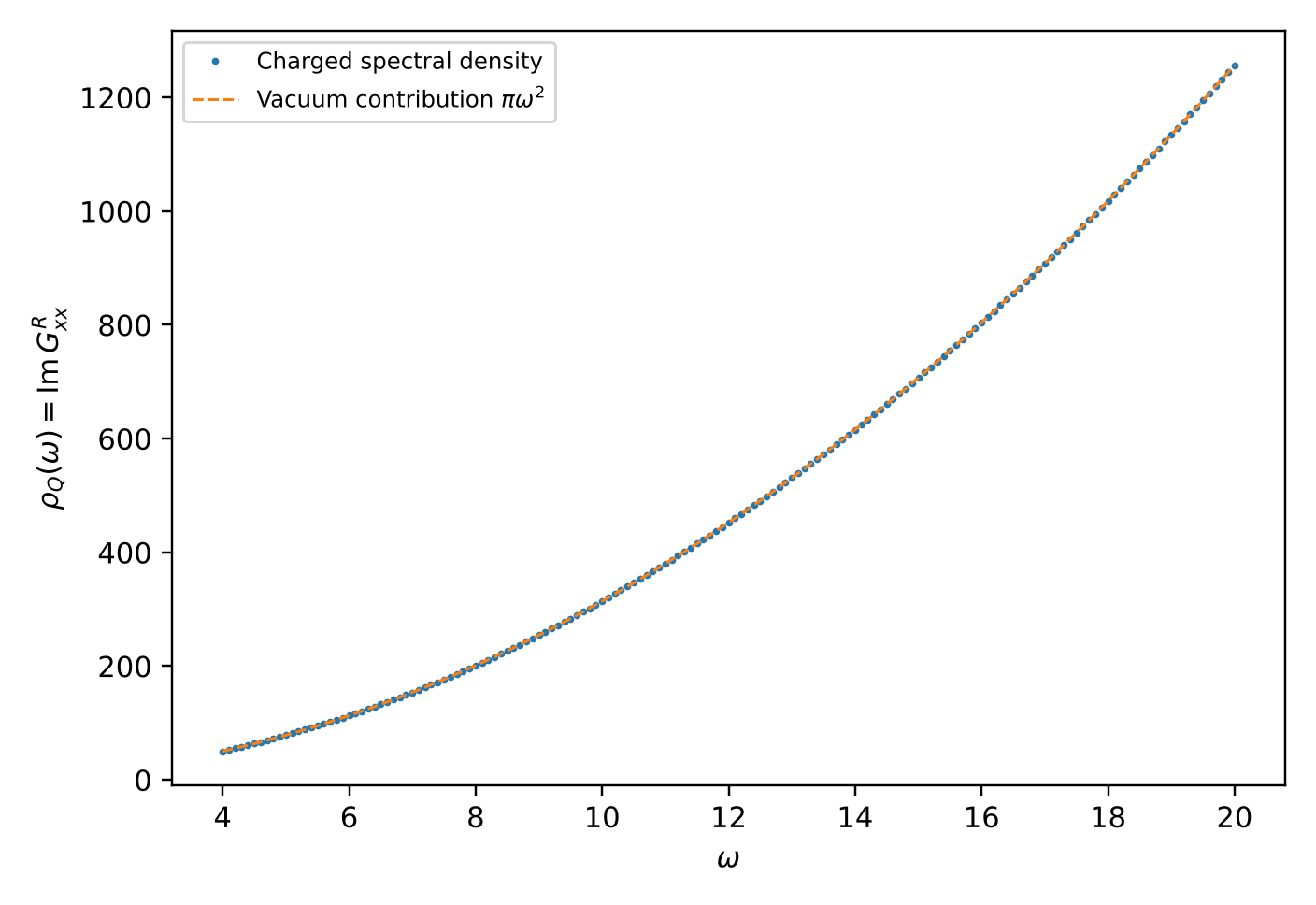}
        \includegraphics[width=0.75\textwidth]{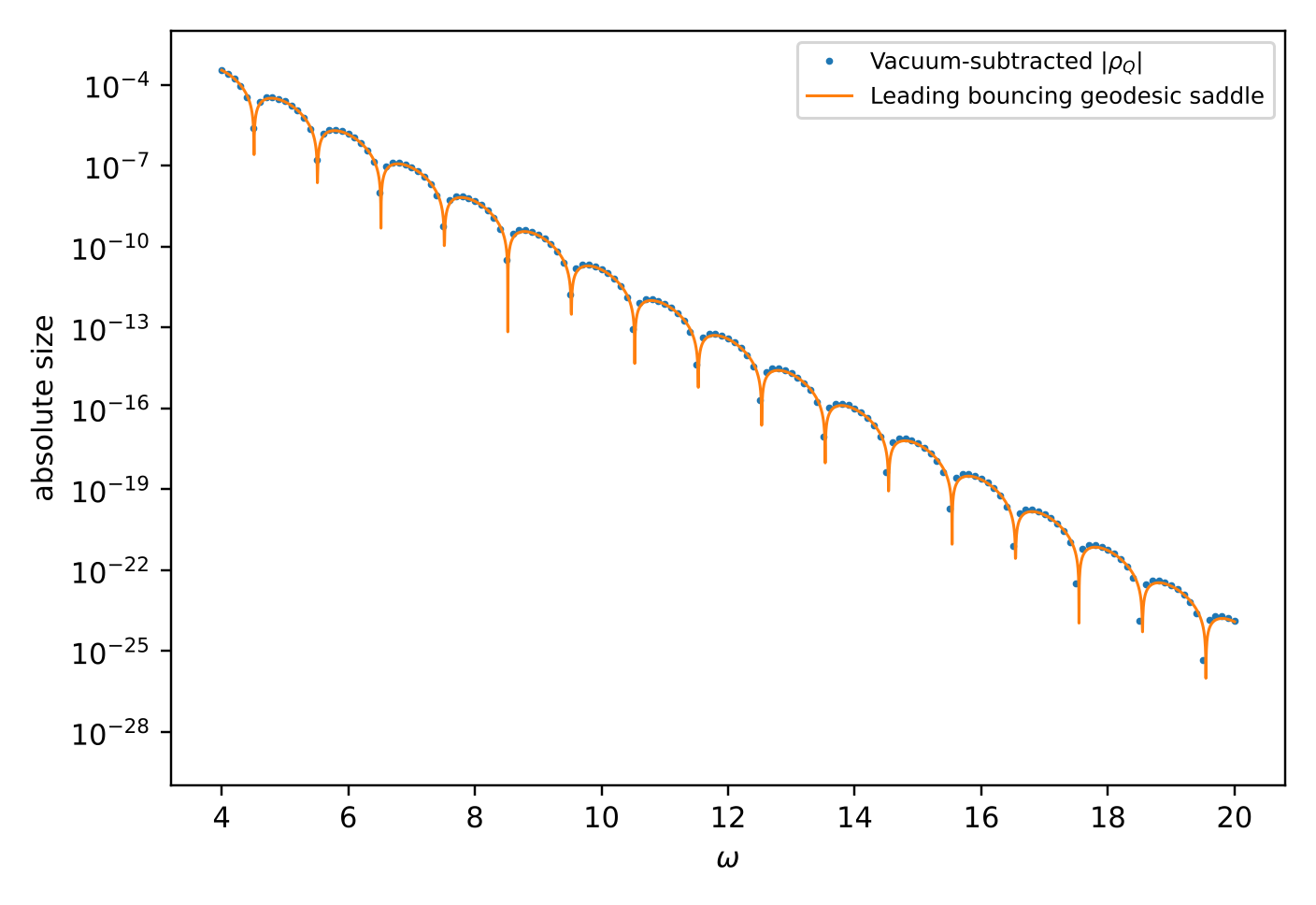}
        \caption{Top: We plot the current spectral density vs.~normalized frequency for the charge parameter $Q=0.05$, compared with the vacuum spectral density. The two curves are nearly indistinguishable on this scale. Bottom: Depicted is the vacuum-subtracted current spectral density against normalized frequency, revealing the exponentially damped oscillations. Analogous oscillations in the uncharged case were observed in \cite{Teaney:2006nc} and studied further in \cite{Myers:2007we,Romatschke:2015gic}; their interpretation in terms of complex bouncing trajectories was developed in \cite{Afkhami-Jeddi:2025wra}.}
    \label{fig:small-charge-current} 
\end{figure}
In particular, we fit the amplitudes using:
\begin{align}
C_J&=C_{J,0}+C_{J,1}\,\omega^{-4/5}+C_{J,2}\,\omega^{-8/5}+C_{J,3}\,\omega^{-12/5}+\cdots~,
  \label{eq:current}
\end{align}
\begin{equation}
     \mathcal{C}_T =C_{T,0}
 +C_{T,1}\omega^{-4/5}+C_{T,2}\omega^{-8/5}+\cdots
 \label{eq:st}
\end{equation}
A physical interpretation of this ansatz in terms of a WKB prescription that involves null geodesics in the classical black brane interior will be explained in section~\ref{sec:reflection-coefficient}.  
 We use Eqs.~(\ref{eq:current}) and (\ref{eq:st}) to obtain the following expressions:
\begin{align}
\frac{\e^{\lambda_Q\omega}\big[\rho_J(\omega,Q)-\pi\omega^2\big]}{2\pi\omega^2}
  &\sim\mathrm{Re}\Big[\e^{\ic\tau_Q\omega}\big(C_{J,0}+C_{J,1}\omega^{-4/5}+C_{J,2}\omega^{-8/5}+\cdots\big)\Big]~,
 \label{eq:charged-current-fit-ansatz}\\
 \frac{\e^{\lambda_Q\omega}
 [\rho_T(\omega,Q)-\pi\omega^4]}{2\pi\omega^4}
 &\sim
 {\rm Re}\!\left[
 \e^{\ic\tau_Q\omega}
 \left(
 C_{T,0}+C_{T,1}\omega^{-4/5}
 +C_{T,2}\omega^{-8/5}+\cdots
 \right)
 \right]~,
 \label{eq:charged-tensor-fit-ansatz}
\end{align}
where the leading coefficients $C_{s,0}$ are predictions rather than fit constraints. In the numerical analysis, all coefficients, including $C_{J,0}$ and $C_{T,0}$ are first fitted freely, which in turn are observed
to be close to the analytical predictions of subsequent sections,  $2\cos(4\pi/5)$, for both channels.
The powers shown in Eqs.~\eqref{eq:charged-tensor-fit-ansatz} describe the leading exponential correction.  Terms proportional to $\e^{-2\lambda_Q\omega}$ and higher exponentials are in general present, but determining precisely their numerical coefficients is beyond the scope of this paper.

\subsubsection{Some explicit finite charge results}
\label{app:fixed-charge-comparison}
Here we summarize a subset of results detailed in Appendix \ref{app:fixed-charge-numerics}, where the spectral densities of the
current channel and the stress tensor channel are evaluated numerically, and take the form of Eq.~(\ref{eq:Gnp-general-structure}). For the current operator, we employ the expansion \eqref{eq:current}, whereas for the stress tensor, we employ Eq.~(\ref{eq:st}).

In particular, for the $Q=1.2$ data shown below, the leading coefficients take the form:
\begin{equation}
 C_{J,0}=-1.61944-0.00088\ic ~~,\qquad C_{T,0}=-1.61054+0.00361886\,\ic ~.
\end{equation}
In section~\ref{sec:reflection-coefficient} and appendices \ref{app:2} and \ref{app:local-to-boundary-matching}, we provide analytical explanations for these answers in terms
of solutions to the wave equations along a contour in the complex $r$-plane, and interpret them as coming from reflection coefficients off the timelike singularity.

\begin{figure}[tbp]
    \centering
    \includegraphics[width=0.8\textwidth]{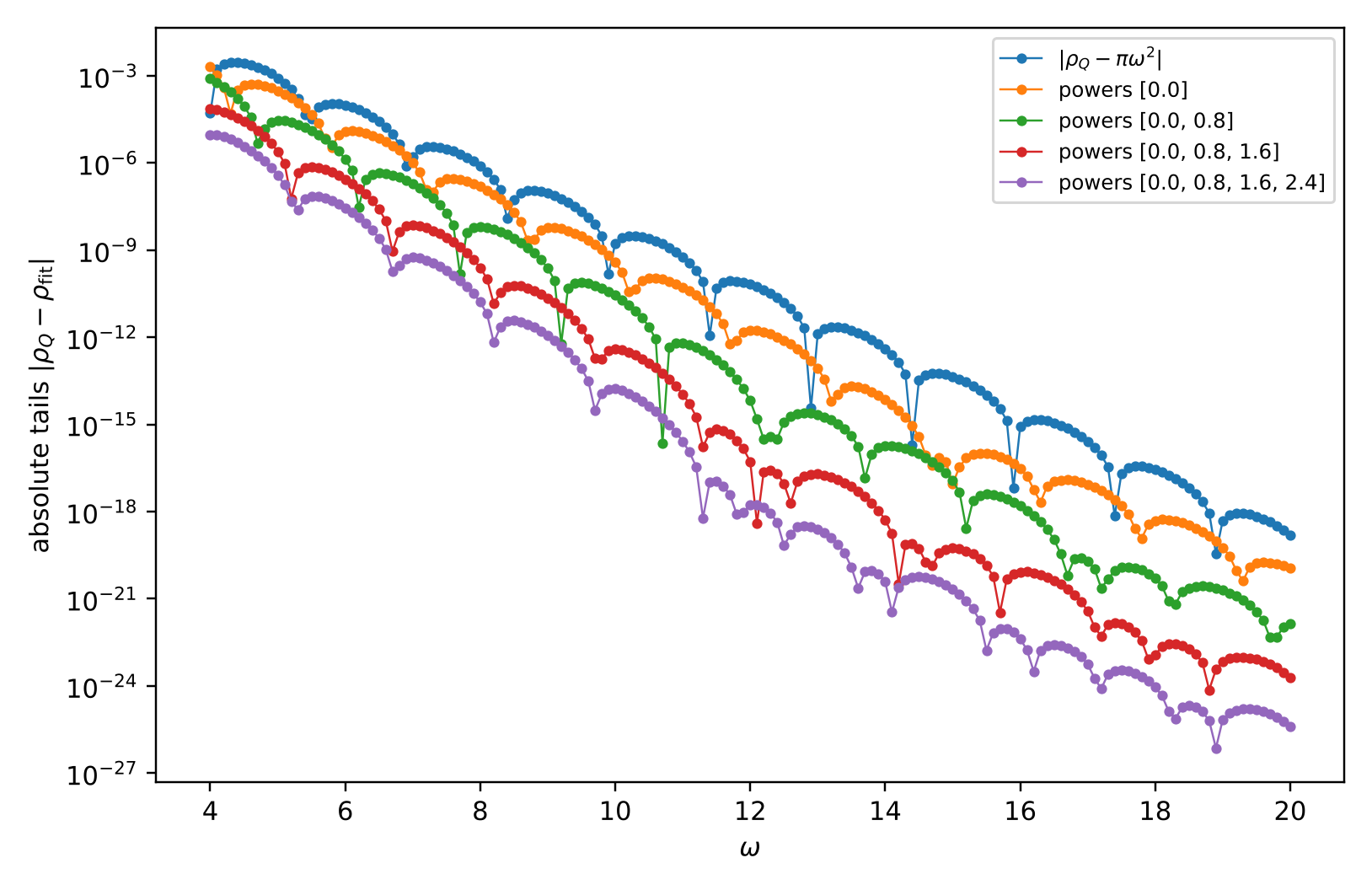}
    \caption{For the current channel, we perform successive extraction of the large-frequency current tail at $Q=1.2$. After subtracting the vacuum contribution and removing the leading exponential envelope, successive subtraction of the fractional-power terms isolates the residual asymptotic corrections. Note that the subtractions become increasingly better at larger frequencies.}
    \label{fig:finite-charge-current-tails} 
\end{figure}
\begin{figure}[tbp]
    \centering
    \includegraphics[width=0.8\textwidth]{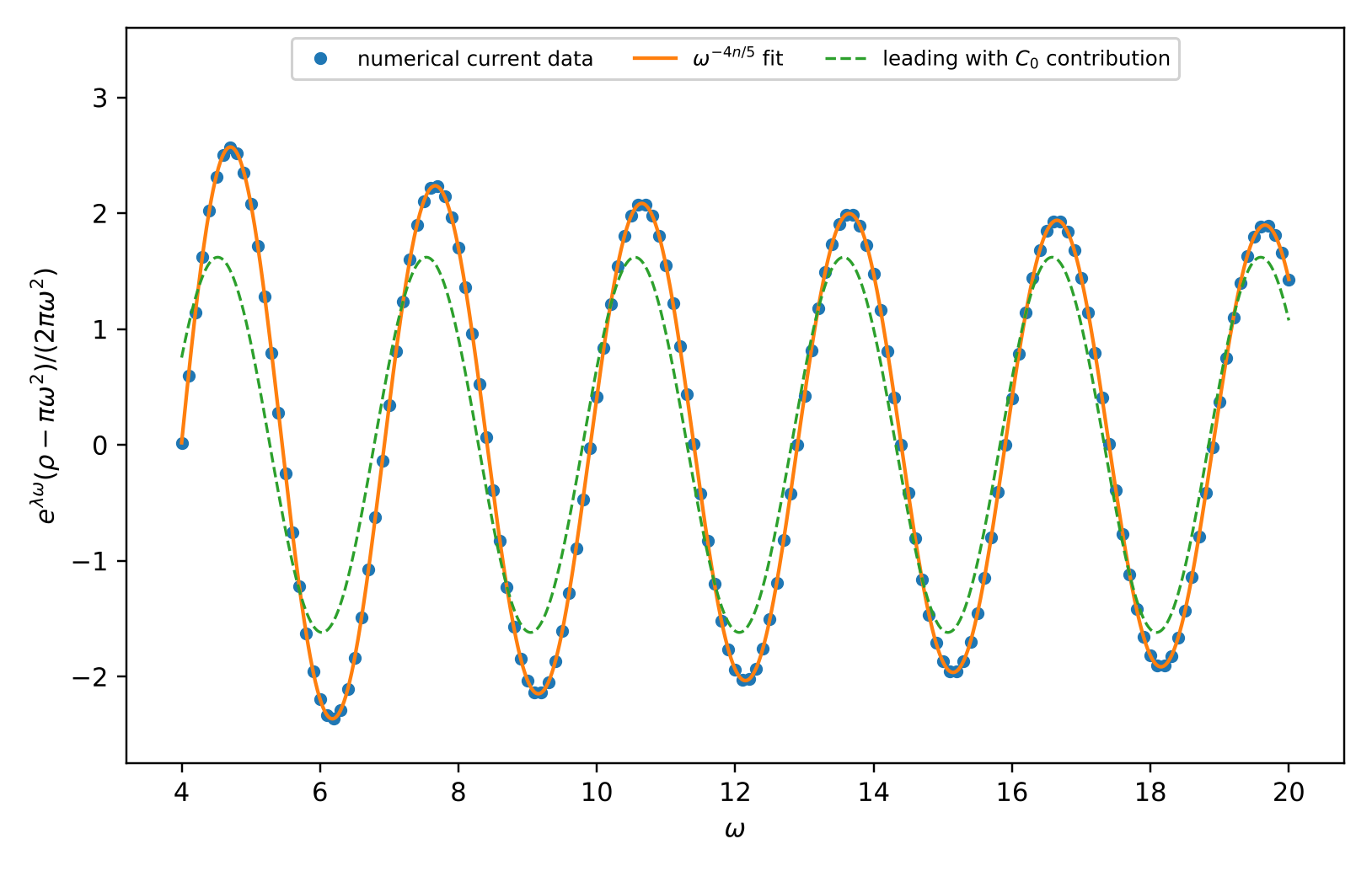}
    \caption{Rescaled current spectral density at $Q=1.2$ after removing the predicted exponential damping. The oscillation period tests $\tau_Q$, while the slowly varying amplitude tests the connection coefficients $C_{J,n}$.}
    \label{fig:finite-charge-current-fit}
\end{figure}

\begin{figure}[h!]
    \centering
    \includegraphics[width=0.8\textwidth]{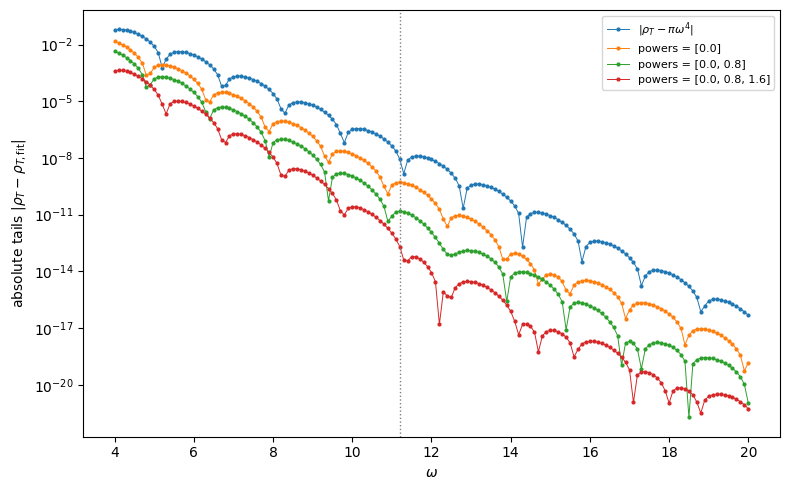}
    \caption{Successive extraction of the large-frequency stress-tensor tail at $Q=1.2$. The residuals after successive subtractions test the fractional-power expansion of the tensor connection coefficient.}
    \label{fig:finite-charge-stress-tails}
\end{figure}
\begin{figure}[h!]
    \centering
    \includegraphics[width=0.8\textwidth]{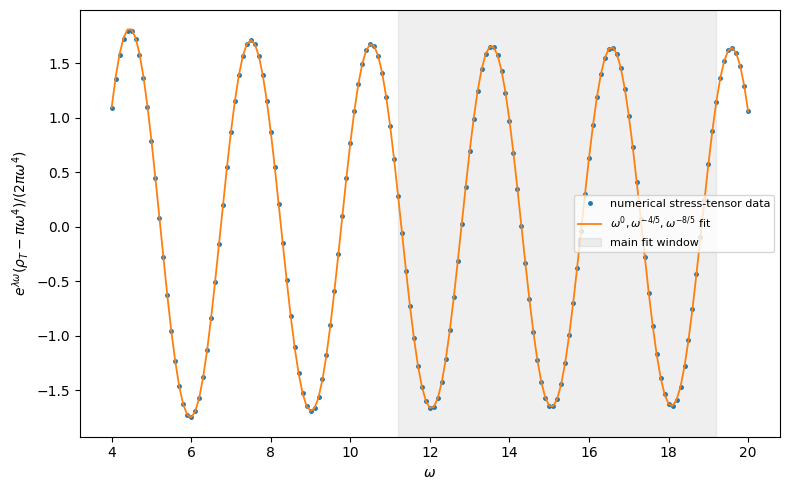}
    \caption{Rescaled stress-tensor spectral density at $Q=1.2$ after removing the predicted exponential damping. The oscillatory phase tests the complex propagation time, while the amplitude tests the tensor connection coefficient.}
     \label{fig:finite-charge-stress-fit}  
\end{figure}
\newpage

\section{Analytic continuation to the extended two-sided geometry}
\label{sec:two-sided-geometry}

The contributions to the retarded Green's functions studied in Section~\ref{sec:retarded-green-function} are obtained solely from the exterior region of the charged black brane. In this section, we show that the exponentially small contributions to these Green's functions can be explained in terms of complex bulk trajectories that probe an analytic continuation of the black brane interior.

\subsection{The WKB approximation}
\label{sec:complex-wkb-solutions}
This subsection uses the WKB approximation to express bulk perturbations at high frequency. 

The bulk wave equations for a scalar perturbation, a transverse-component of a gauge field perturbation, and transverse-transverse graviton, Eqs.~\eqref{eq:scalar-r}-\eqref{eq:tensor-r}, can be rewritten in Schr\"odinger form. Defining
\begin{equation}
    \phi=\frac{\psi_0(r)}{r^{3/2}}~,
    \qquad
    A_x=\frac{\psi_1(r)}{r^{1/2}}~,
    \qquad
    h_{xy}=\frac{\psi_2(r)}{r^{3/2}}~.
    \label{eq:schrodinger-field-redefinitions}
\end{equation}
In terms of the tortoise coordinate defined in Eq.~(\ref{tortoise}), the radial equations take the form
\begin{equation}
    -\frac{\diff^2\psi_s}{\diff r_*^2}
    +V_s(r)\psi_s
    =
    \omega^2\psi_s~,
    \qquad
    s=0,1,2~.
    \label{eq:schrodinger-radial-equation}
\end{equation}
where we use $s$ to label the spin of the fields. The effective potentials governing these fields are
\begin{align}
    V_0(r)
    ={}&
    r^2f(r)
    \left(
    \frac{15}{4}+m^2+\frac{9r_0^4}{4r^4}
    -\frac{21Q^2}{4r^6}
    \right)~,
    \label{eq:scalar-effective-potential}\\
      V_1(r)=&r^2f(r)\left(\frac34+\frac{5r_0^4}{4r^4}+\frac{39\,Q^2}{4r^6}\right)~, 
    \label{eq:current-effective-potential}\\
    V_2(r)
    ={}&
    r^2f(r)
    \left(
    \frac{15}{4}+\frac{9r_0^4}{4r^4}
    -\frac{21Q^2}{4r^6}
    \right)~.
    \label{eq:tensor-effective-potential}
\end{align}
So far, the equations are exact in the context of the quantum field theory on curved spacetime approach~\footnote{The explicit form of $f(r)$ for the charged black brane has been substituted in Eqs. (\ref{eq:scalar-effective-potential})-(\ref{eq:tensor-effective-potential}).}.  At high energies, 
\begin{equation}
    m,\ |\vec{k}|,\ r_0,\ Q^{1/3}\ll\omega~,
    \label{eq:high-frequency-wkb-regime}
\end{equation}
the WKB framework approximates the radial equations,  reducing them to a Hamilton-Jacobi equations to leading order. The WKB expressions are given by
\begin{equation}
    \psi_s(r)
    \approx
    \frac{1}{\sqrt{\omega}}
    \left(1-\frac{V_s(r)}{\omega^2}\right)^{-1/4}
    \exp\!\left[
    \pm \mathrm{i}\omega
    \int^r\mathcal P_s(r')\,\diff r'
    \right]~,
    \label{eq:wkb-ansatz-P}
\end{equation}
where we have defined
\begin{equation}
    \mathcal P_s(r)
    \equiv
    \frac{1}{r^2f(r)}
    \sqrt{1-\frac{V_s(r)}{\omega^2}}~.
    \label{eq:wkb-momentum}
\end{equation}
We evaluate $\mathcal{P}_s(r)$ in the regime of large frequency and obtain the “WKB phase” in the exponential of Eq. (\ref{eq:wkb-ansatz-P}):
\begin{equation}
    \int^r
    \frac{\sqrt{\omega^2-V_s(r')}}{r'^2f(r')}
    \,\diff r'
    =
    \omega\int^r\frac{\diff r'}{r'^2f(r')}
    -\frac{1}{2\omega}
    \int^r\frac{V_s(r')}{r'^2f(r')}\,\diff r'
    \mathcal~+~O(\omega^{-3})~.
    \label{eq:wkb-action-expansion}
\end{equation}
An intermediate conclusion is that the time along radial null geodesic determines the leading-order phase of bulk perturbations in the black brane background.  

In principle, Eq.~(\ref{eq:wkb-ansatz-P}) provides us with a WKB approximation for the bulk perturbations at any real $r\geq 0$. However, when restricting to real $r$, the integrand has singularities at the horizons $r_{+}$ and $r_-$, and furthermore, as we will explain, the WKB approximation breaks down sufficiently near the singularity $r=0$.

\subsubsection{A contour for the retarded Green's function}
In order to obtain a well-defined bulk solution that will give rise to the non-perturbative contributions
to the spectral density obtained in the previous section, we evaluate the bulk fields
along contour ``$\mathcal{C}$'' in complex $r$ plane. Here we define a contour that will be shown to complete this objective.

Before defining this path, it is useful to introduce a complex phase function that features in the leading WKB phase at large frequency, in Eq.~(\ref{eq:wkb-action-expansion}):
\begin{equation}
    \mathcal{S}(r)
    \equiv
    \int^r\frac{\diff r'}{r'^2f(r')}~.
    \label{eq:complex-wkb-time-function}
\end{equation}
\begin{figure}[tbp]
    \centering
    \includegraphics[width=0.8\textwidth]{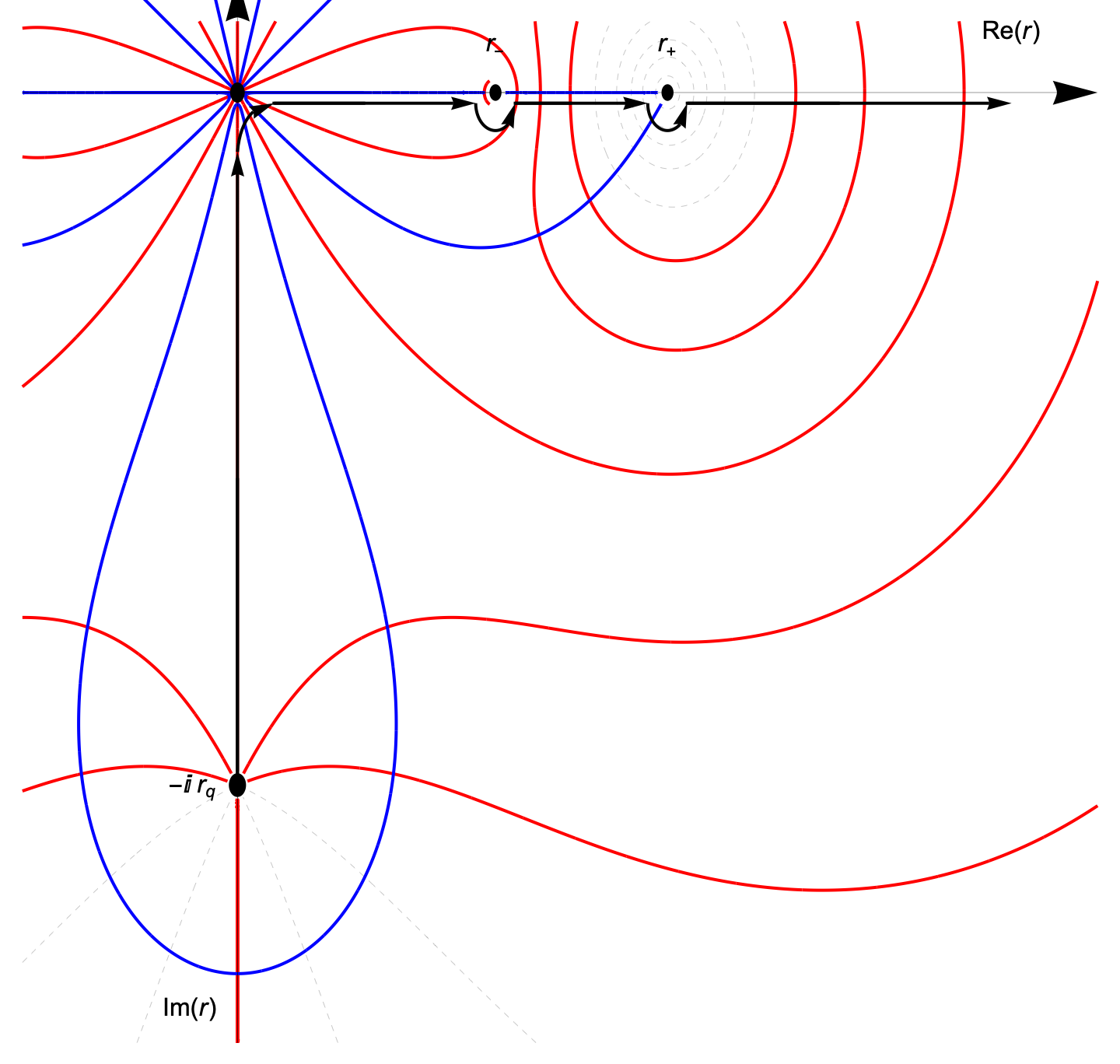}
    \caption{
    Stokes geometry of the leading WKB phase for a charged RN-AdS$_5$
    black brane, shown for $r_+=1$ and $r_-=0.6$. Red curves are constant-$\operatorname{Re}\mathcal{S}$ trajectories, where $\mathcal{S}(r)$ is defined in Eq.~\eqref{eq:complex-wkb-time-function}. Blue curves are Stokes lines on which $\operatorname{Im}\mathcal{S}=0$ and hence the leading WKB solutions are oscillatory.
    Near the timelike singularity at $r=0$, $\mathcal{S}(r)\approx \frac{r^5}{5Q^2}$; this implies that the near-singularity Stokes geometry has ten rays coming out from $r=0$. The black contour is associated with calculating Im $G_{\rm ret}$: The decaying solution is followed from the lower complex pole $r=-\mathrm{i}r_q$, where
    $r_q=\sqrt{r_+^2+r_-^2}$, towards the near-singularity region and is then
    continued to the exterior by passing below the inner and outer horizons.}
    \label{fig:complex-bounce-contour}
\end{figure}
Figure~\ref{fig:complex-bounce-contour} displays the complex $r$ plane, with the black curve being the contour $\mathcal{C}$, red curves representing ``anti-Stokes lines'', namely, constant-$\operatorname{Re}\mathcal S$ trajectories, and blue curves are ``Stokes lines'' with constant-${\rm Im}\mathcal S$. Along anti-Stokes curves, the WKB phase of the bulk fields, $\exp(\mathrm{i}\omega\mathcal S)$, is fixed, providing steepest-descent trajectories. On Stokes curves with ${\rm Im}\mathcal S=0$, the two WKB solutions $\exp(\pm\mathrm{i}\omega\mathcal S)$ are purely oscillatory. 

We find below the Stokes lines near the singularity. In this region, the term proportional to the charge squared dominates the blackening factor,
\begin{equation}
    f(r)\approx\frac{Q^2}{r^6}~,
    \qquad r\to0.
    \label{eq:charged-near-singularity-blackening}
\end{equation}
Consequently, 
\begin{equation}
    \mathcal S(r) \approx \frac{r^5}{5Q^2}~.
\end{equation}
Writing $r=|r|\exp(\mathrm{i}\theta)$, the condition of Stokes lines ${\rm Im}\mathcal S=0$ gives
\begin{equation}
    \sin(5\theta)=0~,
    \qquad
    \theta=\frac{n\pi}{5}~,
    \qquad
    n=0,\ldots,9~.
    \label{eq:charged-local-stokes-rays}
\end{equation}
There are therefore ten Stokes rays dividing the singularity $r=0$ to slices of equal angle $\frac{\pi}{5}$ in the complex $r$-plane; six of which are visible in the lower-half-plane portion of Fig.~\ref{fig:complex-bounce-contour}.

Below, we explain that for the purpose of capturing non-perturbative effects in the frequency, it is necessary to depart from purely anti-Stokes lines. Let $\Phi_s^{\mathrm{in}}(r,\omega)$ denote the physical solution satisfying the purely infalling boundary condition at the outer horizon. We define $\Phi_s^{\mathrm{sd}}(r,\omega)$ as the canonical steepest-descent solution whose large-frequency expansion agrees order by order with that of $\Phi_s^{\mathrm{in}}$ in the exterior region. The two solutions therefore agree to all orders in the ordinary inverse-frequency expansion, but differ by an exponentially small Stokes contribution: 
\begin{equation} 
\Phi_s^{\mathrm{in}}(r,\omega) = \Phi_s^{\mathrm{sd}}(r,\omega) +  O\!\left( e^{\mathrm{i}\omega\Delta t_Q} \right) = \Phi_s^{\mathrm{sd}}(r,\omega) + O\!\left( e^{-\lambda_Q\omega} \right)~. 
\label{eq:infalling-steepest-descent-relation} 
\end{equation}
Our contour $\mathcal C$ runs along the real $r$ axis from the singularity at
$r\approx0$ to the asymptotic region. It passes \emph{below} the poles of
$1/(r^2f)$ at the inner and outer horizons.

Two types of curves leave the timelike singularity. Near $r=0$,
$\mathcal S\simeq r^5/(5Q^2)$, and the anti-Stokes lines with $\mathrm{Re}\,\mathcal{S}=0$ leave the
singularity along the ten rays $\arg r=n\pi/10$ with $n=0,..,9$. Along these trajectories, one
WKB branch is maximally suppressed relative to the other, so the coefficient of the
suppressed branch can change discontinuously when such a trajectory is crossed.
In the lower half-plane, Fig.~8 shows the following:
\begin{enumerate}
\item The trajectory leaving at $\arg r=-\pi/10$ ends on the inner-horizon pole $r_-$.
\item The trajectory leaving along $\arg r=-\pi/2$ follows the negative imaginary axis
  and ends on the complex pole $-\ic r_q$.
\item The trajectory leaving at $\arg r=-3\pi/10$ passes below both horizons and
  returns to the positive real axis at $r_c$, where $\mathrm{Re}\,S(r_c)=0$. For $Q=1.2$
  this happens at $r_c\simeq1.98$, or $r_c\approx1/\mathrm{Re}\,S_\infty$ asymptotically.
\end{enumerate}
The physical infalling solution crosses only the third trajectory on its way from the
outer horizon to the AdS boundary in a way that matters. The exponentially small
contribution to the spectral density is generated there.

\subsection{Time elapsed along bouncing geodesic}
\label{sec:bouncing-geodesic-time}
We determine the time coordinate along the bouncing geodesic represented in red in Fig~\ref{fig:penrose}. For the metric in Eq.~\eqref{metric}, we obtain the following:
\begin{equation}
    \frac{\diff t}{\diff r}
    =
    \pm\frac{1}{r^2f(r)}~.
    \label{eq:radial-null-time}
\end{equation}
The time elapsed along the bouncing geodesic is given by the following integral:
\begin{equation}
    \Delta t_Q
    =
    \int_{ \mathcal C}
    \frac{\diff r}{r^2f(r)}~.
    \label{eq:complex-bounce-time-integral}
\end{equation}
The blackening factor of the black brane can be written as
\begin{equation}
    f(r)
    =
    \frac{
    (r^2-r_+^2)(r^2-r_-^2)(r^2+r_q^2)
    }{r^6},
    \qquad
    r_q^2=r_+^2+r_-^2~.
    \label{eq:blackening-factor-factorized}
\end{equation}
The function $\frac{1}{f(r)}$ in Eq.~(\ref{eq:radial-null-time}) contains poles at the outer and inner horizons, $r=r_+$ and $r=r_-$. In the complexified $r$ plane, it has additional poles at $r=\pm\mathrm{i}r_q$. 
In Appendix~\ref{app:Time}, we evaluate Eq.~\eqref{eq:complex-bounce-time-integral} and obtain
\begin{equation}
    \Delta t_Q
    =
    \frac{
    \pi(r_+^2+r_-^2)^{\frac{3}{2}}
    }{
    (2r_+^2+r_-^2)(r_+^2+2r_-^2)
    }
    +
    \frac{\mathrm{i}}{2}
    \left(\beta_++\beta_-\right)~,
    \label{eq:complex-bounce-time-horizons}
\end{equation}
where $\beta_+$ and $\beta_-$ were defined in Eqs.~(\ref{betap}) and (\ref{betam}). The poles at the two horizons are responsible for the imaginary part of the time along the bouncing geodesic. 
Note here that the quantity $\Delta t_Q$ in Eq.~\eqref{eq:complex-bounce-time-horizons} is expressed in physical boundary-time units and is conjugate to the physical frequency $\omega$. \footnote{The numerical analysis of Section~\ref{sec:numerical-spectral-density} uses the dimensionless frequency 
$ 
\widehat\omega = \frac{\omega}{2r_+}$.
The corresponding time elapsed is therefore 
$\widehat{\Delta t}_Q = 2r_+\Delta t_Q~,  \widehat\tau_Q = 2r_+\tau_Q~, \widehat\lambda_Q = 2r_+\lambda_Q~,$
so that $\omega_{}\Delta t_Q = \widehat\omega\,\widehat{\Delta t}_Q$.  In formulas referring to the numerical analysis, we suppress the hats on the dimensionless quantities.}

The combination appearing in Eq.~\eqref{eq:complex-bounce-time-horizons} is positive. 
The real part of the complex time,
\begin{equation}
    \tau_Q
    =
    \operatorname{Re}\Delta t_Q
    =
    \frac{
    \pi(r_+^2+r_-^2)^{3/2}
    }{
    (2r_+^2+r_-^2)(r_+^2+2r_-^2)
    }~.
    \label{eq:tauQ-horizon-radii}
\end{equation}
The damping coefficient can also be written as
\begin{equation}
    \lambda_Q
    =
    {\rm Im}\Delta t_Q
    =
    \pi
    \frac{
    r_+^4+r_-^4+r_+r_-^3+r_-r_+^3+3r_+^2r_-^2
    }{
    (r_++r_-)(2r_-^2+r_+^2)(2r_+^2+r_-^2)
    }~.
    \label{eq:lambdaQ-horizon-radii}
\end{equation}
In terms of the charge-to-mass parameter
\begin{equation}
    q\equiv\frac{3\sqrt{3}Q^2}{2r_0^6}~,
    \label{eq:charge-to-mass-parameter}
\end{equation}
the real part takes the form
\begin{equation}
    \tau_Q
    =
    \frac{\pi 3^{1/4}}{\sqrt{2}\,r_0}
    \frac{
    \cos^{3/2}\!\left(\frac{\arccos q}{3}\right)
    }{
    \left[
    \cos\!\left(\frac{\pi-\arccos q}{3}\right)
    +
    \cos\!\left(\frac{\arccos q}{3}\right)
    \right]
    \left[
    \cos\!\left(\frac{\pi+\arccos q}{3}\right)
    +
    \cos\!\left(\frac{\arccos q}{3}\right)
    \right]
    }~.
    \label{eq:tauQ-charge-to-mass}
\end{equation}
For $0\leq q\leq1$, $\tau_Q$ decreases monotonically as the charge is increased towards extremality. Thus, at fixed frequency, increasing the charge increases the period of the oscillations in the tail of the spectral density.

To conclude, we note that the complex propagation time $\tau_Q+i\lambda_Q$ also features in the leading non-perturbative contribution to the spectral density. Subleading inverse-frequency corrections to the spectral density are expected to come from higher-bounces, beginning at order $\exp(-2\lambda_Q\omega)$.

\subsubsection{Breakdown of the WKB near the singularity}
\label{sec:wkb-breakdown-near-singularity}
The WKB approximation fails sufficiently close to the timelike singularity. In particular, the steepest-descent solution cannot be continued through the region near the singularity using WKB approximation.  That approximation is valid whenever the effective wavelength in the curved geometry varies slowly: 
\begin{equation}
    \left|
    \frac{1}{\omega}
    \frac{\diff}{\diff r}
    \left(\frac{1}{\mathcal P_s(r)}\right)
    \right|
    \ll1~.
    \label{eq:wkb-validity-condition}
\end{equation}
Using the near-singularity falloff of the metric given by $  f(r)\approx\frac{Q^2}{r^6}$, we find that Eq.~\eqref{eq:wkb-validity-condition} gives a scale where the WKB approximation breaks down
\begin{equation}
    r_{\mathrm{WKB}}
    \sim
    \frac{
    Q^{2/5}
    }{
    \omega^{1/5}
    }~.
    \label{eq:wkb-breakdown-scale}
\end{equation}
Additionally, the two-derivative gravitational description breaks down in the interior due to higher curvature corrections becoming of the same order as the Einstein-Maxwell terms. The Ricci scalar of the charged black brane is given by:
\begin{equation}
R=-20-\frac{2Q^2}{r^6}~.
    \label{eq:ricci-scalar-charged-brane}
\end{equation}
The Kretschmann scalar of the geometry also depends on the mass parameter 
\begin{equation}
    K= 40+\frac{508 Q^4 -360 Q^2 r_0^4 r^2+ 72r_0^8 r^4+8 Q^2 r^6  }{ r^{12}}~.
    \label{Kretch}
\end{equation}
As a result, one scale when string-scale curvature $R\sim \frac{1}{\ell_s^2}$ is reached is given by 
\begin{equation}
    r_{\mathrm{string}}
    \sim
    \left(
    \ell_sQ
    \right)^{1/3}~.
    \label{eq:stringy-curvature-scale}
\end{equation}
For sufficiently nonextremal black branes, this scale is parametrically related to the scale where the uncharged solution has a Kretschmann scalar of order $\frac{1}{\ell_s^4}$. From Eq.~(\ref{Kretch}), that scale is given by 
\begin{equation}
    r_{\mathrm{string}}
    \sim     \sqrt{\ell_s}r_0~.
    \label{eq:stringy-curvature-scale-nonextremal}
\end{equation}
References \cite{Martinec:1994xj}, \cite{Zigdon:2024ljl} wrote similar scales of strong curvature where higher-curvature corrections to Einstein's gravity are important in several examples of classical black hole interiors.
\section{Near-singularity WKB connection problem}
\label{sec:reflection-coefficient}
In this section, we solve a local connection problem near the singularity to obtain the reflection amplitudes $C_{J,0}$, $C_{T,0}$ in Eqs.~(\ref{eq:current})-(\ref{eq:st}).  The subsequent near-singularity calculation should be understood within the two-derivative Einstein-Maxwell approximation. We do not impose a physical boundary
condition at the curvature singularity itself; rather, we solve the wave equations in a region where the $Q^2$-term of $f(r)$ dominates and match its solution to the WKB solution where the WKB approximation is valid.

\subsection{Reflection coefficients from the singularity}
\label{sec:reflection-from-singularity}

\subsubsection{Transverse stress tensor channel}
We wish to connect the following solutions to the wave equations of perturbations in the bulk: (a) the solution near the singularity, (b) the WKB solution associated to the steepest-descent contour and (c) the solution far away that consists of a superposition of two oscillatory WKB waves on the positive real-$r$ axis, with one wave propagating towards infinity and the other to the black brane.

We refer to the ratio of these two local oscillatory waves as a ``near-singularity
reflection coefficient''. This coefficient depends on the choice of local WKB basis of functions and is not, by itself, identical to the coefficient appearing in the boundary spectral density. The latter is obtained only after a propagation to the AdS boundary, an inclusion of the associated matching phases, and a combination with the time-reversed saddle.

We first consider the transverse graviton wave equation. As $r\to0$, the term proportional to $Q^2$ in the blackening factor dominates, and Eq.~\eqref{eq:scalar-r} becomes
\begin{equation}
    h_{xy}''(r)
    -\frac{1}{r}h_{xy}'(r)
    +
    \frac{\omega^2r^8}{Q^4} h_{xy}(r)
    =0~. 
    \label{eq:graviton-near-singularity}
\end{equation}
Introducing the scaling variable
\begin{equation}
    \xi\equiv\frac{\omega r^5}{5Q^2}~,
    \label{eq:scalar-near-singularity-variable}
\end{equation}
the scalar equation takes the form
\begin{equation}
    \frac{\diff^2 h_{xy}}{\diff \xi^2}
    +
    \frac{3}{5\xi}\frac{\diff h_{xy}}{\diff \xi}
    +
    h_{xy}
    =0~.
    \label{eq:scalar-near-singularity-scaled}
\end{equation}
At large $\omega$, the leading local equation has Bessel solutions
\begin{equation}
    h_{xy}(\xi)
    =
    \xi^{1/5}
    \left[
    c_1J_{1/5}(\xi)+c_2Y_{1/5}(\xi)
    \right]~.
    \label{eq:scalar-local-bessel-solution}
\end{equation}
This solution is single-valued as a function of $r$.
We now specify the solution that decays in negative imaginary $r$.  Let
\begin{equation}
    r=e^{3\pi\mathrm{i}/2}\varrho~,
    \qquad
    \varrho>0~,
    \qquad
    X\equiv\frac{\omega\varrho^5}{5Q^2}>0~.
    \label{eq:lower-ray-scaling-variable}
\end{equation}
Along this ray, the scaling variable $\xi$ goes as:
\begin{equation}
    \xi
    =
    \frac{\omega r^5}{5Q^2}
    =
    X\exp\!\left(\frac{3\pi\mathrm{i}}{2}\right)~.
    \label{eq:lower-ray-x-relation}
\end{equation}
The phase relating $\xi^{1/5}$ to $X^{1/5}$ is common to both continued Hankel branches and can be absorbed into the overall normalization. The solution that decays as $X\to+\infty$ is proportional to
\begin{equation}
    h_{xy}^{\mathrm{sd}}
    \left(
    Xe^{3\pi\mathrm{i}/2}
    \right)
    \propto
    X^{1/5}K_{1/5}(X)~.
    \label{eq:scalar-decaying-steepest-solution}
\end{equation}
Here and below, overall $X$-independent normalization factors are not important. Next, we use an identity proven in appendix~\ref{app:2}
\begin{equation}
    K_\nu\!\left(
    Xe^{3\pi\mathrm{i}/2}
    \right)
    =
    -\frac{\pi\mathrm{i}e^{\pi \mathrm{i} \nu/2}}{2}
    \left[
    H_\nu^{(1)}(X)
    +
    2\cos(\pi\nu)e^{-\pi\mathrm{i}\nu}
    H_\nu^{(2)}(X)
    \right]~.
    \label{eq:charged-bessel-connection-formula}
\end{equation}
Following the contour specified in section~\ref{sec:complex-wkb-solutions}, we start from the decaying solution on the ray $\xi=Xe^{3\pi\mathrm{i}/2}$ and analytically continue it to the positive-real-$\xi$ axis.
The continuation prescription fixes the branches of $\xi^\nu$, $K_\nu$, and the Hankel functions. A continuation to a different sheet would change the intermediate phases appearing in the local reflection coefficient. 

In the overlap region where the near-singularity Bessel approximation and the oscillatory WKB approximation are both valid, we define the near-singularity reflection coefficient by
\begin{equation}
    h_{xy}^{\mathrm{sd}}(X)
    \propto
    X^{1/5}
    \left[
    H_{1/5}^{(1)}(X)
    -
    R^{\mathrm{loc}}_s
    H_{1/5}^{(2)}(X)
    \right]~,
    \qquad
    X>0~.
    \label{eq:scalar-local-reflection-definition}
\end{equation}
In this convention, the near-singularity connection formula gives
\begin{equation}
    R^{\mathrm{loc}}(\nu)
    =
    -2\cos(\pi\nu)e^{-\pi\mathrm{i}\nu}~.
    \label{eq:local-reflection-coefficient}
\end{equation}
Hence,
\begin{equation}
    R^{\mathrm{loc}}_s
    =
    R^{\mathrm{loc}}\!\left(\frac{1}{5}\right)
    =-
    \frac{1+\sqrt{5}}{2}
    \exp\!\left(-\frac{\pi\mathrm{i}}{5}\right)~.
    \label{eq:scalar-local-reflection-coefficient}
\end{equation}
The approximate solution is labeled by the index
\begin{equation}
    \nu_T=\frac{1}{5}~.
\end{equation}
Corrections to this leading result arise from the subleading terms neglected in Eq.~\eqref{eq:scalar-near-singularity-scaled} and from matching to the exterior WKB region.

\subsubsection{Transverse-current channel}

We next consider the transverse gauge field perturbation. Near the timelike
singularity, Eq.~\eqref{eq:vector-r} reduces to
\begin{equation}
  A_x''-\frac{3}{r}A_x'+\Big(\frac{\omega^2 r^8}{Q^4}-\frac{12}{r^2}\Big) A_x=0 ~.
    \label{eq:current-near-singularity}
\end{equation}
In terms of the variable $\xi$, Eq.~(\ref{eq:current-near-singularity}) becomes
\begin{equation}
   \frac{\diff^2A_x}{\diff \xi^2}+\frac{1}{5x}\frac{\diff A_x}{\diff \xi}
  +\left(1-\frac{12}{25x^2}\right)A_x~.
    \label{eq:current-near-singularity-scaled}
\end{equation}
At fixed $k$ and large $\omega$, the leading local solutions are
\begin{equation}
    A_x(\xi)
    =
    \xi^{2/5}
    \left[
    a_1J_{4/5}(x)+a_2Y_{4/5}(x)
    \right]~.
    \label{eq:current-local-bessel-solution}
\end{equation}
The solution selected by the property of decay along the lower-half-plane ray is again the corresponding $K_{\frac{4}{5}}$ modified Bessel solution. Applying Eq.~\eqref{eq:charged-bessel-connection-formula} with $\nu=4/5$ gives
\begin{equation}
     R_J^{\rm loc}=R^{\rm loc}\!\left(\tfrac45\right)=-2\cos\Big(\frac{4\pi}{5}\Big)\,\e^{-4\pi\ic/5}
  =\frac{1+\sqrt5}{2}\,\e^{-4\pi\ic/5}~.
    \label{eq:current-local-reflection-coefficient}
\end{equation}

\subsection{From near-singularity connection data to the boundary spectral density}
\label{sec:local-to-boundary-matching}

The coefficients in
Eqs.~\eqref{eq:scalar-local-reflection-coefficient} and
\eqref{eq:current-local-reflection-coefficient} are defined in a local
Hankel basis in the region near the timelike singularity. To convert them to coefficients that appear in the boundary spectral density,
three steps are required. First, the large-$|\xi|$ asymptotics of the local Hankel solutions are matched to the WKB branches in an overlap region where $|\xi|\gg1$, while $r$ remains sufficiently small so that the charge-dominated expansion of the geometry is valid. Second, these WKB branches are transported along the contour $\mathcal{C}$ to the asymptotic AdS region, where the WKB contour passes below both the inner and outer horizon poles, as required by the retarded propagator prescription \cite{Afkhami-Jeddi:2025wra, Chakrabarty:2019aeu, Sharma:2025hbk}. 
Finally, they are matched to the boundary source/response basis and
the conserved flux is evaluated after source normalization.

The complete matching calculation is given in
Appendix~\ref{app:local-to-boundary-matching}. In particular, the
near-singularity WKB phase and its asymptotic boundary behavior are given in Eq. \eqref{eq:appendix-boundary-wkb-action}. The relative transport factor of the reflected branch is derived in
Eq.~\eqref{eq:appendix-relative-propagation-factor},
\begin{equation}
   \e^{2\ic\omega S_\infty}=\e^{\ic\omega\Delta t_Q}=\e^{-\lambda_Q\omega}\e^{\ic\tau_Q\omega}~.
    \label{eq:local-to-boundary-propagation-factor}
\end{equation}
The local Hankel branches are matched to the two WKB branches in Eq.~\eqref{eq:appendix-local-hankel-to-minus-wkb}. Their subsequent matching to the two boundary Hankel branches is given in Eq.~\eqref{eq:appendix-minus-wkb-to-boundary}. The branch $\Phi_s^{(2)}$ is the boundary continuation of the canonical steepest-descent solution. The local connection problem generates an exponentially small nonperturbative contribution proportional to $\Phi_s^{(1)}$, which must be added to reconstruct the physical infalling solution. Combining these relations gives 
\begin{equation} 
  \Phi^{\rm in}_s\propto\Phi^{(1)}_s+E_s(\omega,Q)\,\Phi^{(2)}_s~,
\label{eq:main-text-boundary-admixture} 
\end{equation}
where $\Phi^{(1)}_s$ is the boundary branch continuously connected to the infalling
wave.  The exponentially small reflected contribution is
\begin{equation}
  E_s(\omega,Q)=R_s^{\rm loc}\,\e^{-2\ic(\alpha_s^{\rm loc}+\alpha_s^{\partial})}\,\e^{\ic\omega\Delta t_Q}~. 
\end{equation}
The sources of $\Phi^{(1)}_s$ and $\Phi^{(2)}_s$ have opposite signs, so
\begin{equation}
  \frac{\rho_s(\omega,Q)}{\rho_{s,\rm vac}(\omega)}=1+2\,\mathrm{Re}\,E_s(\omega,Q)+O\big(\e^{-2\lambda_Q\omega}\big)~, 
\end{equation}
with
\begin{equation}
  C_{s,0}=R_s^{\rm loc}\,\e^{-2\ic(\alpha_s^{\rm loc}+\alpha_s^\partial)}
  =-2\cos\big(\pi\nu_s^{\rm loc}\big)\,\e^{-\ic\pi\nu_s^\partial}~.
\end{equation}
For the transverse current ($\nu_J^{\rm loc}=4/5$, $\nu_J^\partial=1$) and the
transverse stress tensor ($\nu_T^{\rm loc}=1/5$, $\nu_T^\partial=2$), Appendix
\ref{app:local-to-boundary-matching} evaluates this expression in
Eqs.~\eqref{eq:appendix-current-boundary-coefficient} and
\eqref{eq:appendix-stress-boundary-coefficient}, giving

\begin{equation}
  C_{J,0}=2\cos\Big(\frac{4\pi}{5}\Big)=-\frac{1+\sqrt5}{2}~~,\qquad
  C_{T,0}=2\cos\Big(\frac{4\pi}{5}\Big)=-\frac{1+\sqrt5}{2}~. 
\end{equation}
The result is derived explicitly in
Eqs.~\eqref{eq:appendix-spectral-density-from-source-normalization}
and \eqref{eq:appendix-linearized-spectral-density}. 

Thus, the phases present in the local Bessel connection coefficients
combine with the WKB and boundary-Hankel phases such that the leading
boundary coefficients are real. These are the quantities compared
with the high-precision numerical spectral densities in
Section~\ref{sec:son-starinets} and
Appendix~\ref{app:fixed-charge-numerics}.

At fixed nonzero charge, subleading corrections to the near-singularity scaling problem and to the matching onto the exterior
WKB region generate fractional large-frequency expansions. For the
transverse current channel, the fractional large-frequency expansion takes the form given in \eqref{eq:current}~\footnote{When an $\omega^{-2/5}$ term is added to the fit, its
coefficient is small and decreases as the fitting window is raised. An analytic
determination of $C_{J,1}$ requires the subleading local connection problem, and we
do not attempt to determine it here.}, whose powers coincide with those in the expansion for the stress tensor channel, in \eqref{eq:st}. The fractional powers follow from the scaling $r\sim\omega^{-1/5}$ in the local connection region. In particular, the near-singularity expansion of the blackening factor is
\begin{equation}
    f(r)
    =
    \frac{Q^2}{r^6}
    \left(
        1-\frac{r_0^4r^2}{Q^2}
        +\frac{r^6}{Q^2}
    \right)~.
\end{equation}
The first correction to the charge-dominated geometry is therefore of relative order $\omega^{-2/5}$ at fixed nonzero charge. The complete local and WKB matching calculation determines which corrections survive in the boundary coefficient, giving the expansions stated in Eqs.~\eqref{eq:current} and \eqref{eq:st}.

The fixed-charge connection problem applies when the scale where the WKB approximation breaks down lies close to the singularity. At sufficiently small charge and finite frequency, the local reflection problem instead crosses over to the Schwarzschild-like regime discussed in the next subsection.

\subsection{Small-charge regime and the Schwarzschild-like reflection problem}
\label{sec:small-charge-regime}

The small-charge regime is qualitatively different from the finite-charge case discussed in Appendix~\ref{app:fixed-charge-numerics}. In order to probe the region behind the inner horizon at small charge, the frequency must be taken sufficiently high.

Quantitatively, the distinction is set by the frequency-dependent radial scale $r_{\rm WKB}$ at which the WKB approximation ceases to be valid. In the region near the timelike singularity, the blackening factor is approximately given by
\begin{equation}
 f(r)\approx\frac{Q^2}{r^6}~,
\end{equation}
and the slow-variation of the effective wavelength required for the validity of the WKB approximation gives
\begin{equation} 
r_{\rm WKB}(\omega,Q) \sim \left(\frac{Q^2}{\omega}\right)^{1/5}~. 
\label{eq:small-charge-WKB-radius} 
\end{equation}
Thus $r_{\rm WKB}$ moves inwards as the frequency is increased. This scale should not be confused with the geometric radius at which the mass and charge terms in the blackening factor become comparable. For a black brane far from extremality, that geometric scale is of the same order as the inner-horizon radius,
\begin{equation}
 r_-\simeq\frac{|Q|}{\sqrt{M}}\ll r_+, \qquad  M\equiv r_0^4~. 
 \label{eq:small-charge-inner-horizon}
\end{equation}
where $r_0$ is defined through the blackening function in Eq.~\eqref{fdef}. The Schwarzschild-like regime is characterized by
\begin{equation}
 r_{\rm WKB}\gg r_-~.
 \label{eq:small-charge-scale-ordering}
\end{equation}
In this regime the WKB approximation breaks down, and the wave is reflected, at a radial scale larger than the inner-horizon scale. The reflected wave therefore does not reach the region near the timelike singularity, and the local reflection problem remains closer to the uncharged AdS-Schwarzschild problem than to the fixed-charge problem. Conversely, increasing the frequency lowers $r_{\rm WKB}$; once $r_{\rm WKB}$ becomes comparable to or smaller than $r_-$, the wave probes behind the inner horizon and crosses over to the fixed-charge connection problem.

It is useful to introduce a dimensionless crossover parameter
\begin{equation}
 \delta(\omega,Q)
 \equiv
 \frac{Q^2\omega^{2/3}}
 {2^{2/3}M^{5/3}}~.
 \label{eq:small-charge-delta}
\end{equation}
Using Eqs.~\eqref{eq:small-charge-WKB-radius} and \eqref{eq:small-charge-inner-horizon}, $\delta$ is related to the ratio of the inner-horizon and WKB-breakdown scales by
\begin{equation}
 \delta
 \sim
 \frac{1}{2^{2/3}}
 \left(\frac{r_-}{r_{\rm WKB}}\right)^{10/3}~.
 \label{eq:small-charge-delta-scales}
\end{equation}
The hierarchy $ \delta\ll1$ is equivalent to $ r_{\rm WKB}\gg r_-.$ Equivalently, up to order-one factors,
\begin{equation}
 \omega\ll\frac{M^{5/2}}{|Q|^3}~.
\end{equation}
Together with the high-frequency condition, the small-charge Goldilocks range is
\begin{equation}
 1\ll\beta_+\omega
 \ll\frac{\beta_+M^{5/2}}{|Q|^3}~.
 \label{eq:small-charge-double-scaling-window}
\end{equation}
Although the local reflection problem is Schwarzschild-like in this regime, the global propagation factor is still determined by the exact charged geometry. We therefore retain the charged values of $\tau_Q$ and $\lambda_Q$. The leading non-perturbative contributions to the current and stress-tensor spectral densities are written as
\begin{align}
 \rho_J(\omega,Q)-\pi\omega^2
 &=2\pi\omega^2\e^{-\lambda_Q\omega}
 {\mathrm Re}\!\left[
 \e^{\ic\tau_Q\omega}\mathcal C_J(\omega,Q)
 \right]
 +\Order\!\left(\e^{-2\lambda_Q\omega}\right)~,
 \label{eq:small-charge-current-leading-nonperturbative}\\
 \rho_T(\omega,Q)-\pi\omega^4
 &=\pi\omega^4\e^{-\lambda_Q\omega}
 {\mathrm Re}\!\left[
 \e^{\ic\tau_Q\omega}\mathcal C_T(\omega,Q)
 \right]
 +\Order\!\left(\e^{-2\lambda_Q\omega}\right)~.
 \label{eq:small-charge-stress-leading-nonperturbative}
\end{align}
The different overall factors reflect the conventions used for the two channels~\footnote{Note that for the small-charge shear channel we adopt a different normalization from the fixed-charge convention, so that the leading uncharged coefficient includes the combined future- and past-bounce contributions and equals $-4$.}. In the Schwarzschild-like regime, the current amplitude approaches its uncharged value, 
\begin{equation} 
\mathcal C_J(\omega,Q) = 1+O(\delta)~.
\label{eq:small-charge-current-amplitude} 
\end{equation} 
up to possible logarithmic enhancements that would require a separate matched-asymptotic analysis.
The shear amplitude, however, retains the uncharged fractional-power series in addition to its charged crossover correction. To write this series compactly, we define
\begin{equation}
 z_Q
 =\frac{4\omega}{2-Q^2}~,~
 \qquad (r_+=1)~.
 \label{eq:small-charge-zQ}
\end{equation}
 In the uncharged limit $z_0=2\omega$. With this convention,
\begin{align}
\mathcal C_T(\omega,Q) = -4 +C_1^{(0)}z_Q^{-4/3} +C_2^{(0)}z_Q^{-8/3} +O(\delta) +\cdots 
 \label{eq:small-charge-stress-amplitude}
\end{align}
In this regime, we make contact with the small-charge regime investigated in \cite{Ceplak:2025dds}. The leading shear coefficient $-4$ is the sum of the future- and past-reflected contributions motivated by the uncharged case, corresponding to a reflection coefficient $-2$ per saddle. We use effective fitting forms rather than attempting to determine the  $O(\delta)$ coefficients separately. Defining
\begin{equation}
 Y_J(\omega,Q)
 =\frac{\e^{\lambda_Q\omega}}{2\pi\omega^2}
 \left[\rho_J(\omega,Q)-\pi\omega^2\right]~, \qquad
 N_T(\omega,Q)
 =\frac{\e^{\lambda_Q\omega}}{\pi\omega^4}
 \left[\rho_T(\omega,Q)-\pi\omega^4\right]~,
\end{equation}
the numerical data are fitted to
\begin{equation}
 Y_J(\omega,Q)
 ={\mathrm Re}\!\left[
 \mathcal R_J(Q)\e^{\ic\tau_Q\omega}
 \right], \qquad
 N_T(\omega,Q)
 ={\mathrm Re}\!\left[
 \e^{\ic\tau_Q\omega}
 \left(C_{T,0}(Q)+C_{T,1}(Q)z_Q^{-4/3}\right)
 \right]~.
 \label{eq:small-charge-stress-fit-main}
\end{equation}
The fitted coefficients should be understood as ``effective coefficients'' for the selected frequency windows. They absorb the leading crossover correction and, in the shear channel, part of the omitted $z_Q^{-8/3}$ contribution.

A specific example is presented:
For $Q=0.05$, the fixed propagation parameters are $ \tau_Q=3.1338,
 \lambda_Q=3.1447,$ which are close to the Schwarzschild values $\tau_0=\lambda_0=\pi$. For intermediate frequencies, the current fit gives us the effective boundary amplitude:
\begin{equation}
 \mathcal R_J(0.05)=1.0082+0.02900\ic~,
 \qquad
 |\mathcal R_J(0.05)|=1.0086~,
 \label{eq:small-charge-current-result-main}
\end{equation}
while the shear fit gives
\begin{equation}
 C_{T,0}(0.05)=-4.0104-0.0732\ic~,
 \qquad
 C_{T,1}(0.05)=-2.6646-5.3547\ic~.
 \label{eq:small-charge-stress-result-main}
\end{equation}
Equivalently, the effective tensor boundary amplitude per saddle is
\begin{equation}
 \mathcal R_T(0.05)
 =\frac{C_{T,0}(0.05)}{2}
 =-2.0052-0.0366\ic~,
 \qquad
 |\mathcal R_T(0.05)|=2.0055~.
 \label{eq:small-charge-stress-reflection-main}
\end{equation}
Thus, the effective boundary amplitudes in both channels are close to their Schwarzschild values. Their magnitudes differ from the uncharged values by approximately $0.86\%$ in the current channel and $0.28\%$ in the tensor channel, while both acquire a small complex phase. These fitted quantities include crossover effects, WKB transport and boundary-matching phases, and finite-window corrections; they should not be identified directly with the bare local reflection coefficients. The detailed residuals, and stability tests are given in Appendix~\ref{app:small-charge-numerics}.

\section{Discussion}
\label{sec:discussion}
In this paper, we have computed contributions to the spectral densities of transverse currents and stress tensors in a grand canonical ensemble, holographically dual to a planar Reissner-Nordstr\"om AdS$_5$ black brane. Although the spectral density is computed from a standard one-sided retarded prescription in the exterior geometry, its large-real-frequency behavior contains
exponentially small corrections that admit a useful interpretation in terms of the analytically continued bulk geometry.

The leading non-perturbative contribution factorizes into two conceptually distinct ingredients. The first is a global propagation factor $\exp\!\left(\mathrm{i}\omega\Delta t_Q\right)$ determined by the complex time along a bouncing null trajectory $\Delta t_Q=\tau_Q+\mathrm{i}\lambda_Q.$
The imaginary part $\lambda_Q$ fixes the exponential damping of the
vacuum-subtracted spectral density, and the real part $\tau_Q$ sets the
oscillation frequency. The second ingredient is a local connection problem near the singularity, whose solution determines the reflection data multiplying this propagation factor. This separation between global WKB propagation and local
near-singularity scattering is of crucial importance to our analysis.

At a finite, nonzero bulk charge parameter and sufficiently large frequency, the region where the WKB approximation is invalid lies near the singularity. The near-singularity equations are approximately Bessel equations, with distinct indices in the current and stress-tensor channels. This produces fractional-power expansions in the inverse frequency, as in \eqref{eq:current} and \eqref{eq:st}.

The leading coefficients extracted from numerical spectral densities agree with the analytic expression given in \eqref{eq:Gnp-general-structure}. The transverse current and the transverse stress tensor
share the same leading coefficient, $2\cos\big(4\pi/5\big)$.
The numerical analysis also confirms that the fractional corrections are essential: a frequency-independent reflection amplitude does not adequately describe the data.

The small-charge regime exhibits a qualitatively different hierarchy of scales. In the frequency window for which the scale of breakdown of the WKB approximation lies outside the inner horizon scale, the reflected wave does not reach the region near the timelike singularity. The local connection problem is then continuously connected to the uncharged AdS-Schwarzschild problem, even though the complex propagation time already
depends non-perturbatively on the charge. In this regime, the transverse current approaches the uncharged Airy reflection coefficient in \cite{Afkhami-Jeddi:2025wra}, while the shear channel retains the uncharged fractional-power series together with charge-dependent crossover corrections. Reaching the genuinely near-singularity connection problem at a small chemical potential requires taking the frequency to a parametrically higher value.

An important qualification concerns the interpretation of the extended geometry. The bouncing trajectories used here arise from an analytic continuation of the bulk WKB problem and organize exponentially small corrections to a retarded correlator defined in the exterior. Our calculation
does not require a physically infalling observer to traverse the outer or inner horizon, nor does it assume that the classical Cauchy horizon and the spacetime beyond it survive unchanged in the exact quantum theory. The
near-singularity connection coefficient computed here should therefore be viewed as a property of the classical saddle of two-derivative Einstein-Maxwell theory, and its analytic continuation in the probe limit.

The leading correction studied here corresponds to a single bounce off the singularity. The full transseries contains further sectors suppressed by additional powers of $\exp(-\lambda_Q\omega)$. These sectors can be interpreted in terms of trajectories, with additional reflections in the analytically continued geometry.

To provide some additional context as to where our story belongs, a rapidly growing body of work has recently developed the relation between thermal correlators, bulk geometry, and black hole singularity. Analytic treatments of strongly coupled plasma perturbations and of Fourier-space thermal correlators have clarified the structure of large-frequency asymptotic expansions and their non-perturbative completions~\cite{Aniceto:2026zmc,Arnaudo:2026der}. The relation between thermal OPE data and quasinormal-mode spectra has also been explored directly~\cite{Arnaudo:2026axe}, including through spin-resolved double-trace thermal data in holographic theories~\cite{Buric:2026cmn}. Recently, analytic thermal-bootstrap methods were also developed significantly in holographic settings~\cite{Buric:2025fye,
Barrat:2025twb} (see also ~\cite{Niarchos:2025cdg, Niarchos:2026vow} that applied deep learning methods to thermal bootstrap and \cite{Alkalaev:2026sha} for thermal conformal partial waves to organize CFT data). Several works have studied the bulk geometric interpretation of these structures. Bouncing geodesics and the associated singularities of thermal correlators were analyzed in holographic black hole backgrounds~\cite{Grozdanov:2026cut}, while the convergence region of quasinormal-mode expansions was tied to bouncing singularities in
Schwarzschild geometry~\cite{Arnaudo:2026tcy}. Complex geodesics and bulk-cone singularities were examined from several perspectives,
including their relation to bulk phase shifts~\cite{Araya:2026shz,
Jia:2026pmv}. Extensions to AdS exotic compact objects have also been considered~\cite{Chen:2025jbf}. Other recent works about the black hole singularity in different spacetimes and using various observables include ~\cite{Rodriguez-Gomez:2021pfh, David:2022nfn, Singhi:2024sdr ,Sorokhaibam:2025nur, Dey:2026ckh, Aalsma:2026kqj,Grozdanov:2026ktq, Saha:2026ows, Shi:2026cfq}. 

It is pointed out that timelike entanglement entropy also displays features where geodesics reflect off the black hole singularity \cite{Doi:2023zaf, Anegawa:2024kdj, Bernamonti:2026pxo}. In this context, the geodesics reflected off the singularity in three bulk spacetime dimensions have both spacelike and timelike portions. In the three-dimensional case, this seems to be an interesting alternative observable that probes the singularity since the original bouncing geodesics story \cite{Fidkowski:2003nf} does not have a bouncing geodesic reflecting off the black hole singularity.

Several generalizations are immediate. First, the present work set the momentum to zero. At nonzero momentum, the WKB phase and the near-singularity scaling problem receive
additional momentum-dependent corrections. Additionally, for charged bulk fields, the background gauge field also contributes to the WKB phase, and the grand-canonical ensemble naturally introduces a dependence on the chemical
potential. It would be interesting to extend the present analysis to charged operators, where the relevant KMS relations and the bulk real-time prescription involve the combination $\omega-\mu q$. Real-time holographic methods at a finite density might prove helpful here \cite{Sharma:2025hbk}.

Second, it would be useful to understand whether the present construction has an analog in other black hole geometries. Bouncing null trajectories and their associated singularities of retarded Green's functions have recently been studied in asymptotically flat and Schwarzschild-de Sitter geometries,
in particular in the presence of a reflecting cavity~\cite{Grozdanov:2026ktq}. The Penrose diagrams of charged AdS black holes and black holes in de Sitter share several qualitative features, including multiple
horizons and nontrivial analytic continuations of null trajectories. However, the appropriate observables differ: In de Sitter space, one must account for the cosmological horizon and replace the standard holographic spectral density by a static-patch, cavity, or other suitably defined observable. It would be interesting to determine whether a local connection coefficient near the singularity can be extracted from
such observables.

Third, it would be interesting to find an alternative interpretation of the non-perturbative corrections to the spectral density in terms of a different contour in complex $r$-plane that avoids the region behind the inner horizon but nonetheless explores the region between the inner horizon and outer horizon; this would be compatible with prior evidence that the Cauchy horizon is singular in the large-$N$ gravity regime. Since the time coordinate along the null geodesic is preserved for contours differed by smooth deformations, we expect such an alternative contour would still capture the time $\tau_Q+\ic\lambda_Q$, although a challenge is to recover precisely the reflection coefficients.  

Fourth, our analysis is restricted to two-derivative Einstein-Maxwell theory. As the frequency increases, the WKB breakdown scale moves towards the singularity. Eventually, it approaches the scale at which higher-derivative or stringy corrections become important. Such corrections are expected to modify the local connection problem, and hence the
reflection coefficient, before the formal Einstein-Maxwell scaling limit is reached. It would be particularly interesting to determine how stringy effects modify the connection problem and whether the resulting
changes can be extracted from the large-frequency spectral density. Recent progress indicating that boundary observables may encode detailed information about the singularity, such as stringy effects (as well as Kasner-like behavior near the singularity), makes this question a natural one~\cite{Dodelson:2025jff,Hartnoll:2026vhu, Coviello:2026hdc, Coviello:2026tol, Xiao:2026pir, Ceplak:2026ard, Gao:2026ieh}.

Particularly relevant for the present work, exact holographic thermal spectral functions were used to relate OPE data, exponentially small large-frequency corrections, and black hole
singularities~\cite{Jia:2025jbi, Jia:2026ryl}, which also probe the case of non-zero spatial momenta. These developments suggest that real-time boundary observables encode the analytic structure of black hole singularities. As in \cite{Afkhami-Jeddi:2025wra}, our present results provide another explicit example in which local reflection data off a timelike singularity can be extracted from the thermal spectral density.

\section*{Acknowledgements}
We thank Nima Afkhami-Jeddi, Nejc \v{C}eplak, Matthew Dodelson, Daniel Jafferis, Diksha Jain, Alex Maloney, Rob Myers and Zixia Wei for useful discussions. In particular, we thank Simon Caron-Huot for many helpful discussions on related topics which vastly improved our understanding of the subject. We utilized OpenAI GPT 5.6 and Claude Opus 5.5 for numerical coding assistance. JC is supported by the NSERC of Canada and the CRC program, reference number CRC-2022-00421, and by J. Cotler's Alfred P. Sloan Research Fellowship.  YZ is supported by the Israel Science Foundation, grant number 1099/24.

\appendix
\section{Detailed numerical analysis at a finite nonzero charge}
\label{app:fixed-charge-numerics}

This appendix provides numerical details underlying the finite-charge fits
quoted in Section~\ref{sec:son-starinets}. We work with
\begin{equation}
    Q=1.2~,\qquad r_+=1~,\qquad \vec{k}=0~.
    \label{eq:app-fixed-charge-parameters}
\end{equation}
All frequencies and spectral density normalizations are those Eq.~(\ref{eq:dimensionless-frequency-momentum}). In particular, the current and shear
vacuum spectral densities are $\pi\omega^2$ and $\pi\omega^4$, respectively.
Further, we fix the damping rate and oscillation frequency so they are not fitting
parameters. They were computed from the time elapsed along the complex bouncing trajectory of
Section~\ref{sec:two-sided-geometry}:
\begin{equation}
    \tau_Q=2.084~,~ \lambda_Q=2.574~.
    \label{eq:appendix-large-charge-path}
\end{equation}
The purpose of the fits below is to test the remaining local connection data.

For reference, the crossover parameter introduced in Section~\ref{sec:small-charge-regime} is
\begin{equation}
    \delta(\omega,Q)=\frac{Q^2}{2^{2/3}(1+Q^2)^{5/3}}\omega^{2/3}~.
    \label{eq:appendix-crossover-variable}
\end{equation}
Over the range $4\leq\omega\leq20$ used in the numerical calculation,
\begin{equation}
    0.5169\lesssim\delta(\omega,1.2)\lesssim1.511~.
    \label{eq:appendix-finite-charge-delta-range}
\end{equation}
Thus, the data are not deep in the asymptotic charge-dominated regime. The
subleading fractional powers predicted by the local connection problem are
therefore important for extracting the leading coefficients reliably.

\subsection{Transverse current}
\label{app:fixed-charge-current}

We remove the known vacuum contribution and exponential envelope by defining
\begin{equation}
    Y_J(\omega,Q)=\frac{\exp(\lambda_Q\omega)}{2\pi\omega^2}
    \left[\rho_J(\omega,Q)-\pi\omega^2\right]~.
    \label{eq:appendix-current-normalized}
\end{equation}
The current connection problem is governed by the
Bessel index $4/5$, which sets the fractional powers in the expression 
\begin{align}
Y_J(\omega,1.2)=\mathrm{Re}\Big[&\exp(\ic\tau_Q\omega)\big(C_{J,0}+C_{J,1}\omega^{-4/5}+C_{J,2}\omega^{-8/5}+C_{J,3}\omega^{-12/5}+\cdots\big)\Big]~, 
 \label{eq:appendix-current-expansion0}
\\
  C^{\rm analytic}_{J,0}&=2\cos\Big(\frac{4\pi}{5}\Big)=-\frac{1+\sqrt5}{2}=-1.618034\ldots ~.
    \label{eq:appendix-current-expansion}
\end{align}
The numerical data have a spacing $\Delta\omega=0.1$ on $4\le\omega\le20$. Our main fitting
window is $11.2\le\omega\le19.2$, which contains 81 points. The most general fit has eight real parameters. We test this prediction by leaving $C_{J,0}$ unconstrained in the fit. We compare the following
nested approximations:
\begin{align}
    Y_J^{(0)}(\omega)&=\mathrm{Re}\!\left[
    \exp(\mathrm{i}\tau_Q\omega)C_{J,0}\right]~, \nonumber\\
    Y_J^{(1)}(\omega)&=\mathrm{Re}\!\left[
    \exp(\mathrm{i}\tau_Q\omega)
    \left(C_{J,0}+C_{J,1}\omega^{-4/5}\right)\right]~, \nonumber\\
    Y_J^{(3)}(\omega)&=\mathrm{Re}\!\left[
    \exp(\mathrm{i}\tau_Q\omega)
    \left(C_{J,0}+C_{J,1}\omega^{-4/5}
    +C_{J,2}\omega^{-8/5}\right)\right]~, \nonumber\\
        Y_J^{(2)}(\omega)&=\mathrm{Re}\!\left[
    \exp(\mathrm{i}\tau_Q\omega)
    \left(C_{J,0}+C_{J,1}\omega^{-4/5}
    +C_{J,2}\omega^{-8/5}+C_{J,3} \omega^{-\frac{12}{5}}\right)\right]
    \label{eq:appendix-current-nested-models}
\end{align}

\subsubsection{Fitting methodology}
Eq.~(\ref{eq:current-u-equation}) for the gauge field perturbation is solved at 70-digit precision. We use the near-horizon series (\ref{eq:numerical-horizon-series}) at $u=1$, to continue
through the exterior, and the two Frobenius solutions at $u=0$, including the
logarithm one. Results are stable to better than $10^{-50}$ under changes of the series
orders and matching points. The same code reproduces the exact uncharged result,
$\rho/\rho_{\rm vac}=\sinh\pi\omega/(\cosh\pi\omega-\cos\pi\omega)$, to $10^{-47}$. 

We perform the fits by calculating root-mean-square residual (RMS), and a leave-one-out cross-validation (LRMS)
test. While the former is the usual RMS; in the latter each point is omitted in turn, the remaining points are
fitted, and the omitted point is predicted. Table~\ref{tab:1} shows both the ordinary root-mean-square residual and a leave-one-out cross-validation
test. 

\begin{table}[h]
\centering
\begin{tabular}{lcc}
\toprule
Model & In-sample RMS & LRMS\\
\midrule
$C_{J,0}$ & $4.272\times10^{-2}$ & $4.380\times10^{-2}$\\
$C_{J,0}+C_{J,1}\omega^{-4/5}$ & $7.587\times10^{-4}$ & $8.111\times10^{-4}$\\
$C_{J,0}+C_{J,1}\omega^{-4/5}+C_{J,2}\omega^{-8/5}$ & $4.324\times10^{-6}$ & $4.765\times10^{-6}$\\
$C_{J,0}+\dots+C_{J,3}\omega^{-12/5}$ & $5.760\times10^{-8}$ & $6.988\times10^{-8}$\\
\bottomrule
\end{tabular}
\caption{Fit quality for the normalized transverse-current spectral density at
$Q=1.2$ in the interval $11.2\le\omega\le19.2$.}
\label{tab:1}
\end{table}

The successive corrections lower the normalized residual by factors of approximately
$56$, $1.8\times10^2$ and $75$. The leave-one-out residuals improve comparably, which
shows that the fractional-power terms are not merely accommodating the in-sample
data. The four-coefficient fit gives
\begin{equation}
\begin{aligned}
  C_{J,0}&=-1.61944-0.00088\,\ic, & C_{J,1}&=-2.889+2.259\,\ic,\\
  C_{J,2}&=0.313+4.483\,\ic, & C_{J,3}&=3.22+0.52\,\ic .
\end{aligned}\tag{A.9}
\end{equation}
For $\omega=15$, the corresponding terms in the prefactor have magnitudes
\begin{equation}
  |C_{J,0}|\simeq1.619,\quad |C_{J,1}|\omega^{-4/5}\simeq0.420,\quad
  |C_{J,2}|\omega^{-8/5}\simeq0.059,\quad |C_{J,3}|\omega^{-12/5}\simeq0.0049. \tag{A.10}
\end{equation}
Table~2 shows that the three- and four-coefficient fits approach the
analytic value from opposite sides. In the main window
\begin{equation}
  \frac{\big|\mathrm{Re}\,C_{J,0}-C^{\rm analytic}_{J,0}\big|}{\big|C^{\rm analytic}_{J,0}\big|}
  =8.7\times10^{-4}\ \text{(four coefficients)},\qquad
  2.3\times10^{-3}\ \text{(three coefficients)}, \tag{A.11}
\end{equation}
while $|\mathrm{Im}\,C_{J,0}/\mathrm{Re}\,C_{J,0}|\simeq5.5\times10^{-4}$. Extending the dataset
to $\omega=30$ and fitting five coefficients on $18\le\omega\le30$ gives
$C_{J,0}=-1.61851-0.00040\,\ic$, which agrees at the $3\times10^{-4}$ level. The same
fit at $Q=1.35$ gives $C_{J,0}=-1.61847-0.00039\,\ic$, confirming that the leading
coefficient does not depend on $Q$.
A useful check of the above coefficients is the stability of the leading coefficient as the lower endpoint of the fit is raised. The result is shown in Table~\ref{tab:2}.
\begin{table}[h]
\centering
\begin{tabular}{lcc}
\toprule
Fit window & $C_{J,0}$ (three coefficients) & $C_{J,0}$ (four coefficients)\\
\midrule
$6.4\le\omega\le19.2$ & $-1.60856+0.00045\,\ic$ & $-1.62039-0.00086\,\ic$\\
$8.0\le\omega\le19.2$ & $-1.61154+0.00028\,\ic$ & $-1.61990-0.00091\,\ic$\\
$9.6\le\omega\le19.2$ & $-1.61325+0.00011\,\ic$ & $-1.61962-0.00091\,\ic$\\
$11.2\le\omega\le19.2$ & $-1.61435-0.00003\,\ic$ & $-1.61944-0.00088\,\ic$\\
\bottomrule
\end{tabular}
\caption{Dependence of the leading current coefficient on the lower endpoint of
the fitting window at $Q=1.2$.}
\label{tab:2}
\end{table}
\\ Both the deviation of the real part from the analytic value and the small
imaginary part decrease as lower-frequency points are removed. In the upper
window,
\begin{equation}
    \frac{\mathrm{Re}\,C_{J,0}-C_{J,0}^{\mathrm{analytic}}}
    {C_{J,0}^{\mathrm{analytic}}}=1.11\times10^{-3}~.
    \label{eq:app-fixed-current-leading-relative-difference}
\end{equation}
while $\arg C_{J,0}\simeq5.57\times10^{-4}\,\mathrm{rad}$. Thus, the leading
coefficient agrees with the analytic prediction at the $0.11\%$ level. For
completeness, the fitted spectral density is
\begin{equation}
    \rho_{J,\mathrm{fit}}(\omega)=\pi\omega^2
    +2\pi\omega^2\exp(-\lambda_Q\omega)Y_J^{(2)}(\omega)~.
    \label{eq:appendix-current-full-fit}
\end{equation}
Representative residuals in physical units are listed in Table~3.
\begin{table}[h]
\centering
\begin{tabular}{ccc}
\toprule
$\omega$ & $|\rho_J-\pi\omega^2|$ & $|\rho_J-\rho_{J,\rm fit}|$\\
\midrule
10.0 & $1.719\times10^{-9}$ & $1.677\times10^{-14}$\\
12.0 & $6.750\times10^{-11}$ & $1.689\times10^{-18}$\\
14.0 & $4.064\times10^{-13}$ & $3.795\times10^{-21}$\\
16.0 & $8.411\times10^{-16}$ & $4.257\times10^{-23}$\\
18.0 & $2.796\times10^{-17}$ & $9.083\times10^{-25}$\\
19.2 & $8.752\times10^{-19}$ & $1.440\times10^{-25}$\\
20.0 & $1.573\times10^{-19}$ & $3.881\times10^{-26}$\\
\bottomrule
\end{tabular}
\caption{Representative residuals for the transverse-current spectral density at
$Q=1.2$, using the four-coefficient fit. The points at $\omega=10$ and $\omega=20$ lie
outside the main fitting window and are therefore extrapolations.
}
\label{tab:3}
\end{table}
The residual is not monotonic because both the signal and the
difference between the data and the fit are oscillatory. Taken together, the
fits test the independently predicted damping and oscillation parameters, the
fractional powers of frequency, and the leading Bessel connection coefficient.

\subsection{Transverse stress tensor}
\label{app:fixed-charge-stress}

For the shear channel, we use
\begin{equation}
    Y_T(\omega,Q)=\frac{\exp(\lambda_Q\omega)}{2\pi\omega^4}
    \left[\rho_T(\omega,Q)-\pi\omega^4\right]~.
    \label{eq:appendix-stress-normalized}
\end{equation}
The finite-charge connection problem has the Hankel structure appropriate to
the transverse graviton. The two local phases combine to give
\begin{equation}
    C_{T,0}^{\mathrm{analytic}}=
    \exp\!\left(\frac{4\pi\mathrm{i}}{5}\right)
    +\exp\!\left(-\frac{4\pi\mathrm{i}}{5}\right)
    =2\cos\!\left(\frac{4\pi}{5}\right)
    =-\frac{1+\sqrt{5}}{2}=-1.61803\ldots
    \label{eq:appendix-stress-leading-value}
\end{equation}
The phase $4\pi/5$ is part of the local connection coefficient and is not
included in the propagation phase $\tau_Q\omega$. The finite-charge expansion is
\begin{equation}
    Y_T(\omega,1.2)\sim\mathrm{Re}\!\left[
    \exp(\mathrm{i}\tau_Q\omega)
    \left(C_{T,0}+C_{T,1}\omega^{-4/5}
    +C_{T,2}\omega^{-8/5}+\cdots\right)\right]~.
    \label{eq:appendix-stress-expansion}
\end{equation}
The dataset contains 161 frequencies with spacing $\Delta\omega=0.1$ over
$4\leq\omega\leq20$. As in the current channel, the main fit uses
$11.2\leq\omega\leq19.2$, with 80 data points:
\begin{equation}
    Y_{T,\mathrm{fit}}(\omega)=\mathrm{Re}\!\left[
    \exp(\mathrm{i}\tau_Q\omega)
    \left(C_{T,0}+C_{T,1}\omega^{-4/5}
    +C_{T,2}\omega^{-8/5}\right)\right]~.
    \label{eq:appendix-stress-fit}
\end{equation}

In addition to the ordinary RMS residual and LRMS cross-validation, we use blocked cross-validation (BRMS): A contiguous segment of the oscillatory
curve is omitted and then predicted from the remaining data. We also quote
the condition number $\kappa$ as a measure of parameter correlation in the
linear fit.

The fitted coefficients are
\begin{align}
    C_{T,0}=-1.61054+0.00361886\,\mathrm{i}~, \quad 
    C_{T,1}=-0.2091-0.3512\,\mathrm{i}~, \quad
    C_{T,2}=-1.444-1.467\,\mathrm{i}~.
    \label{eq:app-fixed-stress-coefficients}
\end{align}
The residuals are
\begin{align}
    \mathrm{RMS}&=3.514\times10^{-6}~,\quad &
    \mathrm{LRMS}=3.836\times10^{-6}~,\\
    \mathrm{BRMS}&=1.043\times10^{-5}~, \quad
    &\kappa=3.828\times10^2~.
    \label{eq:appendix-stress-fit-diagnostics}
\end{align}
The agreement between the RMS and LRMS residuals indicates that
the fit is not simply following pointwise fluctuations. The BRMS residual
is larger, as expected, but remains of order $10^{-5}$ in the normalized
amplitude.

For $\omega=15$, the magnitudes of the successive terms are
\begin{equation}
    |C_{T,0}|\simeq1.611~,\qquad
    |C_{T,1}|\omega^{-4/5}\simeq4.69\times10^{-2}~,\qquad
    |C_{T,2}|\omega^{-8/5}\simeq2.70\times10^{-2}~.
    \label{eq:app-fixed-stress-term-hierarchy}
\end{equation}
The leading term is dominant, although the second correction is still visible. This is consistent with the fact that the data lie in the crossover
region rather than in a parametrically deep finite-charge asymptotic regime.
Table~\ref{tab:app-fixed-stress-window-scan} shows the dependence on the lower endpoint of the fitting window.

\begin{table}[t]
    \centering
    \small
    \begin{tabular}{@{}ccccc@{}}
        \toprule
        Fit window & RMS & LRMS &  BRMS & $C_{T,0}$ \\
        \midrule
        $6.4\leq\omega\leq19.2$
        & $4.469\times10^{-5}$
        & $4.934\times10^{-5}$
        & $1.240\times10^{-4}$
        & $-1.607+0.003684\,\mathrm{i}$ \\
        $8.0\leq\omega\leq19.2$
        & $1.829\times10^{-5}$
        & $1.991\times10^{-5}$
        & $4.972\times10^{-5}$
        & $-1.609+0.003797\,\mathrm{i}$ \\
        $9.6\leq\omega\leq19.2$
        & $8.030\times10^{-6}$
        & $8.693\times10^{-6}$
        & $2.260\times10^{-5}$
        & $-1.610+0.003741\,\mathrm{i}$ \\
        $11.2\leq\omega\leq19.2$
        & $3.514\times10^{-6}$
        & $3.836\times10^{-6}$
        & $1.043\times10^{-5}$
        & $-1.611+0.003619\,\mathrm{i}$ \\
        \bottomrule
    \end{tabular}
        \caption{Stability of the transverse stress-tensor fit at $Q=1.2$ under
    variation of the lower endpoint of the fitting window.}
        \label{tab:app-fixed-stress-window-scan}
\end{table}
The imaginary part of $C_{T,0}$ remains small throughout the scan, while the real part changes by only $3.2\times10^{-3}$. For the main fitting window,
\begin{align}
    C_{T,0}^{\mathrm{fit}}-C_{T,0}^{\mathrm{analytic}}
    &=0.007493+0.003619\,\mathrm{i}~, \quad
    \left|C_{T,0}^{\mathrm{fit}}-C_{T,0}^{\mathrm{analytic}}\right|
    =8.321\times10^{-3}~,\nonumber\\
    \frac{\left|C_{T,0}^{\mathrm{fit}}-C_{T,0}^{\mathrm{analytic}}\right|}
    {\left|C_{T,0}^{\mathrm{analytic}}\right|}
    &=5.143\times10^{-3}~.
    \label{eq:app-fixed-charge-stress-analytic-comparison}
\end{align}
The agreement with $2\cos\big(\frac{4\pi}{5}\big)$ is therefore at the $0.51\%$ level. The corresponding fit in physical units is
\begin{equation}
    \rho_{T,\mathrm{fit}}(\omega)=\pi\omega^4
    +2\pi\omega^4\exp(-\lambda_Q\omega)Y_{T,\mathrm{fit}}(\omega)~.
    \label{eq:appendix-stress-full-fit}
\end{equation}
The associated physical residual is
\begin{equation}
    \left|\rho_T-\rho_{T,\mathrm{fit}}\right|
    =2\pi\omega^4\exp(-\lambda_Q\omega)
    \left|Y_T-Y_{T,\mathrm{fit}}\right|~.
    \label{eq:app-fixed-stress-physical-residual}
\end{equation}
As for the current, pointwise monotonicity of the residual is not expected.
The successive subtractions instead test whether the predicted hierarchy
improves the description of the full spectral density over the fitting range.
\section{Time elapsed along the bouncing null geodesic}
\label{app:Time}
We consider a radial, null geodesic that starts from one boundary of AdS, falls directly into the black hole timelike singularity, and then reflects off to a different AdS boundary. This appendix computes the coordinate time along this bouncing null trajectory. This quantity is not well-defined on a real integration contour because of singularities corresponding to the presence of horizons; however, by selecting a different contour that detours in complex $r$-plane, the answer is finite. One choice of such a contour is along the real line from $r\to \infty$ towards $r=0$, except for two semicircular indentations in the upper complex plane around $r=r_+$ and $r=r_-$. The contour then goes back from $r=0$ towards $r\to \infty$ mostly on the real line, except for two semicircles in the lower imaginary plane at $r=r_+$ and $r=r_-$. We call this contour ``$C_0$'' and use a different contour in the main text ``$\mathcal C$'', with the same result for the time elapsed in both contours.     The integral about the contour $C_0$ gives an identical result to the one associated with $\mathcal C$ because they can be smoothly deformed to each other without encountering any singularities.  The time along the bouncing geodesic is given by
	\begin{align}
    \label{dt}
		\Delta t_Q =\int_{C_0} \frac{\diff r}{r^2 f(r)}~,
	\end{align}
	Plugging the expression $f(r)=1 - \frac{r_0^4}{r^4} + \frac{Q^2}{r^6}$ into Eq.~(\ref{dt}), one obtains
			\begin{align}
		\Delta t_Q = \int_{C_0} \frac{\diff r  \ r^4}{(r^2 - r_+ ^2) (r^2 - r_- ^2) (r^2 + r_q ^2)}~.
	\end{align}
	The method of partial fraction  decomposition is apt to solve the relevant integral:
	\begin{align}
		&\frac{r^4}{(r^2 - r_+ ^2) (r^2 - r_- ^2) (r^2 + r_q ^2)} = \frac{A}{r^2 - r_+ ^2} + \frac{B}{r^2 - r_- ^2}  + \frac{C}{r^2 + r_q ^2}~.
	\end{align}
	The following equations determine the constants $A,B$ and $C$:
	\begin{align}
		&~A + B + C = 1~, \nonumber\\ 
		&~A( r_q ^2 -r_- ^2) + B(r_q ^2 - r_+ ^2) - C( r_+ ^2 + r_- ^2) = 0~, \nonumber \\
		 &- A r_- ^2 r_q ^2 - B r_+ ^2 r_q ^2 + C r_+ ^2 r_- ^2 =0 ~.
	\end{align}
	These are three equations in the three variables $A,B,C$. The following is their unique solution:
	\begin{align}
    \label{ABC}
		A &= \frac{r_+ ^4}{(r_+ ^2 - r_- ^2) (r_+ ^2 + r_q ^2)}~, \nonumber\\ 
		 B &= -\frac{r_- ^4}{ (r_+ ^2 - r_- ^2) (r_- ^2 + r_q ^2)}~, \nonumber\\ 
		 C &= \frac{r_q ^4}{(r_+ ^2 + r_q ^2) (r_- ^2 + r_q ^2)}~.
	\end{align}
	A further partial fraction decomposition allows one to solve for the integral $\frac{A}{r^2 - r_+^2}$:
	\begin{equation}
    \label{intermediate}
		\frac{A}{ r^2 -r_+ ^2} = \frac{A}{2r_+} \Big( \frac{1}{r-r_+} - \frac{1}{r+r_+}\Big)~,
	\end{equation}
    and therefore,
    \begin{equation}
    	\int_{C_0} \frac{\diff r  \ A}{r^2 - r_+ ^2} =\frac{A}{2r_+} \times 2\pi \ic  = \frac{2\ic\pi r_+ ^3}{2(r_+ ^2 - r_- ^2) (r_+ ^2 + r_q^2)}=\frac{\ic \beta_+}{2}~.
    \end{equation}
    In the first transition, Eq.~(\ref{intermediate}) was used and an anti-clockwise winding of the contour produced the factor of $2\pi \ic$; in the second transition, $A$ from Eq.~(\ref{ABC}) was substituted, and in the third transition, Eq.~(\ref{betap}) was used. Likewise,
    \begin{equation}
   \int_{C_0} \frac{\diff r  \ B}{r^2-r_- ^2}=- \frac{\ic\pi r_-  ^3}{(r_+ ^2 - r_- ^2) (r_- ^2 + r_q ^2)}=\frac{\ic \beta_-}{2}~.
    \end{equation}
    The sum of the two expressions above is
    \begin{align}
    	\int_{\infty} ^0 \frac{\diff r  \ A}{r^2 - r_+ ^2} +\int_{\infty} ^0 \frac{\diff r  \ B}{r^2 - r_- ^2} = \frac{\ic\pi}{2(r_+ + r_-)} \frac{r_+ ^4 + r_- ^4 + r_+ r_- ^3 + r_- r_+ ^3 + 3 r_+ ^2 r_-^2}{(2r_- ^2 + r_+ ^2) (2r_+ ^2 + r_- ^2)}~. 
    \end{align}
    Another piece comes from $\frac{C}{r^2 + r_q ^2}$:
    \begin{align}
    	\int_{C_0}  \frac{\diff r  \ C}{r^2 + r_q^2} = \frac{\pi C}{r_q}= \frac{\pi (r_+ ^2 + r_- ^2)^{\frac{3}{2}}}{(r_+ ^2 + r_q ^2) (r_- ^2 + r_q ^2)}~.
    \end{align}
    Collecting the various contributions to $\Delta t_Q$, the result is
    \begin{align}
    	\Delta t_Q=  \frac{\pi (r_+ ^2 + r_- ^2)^{\frac{3}{2}}}{(2r_+ ^2 + r_- ^2) (2r_- ^2 + r_+ ^2)} +\ic\pi \frac{r_+ ^4 + r_- ^4 + r_+ r_- ^3 + r_- r_+ ^3 + 3 r_+ ^2 r_-^2}{(r_+ + r_-)(2r_- ^2 + r_+ ^2) (2r_+ ^2 + r_- ^2)}~.  
    \end{align}
Below, we express the result directly in terms of the mass density and charge density parameters $r_0^4$ and $Q$. 
    
    The equation that determines where the horizons are is
    \begin{equation}
    	r^6 - r_0 ^4 r^2 + Q^2 = 0~.
    \end{equation}
    Denote 
    \begin{equation}
    	x \equiv \frac{r^2}{r_0 ^2}~,~ q\equiv \frac{3\sqrt{3}Q^2}{2r_0 ^6}~.
    \end{equation}
    the cubic equation is
    \begin{equation}
    	x^3 - x+ \frac{2}{3\sqrt{3}}q=0~. 
    \end{equation}
    The discriminant of this equation is
    \begin{equation}
    	\Delta = 4- 4 q^2~. 
    \end{equation}
    For $\Delta\geq 0$, the cubic equation has three real root, while for $\Delta<0$, one root is real and the two others are complex, and complex conjugate of each other. Focus on the case $\Delta \geq 0$. This means $|q|\leq 1$.  The roots in this case are given by
     \begin{equation}
     	x_1 = \frac{2}{\sqrt{3}} \cos\Big(\frac{\pi - \arccos(q)}{3}\Big)~, 
     \end{equation}
          \begin{equation}
     	x_2 = \frac{2}{\sqrt{3}} \cos\Big(\frac{\pi + \arccos(q)}{3}\Big)~, 
     \end{equation}
          \begin{equation}
     	x_3 = -\frac{2}{\sqrt{3}} \cos\Big(\frac{ \arccos(q)}{3}\Big)~. 
     \end{equation}
     When $q=0$, $x_1=1$, $x_2= 0$ and $x_3 = -1$, which translates into one horizon at $r=r_0$. When $q=1$, then $x_1 = \frac{1}{\sqrt{3}}, ~ x_2 =\frac{1}{\sqrt{3}}, x_3 = -\frac{2}{\sqrt{3}} $, which means an extremal horizon at $r_{\rm ext}= \frac{r_0}{ 3^{\frac{1}{4}}}$~.
     Identifying the outer horizon with $r_0 \sqrt{x_1}$ and the inner horizon with $r_0 \sqrt{x_2}$, the following formulas apply:
     \begin{align}
     	r_+ = r_0 \sqrt{ \frac{2}{\sqrt{3}} \cos\Big(\frac{\pi - \arccos(q)}{3}\Big)} ~.
     \end{align}
      \begin{align}
     	r_- = r_0 \sqrt{ \frac{2}{\sqrt{3}} \cos\Big(\frac{\pi + \arccos(q)}{3}\Big)} ~.
     \end{align}
     The real part of the time along the bouncing geodesic is
     \begin{align}
     	&\text{Re}(\Delta t_Q)=\frac{\pi (r_+ ^2 + r_- ^2)^{\frac{3}{2}}}{(2r_+ ^2 + r_- ^2) (2r_- ^2 + r_+ ^2)}=\frac{\pi 3^{\frac{1}{4}}}{\sqrt{2}r_0}\times 
     	\nonumber\\
     	&\frac{ \Big(\cos\Big(\frac{\pi - \arccos(q)}{3}\Big) + \cos\Big(\frac{\pi + \arccos(q)}{3}\Big)\Big)^{\frac{3}{2}}}{\Big(2\cos\Big(\frac{\pi - \arccos(q)}{3}\Big) + \cos\Big(\frac{\pi + \arccos(q)}{3}\Big)\Big) \Big(2\cos\Big(\frac{\pi + \arccos(q)}{3}\Big) + \cos\Big(\frac{\pi - \arccos(q)}{3}\Big) \Big)\Big)}~.
     \end{align}
\section{Small-charge numerical analysis}
\label{app:small-charge-numerics}

We next consider a small charge parameter,
\begin{equation}
    Q=0.05~,\qquad r_+=1~,\qquad \vec{k}=0~.
    \label{eq:app-small-charge-parameters}
\end{equation}
In the frequency range available numerically, the WKB-breakdown scale lies
outside the inner-horizon scale,
\begin{equation}
    r_{\rm WKB}\gg r_-.
    \label{eq:app-small-charge-wkb-scale}
\end{equation}
The local reflection problem is therefore Schwarzschild-like rather than
charge-dominated. The propagation factor, however, is still evaluated using
the exact charged geometry. With the conventions of
Section~\ref{sec:small-charge-regime},
\begin{equation}
    z_Q=\frac{\beta_{\mathrm{eff}}\omega}{\pi}
    =\frac{4\omega}{2-Q^2}~,\qquad
    \beta_{\mathrm{eff}}=\frac{4\pi}{2-Q^2}~.
    \label{eq:app-small-charge-effective-variables}
\end{equation}
For $Q=0.05$, we have
\begin{equation}
\tau_Q=3.1338~,\qquad
    \lambda_Q=3.1447~,\qquad
    \beta_{\mathrm{eff}}=6.2910~.
    \label{eq:small-appendix-path}
\end{equation}
These values are close to the uncharged result
$\tau_0=\lambda_0=\pi$.

     \begin{figure}[t!]
    \centering
    \includegraphics[width=0.8\textwidth]{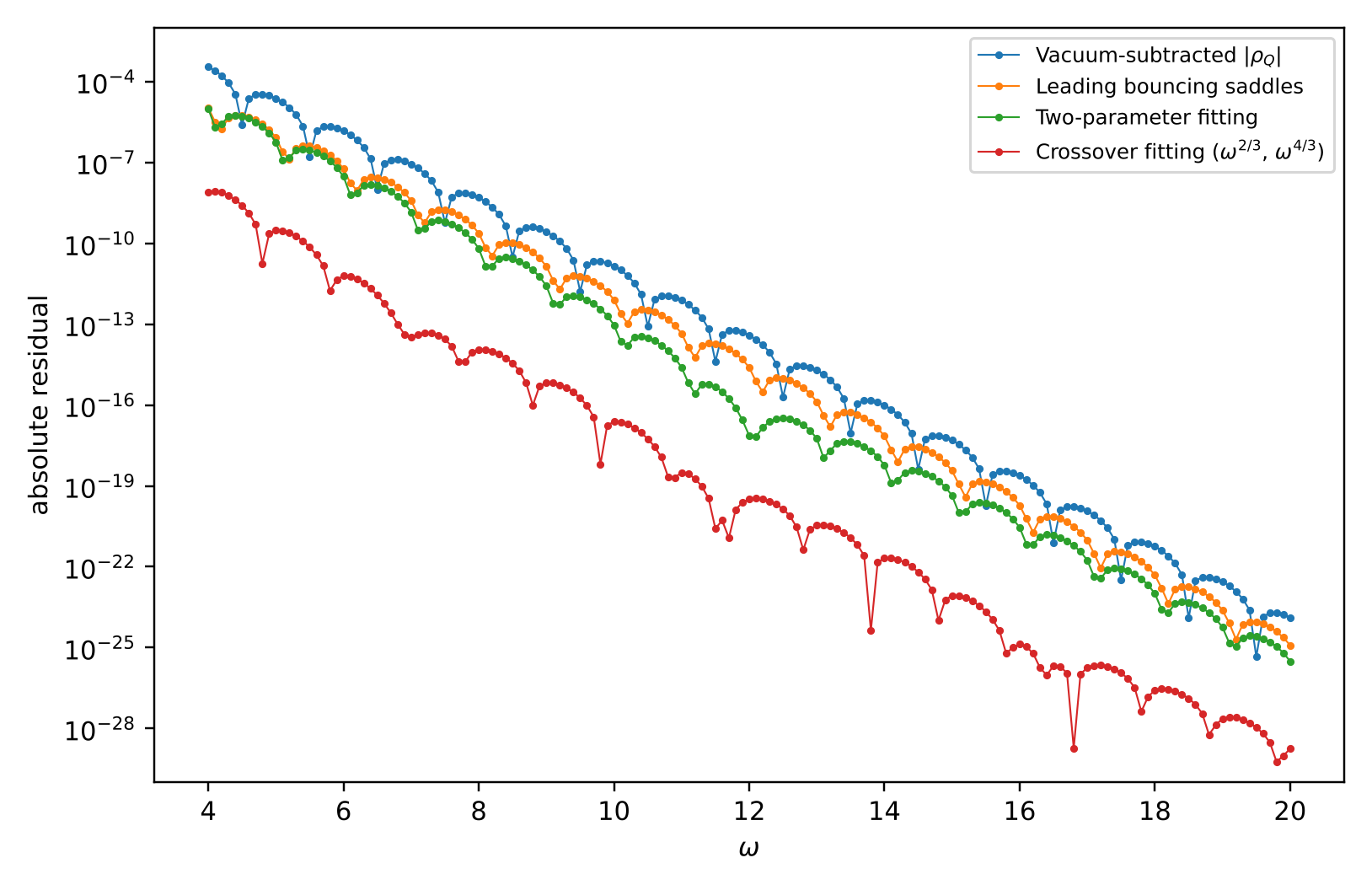}
    \caption{Successive subtractions from the transverse-current spectral
    density at $Q=0.05$.}
    \label{fig:small-charge-current-tails}
\end{figure}

\begin{figure}[h!]
    \centering
    \includegraphics[width=0.8\textwidth]{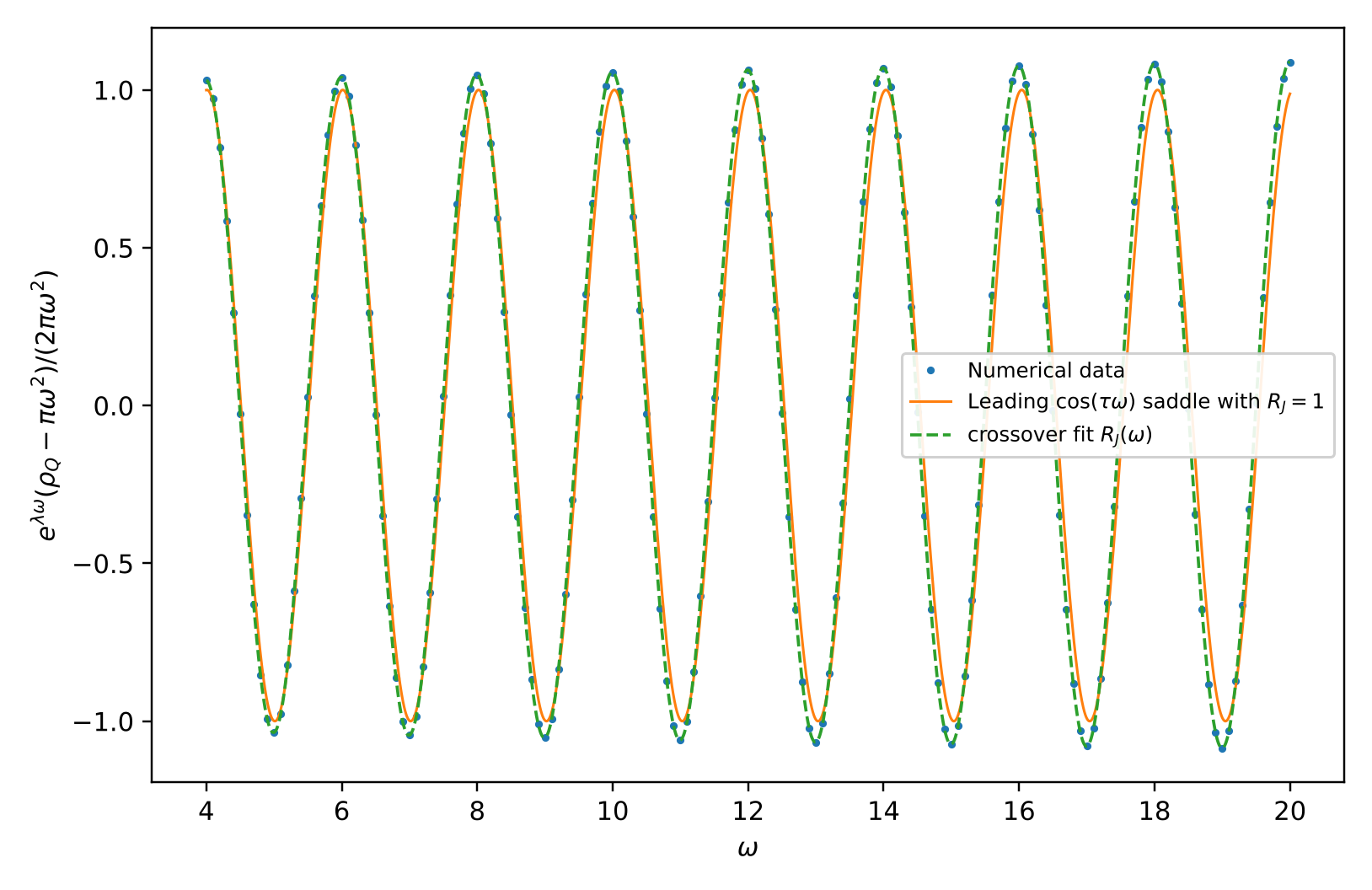}
    \caption{Normalized transverse-current spectral density at $Q=0.05$ after
    removing the leading exponential envelope.}
    \label{fig:small-charge-current-fit}
\end{figure}

\subsection{Transverse current}
\label{app:small-charge-current}

We define
\begin{equation}
    Y_J(\omega,Q)=\frac{\exp(\lambda_Q\omega)}{2\pi\omega^2}
    \left[\rho_J(\omega,Q)-\pi\omega^2\right]~.
    \label{eq:app-small-current-normalized}
\end{equation}
The leading Schwarzschild-like approximation retains the charged propagation
time but uses the uncharged Airy connection coefficient:
\begin{equation}
    \rho_J^{(1)}(\omega)=\pi\omega^2
    +2\pi\omega^2\exp(-\lambda_Q\omega)\cos(\tau_Q\omega)~.
    \label{eq:small-appendix-current-leading}
\end{equation}

To quantify the small departure from this parameter-free result, we fit the
data in the interval $6\leq\omega\leq18$ to
\begin{equation}
    Y_J(\omega,0.05)=A_J\cos(\tau_Q\omega)+B_J\sin(\tau_Q\omega)
    =\mathrm{Re}\!\left[\mathcal{R}_J\exp(\mathrm{i}\tau_Q\omega)\right],
    \qquad \mathcal{R}_J=A_J-\mathrm{i}B_J~.
    \label{eq:app-small-current-complex-amplitude}
\end{equation}
The fit gives
\begin{align}
    A_J&=1.0082,& B_J&=-0.02900,\nonumber\\
    |\mathcal{R}_J|&=1.0086,& \arg\mathcal{R}_J&=0.02875~,
    \label{eq:app-small-current-fit-data}
\end{align}
with normalized residual $    \mathrm{RMS}_Y=3.940\times10^{-3}.$ Equivalently,
\begin{equation}
    \mathcal{R}_J(0.05)=1.0082+0.02900\,\mathrm{i}~.
    \label{eq:small-appendix-current-result}
\end{equation}
The magnitude differs from the uncharged Airy value by $0.86\%$. At this
accuracy, the small displacement and phase should be regarded as effective
finite-window corrections to the Schwarzschild-like connection problem, rather
than as separate determinations of individual charged subleading terms.

\subsection{Transverse stress tensor}
\label{app:small-charge-stress}

The shear channel is normalized so that the leading uncharged coefficient
contains the combined future- and past-reflected contributions. In the
uncharged limit,
\begin{equation}
    \rho_T(\omega,0)-\pi\omega^4=\pi\omega^4\exp(-\pi\omega)
    \mathrm{Re}\!\left[\exp(\mathrm{i}\pi\omega)
    \left(C_{T,0}^{(0)}+C_{T,1}^{(0)}z_0^{-4/3}+\cdots\right)\right]~,
    \label{eq:app-small-stress-uncharged-expansion}
\end{equation}
where
\begin{equation}
    z_0=2\omega,\qquad
    C_{T,0}^{(0)}=-4,\qquad
    C_{T,1}^{(0)}=6.4664\,\exp\!\left(-\frac{2\pi\mathrm{i}}{3}\right)
    =-3.2332-5.6001\,\mathrm{i}~.
    \label{eq:app-small-stress-uncharged-coefficients}
\end{equation}
The leading coefficient is the sum of the two reflected saddles:
\begin{equation}
    C_{T,0}^{(0)}=2\mathcal{R}_T^{(0)},\qquad
    \mathcal{R}_T^{(0)}=-2~.
    \label{eq:app-small-stress-uncharged-reflection}
\end{equation}

At small nonzero $Q$, we use
\begin{equation}
    N_T(\omega,Q)=\frac{\exp(\lambda_Q\omega)}{\pi\omega^4}
    \left[\rho_T(\omega,Q)-\pi\omega^4\right]~.
    \label{eq:app-small-stress-normalized}
\end{equation}
The normalization differs by a factor of two from that used for the finite-charge shear analysis precisely so that the uncharged leading coefficient
is $-4$.

The available data are sufficient to test the leading reflection coefficient,
but not to resolve independently all subleading charged corrections. We
therefore retain the known uncharged first correction and fit only the leading
complex coefficient:
\begin{equation}
    N_T(\omega,Q)=\mathrm{Re}\!\left[
    \exp(\mathrm{i}\tau_Q\omega)
    \left(C_{T,0}^{\mathrm{eff}}(Q)
    +C_{T,1}^{(0)}z_Q^{-4/3}\right)\right]~.
    \label{eq:app-small-stress-constrained-fit}
\end{equation}
The superscript ``eff'' emphasizes that this is the coefficient determined
over the finite frequency range of the fit.

\subsubsection{Numerical methodology}
To motivate our numerics, suppose we first use the numerics for frequencies within $\omega =4.25$ to $\omega = 7.5$. Fitting them gives
\begin{equation}
    C_{T,0}^{\mathrm{eff}}(0.05)=-4.0104-0.07324\,\mathrm{i}~.
    \label{eq:app-small-stress-constrained-full-fit}
\end{equation}
We essentially do a similar  fit on three different frequency domains: lower-, intermediate-, and higher-frequency selections. On each of these, the fit gives us:
\begin{equation}
    C_{T,0}^{\mathrm{eff}}(0.05)
    =-4.0083-0.07174\,\mathrm{i},\qquad
    -4.0115-0.07185\,\mathrm{i},\qquad
    -4.0135-0.07406\,\mathrm{i}~,
    \label{eq:app-small-stress-constrained-window-scan}
\end{equation}
respectively. The leading coefficient is therefore stable at the precision relevant for the present comparison. Since $C_{T,0}^{\mathrm{eff}}$ includes both reflected saddles, the effective
reflection coefficient associated with a single saddle is $    \mathcal{R}_T^{\mathrm{eff}} =C_{T,0}^{\mathrm{eff}}/ 2$. The full result for the reflection coefficient is thus
\begin{equation}
    \mathcal{R}_T^{\mathrm{eff}}(0.05)
    =-2.0052-0.03662\,\mathrm{i}~,
    \label{eq:app-small-stress-effective-reflection}
\end{equation}
while the variation across the three frequency selections is
\begin{equation}
    \mathcal{R}_T^{\mathrm{eff}}(0.05)\simeq
    -(2.004\text{--}2.007)-\mathrm{i}(0.036\text{--}0.037)~.
    \label{eq:app-small-stress-effective-reflection-range}
\end{equation}
Thus, the leading shear reflection data remain close to the uncharged
AdS-Schwarzschild value,
\begin{equation}
    \mathcal{R}_T^{(0)}=-2~,
\end{equation}
with a small phase induced by the finite charge parameter. The purpose of this fit is to test the continuity of the leading
Schwarzschild-like reflection coefficient at small $Q$. A separate extraction
of charge-dependent subleading coefficients would require a broader
frequency range and additional analytic control of the higher terms in the
uncharged expansion.

\section{Proof of formula (\ref{eq:charged-bessel-connection-formula})}
\label{app:2}
In this appendix, we use identities appearing in the mathematical textbook due to Abramowitz and Stegun \cite{AS} to demonstrate Eq.~(\ref{eq:charged-bessel-connection-formula}) in the main text.
	Eqs.~(9.6.4) of this reference tie the modified Bessel functions and the Hankel functions:
	\begin{align}
		\label{AS964}
		K_{\nu}(z) &= \frac{\pi \ic}{2} e^{\frac{1}{2}\nu \pi \ic}  H_{\nu} ^{(1)} (z e^{\frac{\pi}{2} \ic })~,~-\pi < \text{arg}(z)\leq \frac{\pi}{2}~.
	\end{align}
Eq.~(9.6.31) of \cite{AS} is useful to transition the modified Bessel function of the second kind between sheets in the complex plane:
	\begin{equation}
		\label{AS9631}
		K_{\nu} (z e^{m\pi \ic})=e^{-m\nu \pi \ic } K_{\nu} (z)-\pi \ic \sin(\pi m \nu)\csc(\nu \pi) I_{\nu} (z)~.
	\end{equation}
    Setting $m=2$ in Eq.~(\ref{AS9631}) yields
	\begin{equation}
		K_{\nu} (z e^{2\pi \ic}) = e^{-2\pi \ic  \nu} K_{\nu} (z) - 2\pi \ic \cos (\pi \nu) I_{\nu} (z)~.
	\end{equation}
	Write $z= e^{-\frac{\pi \ic}{2}}x$ with $x>0$:
		\begin{equation}
        \label{ingredient}
		K_{\nu} \big(x e^{\frac{3}{2}\pi \ic}\big) = e^{-2\pi \ic \nu} K_{\nu} (x e^{-\frac{\pi \ic}{2}}) - 2\pi \ic \cos (\pi \nu) I_{\nu} \big(x e^{-\frac{i \pi}{2}}\big)~.
	\end{equation}
    We want to evaluate the R.H.S. of the last equations.
    	Plugging $z=e^{-\frac{\pi \ic}{2}} x$ into Eq.~(\ref{AS964}) gives
	\begin{equation}
		\label{AS3}
		K_{\nu}\big(e^{-\frac{\pi \ic}{2}}x\big) = -\frac{\pi}{2\ic} e^{\frac{1}{2}\pi \ic \nu} H_{\nu} ^{(1)} (x)~.
	\end{equation}
        Eq.~(9.6.3) of \cite{AS} ties the modified Bessel function of the first kind and an analytical continuation of the conventional Bessel function of the first kind: 
	\begin{equation}
		\label{AS963}
		I_{\nu} (z) = e^{-\frac{\pi \ic \nu}{2}} J_{\nu} \big(ze^{\frac{\pi \ic}{2}}\big) ~,~ -\pi < \text{arg}(z)\leq \frac{\pi}{2}~.
	\end{equation}
		Now use Eqs.~(\ref{AS3}) and (\ref{AS963}) in the R.H.S. of Eq.~(\ref{ingredient}): 
			\begin{equation}
				\label{Knu}
		K_{\nu} \big(x e^{\frac{3}{2}\pi \ic}\big) = -\frac{\pi }{2i}e^{-\frac{3}{2}\pi \ic \nu} H_{\nu} ^{(1)}(x)  - 2\pi \ic \cos (\pi \nu) e^{-\frac{\pi \ic \nu}{2}} J_{\nu}\big(x \big)~.
	\end{equation}
	Next, the sum of Eqs.~(9.1.3) and (9.1.4) of \cite{AS} gives
 	\begin{equation}
		J_{\nu} (x) = \frac{1}{2} \Big( H_{\nu} ^{(1)} (x) + H_{\nu} ^{(2)} (x)\Big)~,
	\end{equation}
	and plugging this last equation into equation~(\ref{Knu}), the result is
				\begin{align}
		K_{\nu} \big(x e^{\frac{3}{2}\pi \ic}\big) &= \frac{\pi \ic }{2}e^{-\frac{3}{2}\pi \ic \nu} H_{\nu} ^{(1)}(x)  - \pi \ic \cos (\pi \nu) e^{-\frac{\pi \ic \nu}{2}} \Big(H_{\nu} ^{(1)} (x)+ H_{\nu} ^{(2)} (x)\Big)\nonumber\\
	&= \frac{\pi \ic }{2}e^{-\frac{3}{2}\pi \ic \nu} \Big( H_{\nu} ^{(1)}(x)  - 2 \cos (\pi \nu) e^{\pi \ic \nu} \Big(H_{\nu} ^{(1)} (x)+ H_{\nu} ^{(2)} (x)\Big)\Big)~.
	\end{align}
	Simplifying,
			\begin{equation}
		K_{\nu} \big(x e^{\frac{3}{2}\pi \ic}\big) = -\frac{\pi \ic e^{\frac{1}{2}\pi \ic \nu}}{2}  \Bigg( H_{\nu} ^{(1)}(x)  + 2 \cos (\pi \nu) e^{-\pi \ic \nu}  H_{\nu} ^{(2)} (x)\Bigg)~.
	\end{equation}
    This proves that Eq.~(\ref{eq:charged-bessel-connection-formula}) is correct, which is one of the technical results of the paper. 

    \section{Matching solutions from near the singularity to the boundary}
\label{app:local-to-boundary-matching}

In this appendix, we derive a relation between the near-singularity reflection coefficients and the leading coefficients of the boundary spectral density transseries. To this end, three steps follow: (a) The solution near the singularity, given as a superposition of Hanekl functions, is matched onto a WKB-approximated solution away from the singularity; (b) The WKB-approximated solution is then propagated to the AdS boundary;  (c) We normalize sources and evaluate the conserved radial flux to convert the latter near-boundary solution into the leading non-perturbative correction to the vacuum spectral density.

\subsection*{WKB phases and the one-bounce contour}
\label{appsec:wkb-phases-one-bounce}
The radial equations that govern bulk perturbations take the Schr\"odinger form as in \eqref{eq:schrodinger-radial-equation}. To leading order at large $\omega$, the WKB approximation gives two independent branches given by
\begin{equation}
    \psi_{s,\pm}^{\rm WKB}(r)
    =
    \frac{1}{\sqrt{\omega}}
    \exp\!\left[
        \pm \mathrm{i}\omega \mathcal{S}(r)
    \right]
    \left[
        1+O(\omega^{-1})
    \right]~,
    \label{eq:appendix-wkb-branches}
\end{equation}
where $     \mathcal{S}(r)
    \equiv    \int^r\frac{\diff r'}{r'^2f(r')}.$
The integral here is understood on the branch that starts at the positive real axis near the singularity and is analytically continued along the retarded contour of Section~\ref{sec:two-sided-geometry}.

Near the timelike singularity, the term in $f(r)$ proportional to $Q^2$ dominates,
\begin{equation}
    f(r)
    \approx 
    \frac{Q^2}{r^6}~,
    \label{eq:appendix-near-singularity-blackening}
\end{equation}
and hence $\mathcal{S}(r) \approx \frac{r^5}{5Q^2}.$ Thus, on the positive real $r$ axis,
\begin{equation}
    X
    \equiv
    \omega \mathcal{S}(r)
    \approx 
    \frac{\omega r^5}{5Q^2}
    ~.
    \label{eq:appendix-local-wkb-variable}
\end{equation}
Near the asymptotic AdS boundary, the same WKB action has the form
\begin{equation}
    \mathcal{S}(r)
    =
    \mathcal{S}_\infty-\frac{1}{r}
    + O(r^{-5})~,
    \qquad
    r\rightarrow\infty~,
    \label{eq:appendix-boundary-wkb-action}
\end{equation}
where $\mathcal{S}_\infty$ is a complex constant determined by the contour. Writing
\begin{equation}
    z\equiv\frac{\omega}{r}~,
    \label{eq:appendix-boundary-variable}
\end{equation}
the two WKB branches therefore behave as
\begin{align}
    \psi_{s,+}^{\rm WKB}
    &\propto
    e^{+\mathrm{i}\omega \mathcal{S}_\infty}
    e^{-\mathrm{i}z}~,
    \label{eq:appendix-plus-wkb-boundary}\\
    \psi_{s,-}^{\rm WKB}
    &\propto
    e^{-\mathrm{i}\omega \mathcal{S}_\infty}
    e^{+\mathrm{i}z}~.
    \label{eq:appendix-minus-wkb-boundary}
\end{align}
The one-bounce contour consists of two WKB branches related by the reflection from the near-singularity region. With the orientation used in this paper, the full bounce contour is homologous to minus twice the one-leg contour defining $\mathcal{S}_\infty$. Consequently,
\begin{equation}
    \Delta t_Q
    =
    \int_{\mathcal C}
    \frac{\diff r'}{(r')^2f(r')}
    =
    -2\mathcal{S}_\infty~.
    \label{eq:appendix-bounce-action-identity}
\end{equation}
It follows that the relative propagation factor of the reflected WKB branch is
\begin{equation}
    e^{-2\mathrm{i}\omega \mathcal{S}_\infty}
    =
    e^{\mathrm{i}\omega\Delta t_Q}
    =
    e^{-\lambda_Q\omega}
    e^{\mathrm{i}\tau_Q\omega}~.
    \label{eq:appendix-relative-propagation-factor}
\end{equation}
Equation~\eqref{eq:appendix-relative-propagation-factor} is the precise sense in which the global part of the one-bounce contribution is controlled by the complex null coordinate time in the bouncing geodesic.
\subsection*{Near-singularity Hankel basis and overlap with WKB}
\label{appsec:local-hankel-overlap}

The near-singularity Bessel indices are
\begin{equation}
    \nu_J^{\rm loc}=\frac{4}{5}~,
    \qquad
    \nu_T^{\rm loc}=\frac{1}{5}~.
    \label{eq:appendix-local-indices}
\end{equation}
The steepest-descent solution, continued from the decaying ray in the lower-half-plane, takes the form
\begin{equation}
    \Phi_s^{\rm sd}(X)
    \propto
    X^{\nu_s}
    \left[
        H^{(1)}_{\nu_s^{\rm loc}}(X)
        -
        R_s^{\rm loc}
        H^{(2)}_{\nu_s^{\rm loc}}(X)
    \right]~,
    \label{eq:appendix-local-steepest-descent-solution}
\end{equation}
where the near-singularity reflection coefficient is:
\begin{equation}
    R_s^{\rm loc}
    =
    -2\cos\!\left(\pi\nu_s^{\rm loc}\right)
    e^{-\mathrm{i}\pi\nu_s^{\rm loc}}~.
    \label{eq:appendix-general-local-reflection-coefficient}
\end{equation}

In the overlap region, where $X\gg1$ while the near-singularity
 approximation remains valid, the large-$X$ Hankel
asymptotics are
\begin{align}
    H^{(1)}_{\nu_s^{\rm loc}}(X)
    \sim
    \sqrt{\frac{2}{\pi X}}\,
    e^{+\mathrm{i}(X-\alpha_s^{\rm loc})}, \qquad 
    H^{(2)}_{\nu_s^{\rm loc}}(X)
    \sim
    \sqrt{\frac{2}{\pi X}}\,
    e^{-\mathrm{i}(X-\alpha_s^{\rm loc})}~,
    \label{eq:appendix-local-hankel-two}
\end{align}
with
\begin{equation}
    \alpha_s^{\rm loc}
    \equiv
    \frac{\pi\nu_s^{\rm loc}}{2}
    +\frac{\pi}{4}~.
    \label{eq:appendix-local-hankel-phase}
\end{equation}
Using Eq.~\eqref{eq:appendix-local-wkb-variable}, the near-singularity Hankel
branches are related to the WKB branches by
\begin{align}
   X^{\nu_s^{\rm loc}} H^{(1)}_{\nu_s^{\rm loc}}(X)
    \longrightarrow
    e^{-\mathrm{i}\alpha_s^{\rm loc}}
    \Phi_{s,+}^{\rm WKB}, \qquad
    X^{\nu_s^{\rm loc}} H^{(2)}_{\nu_s^{\rm loc}}(X)
    \longrightarrow
    e^{+\mathrm{i}\alpha_s^{\rm loc}}
    \Phi_{s,-}^{\rm WKB}~.
    \label{eq:appendix-local-hankel-to-minus-wkb}
\end{align}
The overall real normalization common to both terms will not be needed.

\subsection*{Boundary Hankel basis}
\label{appsec:boundary-hankel-basis}

For large $r$, the blackening factor $f(r)$ approaches unity and the transverse gauge field and transverse graviton equations reduce to their pure-AdS forms. The two independent solutions near the boundary can be written as
\begin{align}
    \Phi_s(z)=
    z^s H_s ^{(1)}(z)~~, ~~
    z^s H_s ^{(2)}(z)~.
    \label{eq:appendix-stress-boundary-basis}
\end{align}
Thus, the boundary Hankel indices are
\begin{equation}
    \nu_J^\partial=1~,
    \qquad
    \nu_T^\partial=2~.
    \label{eq:appendix-boundary-hankel-indices}
\end{equation}
The large-$z$ asymptotics of the Hankel functions are given by
\begin{align}
    H^{(1)}_{\nu_s^\partial}(z)\sim
    \sqrt{\frac{2}{\pi z}}~\,
    e^{+\mathrm{i}(z-\alpha_s^\partial)}, \qquad
    H^{(2)}_{\nu_s^\partial}(z)\sim
    \sqrt{\frac{2}{\pi z}}~\,
    e^{-\mathrm{i}(z-\alpha_s^\partial)}~,
    \label{eq:appendix-boundary-hankel-two}
\end{align}
where
\begin{equation}
    \alpha_s^\partial
    \equiv
    \frac{\pi\nu_s^\partial}{2}
    +\frac{\pi}{4}~.
    \label{eq:appendix-boundary-hankel-phase}
\end{equation}
The field-dependent powers in
Eq.~\eqref{eq:appendix-stress-boundary-basis} reproduce the WKB prefactors. Comparing
Eqs.~\eqref{eq:appendix-plus-wkb-boundary}--
\eqref{eq:appendix-minus-wkb-boundary} with
Eq.~\eqref{eq:appendix-boundary-hankel-two} gives
\begin{align}
    \Phi_{s,+}^{\rm WKB}\longrightarrow
    e^{+\mathrm{i}\omega \mathcal{S}_\infty}
    e^{-\mathrm{i}\alpha_s^\partial}
    \Phi_s^{(2)}~, \qquad
    \Phi_{s,-}^{\rm WKB}\longrightarrow
    e^{-\mathrm{i}\omega \mathcal{S}_\infty}
    e^{+\mathrm{i}\alpha_s^\partial}
    \Phi_s^{(1)}~.
    \label{eq:appendix-minus-wkb-to-boundary}
\end{align}
In particular, the branch with near-singularity phase $e^{+\mathrm{i}X}$ becomes
the boundary $H^{(2)}$ branch, whereas the branch with near-singularity phase
$e^{-\mathrm{i}X}$ becomes the boundary $H^{(1)}$ branch.

The decaying steepest-descent branch supplies the canonically resummed perturbative solution. Its continuation through the local near-singularity connection region generates an exponentially small reflected branch. After WKB transport to the asymptotic region, this reflected branch provides the Stokes admixture required to reconstruct the physical solution satisfying the infalling outer-horizon boundary condition. Combining the local connection formula with the WKB and boundary matching relations gives 
\begin{equation} 
\Phi_s^{\mathrm{in}} \propto e^{\mathrm{i}\omega\mathcal S_\infty} e^{\mathrm{i} (\alpha_s^{\mathrm{loc}}+\alpha_s^\partial)} \Phi_s^{(2)} - R_s^{\mathrm{loc}} e^{-\mathrm{i}\omega\mathcal S_\infty} e^{-\mathrm{i} (\alpha_s^{\mathrm{loc}}+\alpha_s^\partial)} \Phi_s^{(1)}~. 
\label{eq:appendix-full-boundary-matching} 
\end{equation} 
Factoring out the coefficient of $\Phi_s^{(2)}$, the physical infalling solution may be written as 
\begin{equation} 
\Phi_s^{\mathrm{in}} \propto \Phi_s^{(2)} + \mathcal E_s(\omega,Q)\Phi_s^{(1)}~, 
\label{eq:appendix-boundary-admixture} 
\end{equation} 
where  we have 
\begin{equation} 
\mathcal E_s(\omega,Q) = -R_s^{\mathrm{loc}} e^{2\mathrm{i} (\alpha_s^{\mathrm{loc}}+\alpha_s^\partial)} e^{-2\mathrm{i}\omega\mathcal S_\infty} = -R_s^{\mathrm{loc}} e^{2\mathrm{i} (\alpha_s^{\mathrm{loc}}+\alpha_s^\partial)} e^{\mathrm{i}\omega\Delta t_Q}~. 
\label{eq:appendix-boundary-admixture-coefficient} 
\end{equation}
This is the matching formula between the solution near the timelike singularity and the solution near the AdS boundary.

\subsection*{Source normalization and spectral density}
\label{appsec:source-normalization-spectral-density}

We now show how the small boundary mixing
$\mathcal E_s$ produces the leading correction to the spectral density.
At small $z$, the source terms in the two boundary Hankel functions
have opposite signs. More explicitly,
\begin{align}
    zH_1^{(1)}(z)
    &=
    -\frac{2\mathrm{i}}{\pi}+
 O(z^2\log z)~,
    &
    zH_1^{(2)}(z)
    &=
    +\frac{2\mathrm{i}}{\pi}+
 O(z^2\log z)~,
    \label{eq:appendix-current-boundary-source-asymptotics}\\
    z^2H_2^{(1)}(z)
    &=
    -\frac{4\mathrm{i}}{\pi}+
O(z^2)~,
    &
    z^2H_2^{(2)}(z)
    &=
    +\frac{4\mathrm{i}}{\pi}+
   O(z^2)~.
    \label{eq:appendix-stress-boundary-source-asymptotics}
\end{align}
Consequently, for both channels, the boundary source of the unnormalized solution is proportional to
\begin{equation}
    1-\mathcal E_s(\omega,Q)~.
    \label{eq:appendix-boundary-source-factor}
\end{equation}

The imaginary part of the retarded Green's function can be computed from
the conserved radial flux. The flux is fixed by the infalling normalization at the outer horizon. The small-$z$ source coefficients of the two boundary Hankel branches have opposite signs. Consequently, the source of the unnormalized physical infalling solution in Eq.~\eqref{eq:appendix-boundary-admixture} is proportional to $1-\mathcal E_s$. Formally, normalizing this source produces the factor $|1-\mathcal E_s|^{-2}$. However, terms quadratic in $\mathcal E_s$ are of order $e^{-2\lambda_Q\omega}$, at which independent two-bounce sectors also contribute. The controlled one-bounce result is therefore 
\begin{equation} 
\frac{\rho_s(\omega,Q)} {\rho_{s,\mathrm{vac}}(\omega)} = 1 + 2\,\operatorname{Re} \mathcal E_s(\omega,Q) + O\!\left(e^{-2\lambda_Q\omega}\right)~. 
\label{eq:appendix-spectral-density-from-source-normalization} 
\end{equation} 
Substituting Eq.~\eqref{eq:appendix-boundary-admixture-coefficient} gives 
\begin{equation} 
\frac{\rho_s(\omega,Q)} {\rho_{s,\mathrm{vac}}(\omega)} = 1 + 2e^{-\lambda_Q\omega} \operatorname{Re}\!\left[ e^{\mathrm{i}\tau_Q\omega} C_{s,0} \right] +  O\!\left(e^{-2\lambda_Q\omega}\right), 
\label{eq:appendix-linearized-spectral-density} 
\end{equation} 
where $C_{s,0}$  takes the following form:
\beq
C_{s,0}= -R_s^{\mathrm{loc}} e^{2\mathrm{i} (\alpha_s^{\mathrm{loc}}+\alpha_s^\partial)}~.
\label{eq:appendix-leading-boundary-coefficient-general}
\eeq

\subsection*{Evaluation of the current and stress-tensor coefficients}
\label{appsec:evaluation-boundary-coefficients}

\textbf{Current:} For the transverse current, we have the following values of $\nu$:
\begin{equation}
  \nu_J^{\rm loc}=\tfrac45,\quad \nu_J^\partial=1,\qquad
  ~, 
\end{equation}
and hence using this, we have the following values for $\alpha$:
\begin{equation}
\alpha_J^{\rm loc}=\tfrac{13\pi}{20},\quad \alpha^\partial_J=\tfrac{3\pi}{4},\quad
  2\big(\alpha_J^{\rm loc}+\alpha_J^\partial\big)=\tfrac{14\pi}{5}~.
    \label{eq:appendix-current-total-phase}
\end{equation}
Substitution into Eq.~\eqref{eq:appendix-leading-boundary-coefficient-general}
gives
\begin{align}
  C_{J,0}=-\Big[-2\cos \Big(\frac{4\pi}{5}\Big) \,\e^{-4\pi\ic/5}\,\e^{1 4\pi\ic/5}\Big]=2\cos\Big(\frac{4\pi}{5}\Big)~. 
    \label{eq:appendix-current-boundary-coefficient}
\end{align}

\textbf{Stress-tensor:} For the transverse stress tensor, we have the following values of $\nu$:
\begin{equation}
    \nu_T^{\rm loc}=\frac{1}{5}~,
    \qquad
    \nu_T^\partial=2~,
    \label{eq:appendix-stress-indices}
\end{equation}
and therefore, we have the following values for $\alpha$:
\begin{equation}
    \alpha_T^{\rm loc}
    =
    \frac{7\pi}{20}~,
    \qquad
    \alpha_T^\partial
    =
    \frac{5\pi}{4}~,
    \qquad
    2\left(
        \alpha_T^{\rm loc}+\alpha_T^\partial
    \right)
    =
    \frac{16\pi}{5}~.
    \label{eq:appendix-stress-total-phase}
\end{equation}
Similar to the current case, the near-singularity reflection coefficient is given by:
\begin{equation}
    R_T^{\rm loc}
    =
    -2\cos\!\left(\frac{\pi}{5}\right)
    e^{-\pi\mathrm{i}/5}~.
    \label{eq:appendix-stress-local-reflection}
\end{equation}
Equation~\eqref{eq:appendix-leading-boundary-coefficient-general}
then yields
\begin{align}
    C_{T,0}
    &=
    -\left[
        -2\cos\!\left(\frac{\pi}{5}\right)
        e^{-\pi\mathrm{i}/5}
    \right]
    e^{16\pi\mathrm{i}/5}=
    2\cos\!\left(\frac{4\pi}{5}\right)~.
    \label{eq:appendix-stress-boundary-coefficient}
\end{align}

Equations~\eqref{eq:appendix-current-boundary-coefficient} and
\eqref{eq:appendix-stress-boundary-coefficient} show explicitly how the phases in the near-singularity Bessel connection coefficients combine with the WKB and boundary-Hankel phases. Finally, note that both these leading coefficients in the boundary spectral density are real.

\bibliographystyle{JHEP}
\bibliography{citation.bib}

@article{Fidkowski:2003nf,
    author = "Fidkowski, Lukasz and Hubeny, Veronika and Kleban, Matthew and Shenker, Stephen",
    title = "{The Black hole singularity in AdS / CFT}",
    eprint = "hep-th/0306170",
    archivePrefix = "arXiv",
    reportNumber = "SU-ITP-03-16",
    doi = "10.1088/1126-6708/2004/02/014",
    journal = "JHEP",
    volume = "02",
    pages = "014",
    year = "2004"
}

@article{Brecher:2004gn,
    author = "Brecher, Dominic and He, Jianyang and Rozali, Moshe",
    title = "{On charged black holes in anti-de Sitter space}",
    eprint = "hep-th/0410214",
    archivePrefix = "arXiv",
    doi = "10.1088/1126-6708/2005/04/004",
    journal = "JHEP",
    volume = "04",
    pages = "004",
    year = "2005"
}

@article{Festuccia:2005pi,
    author = "Festuccia, Guido and Liu, Hong",
    title = "{Excursions beyond the horizon: Black hole singularities in Yang-Mills theories. I.}",
    eprint = "hep-th/0506202",
    archivePrefix = "arXiv",
    reportNumber = "MIT-CTP-3641",
    doi = "10.1088/1126-6708/2006/04/044",
    journal = "JHEP",
    volume = "04",
    pages = "044",
    year = "2006"
}

@article{Festuccia:2008zx,
    author = "Festuccia, Guido and Liu, Hong",
    title = "{A Bohr-Sommerfeld quantization formula for quasinormal frequencies of AdS black holes}",
    eprint = "0811.1033",
    archivePrefix = "arXiv",
    primaryClass = "gr-qc",
    reportNumber = "MIT-CTP-3995, SCIPP-08-11",
    doi = "10.1166/asl.2009.1029",
    journal = "Adv. Sci. Lett.",
    volume = "2",
    pages = "221--235",
    year = "2009"
}

@article{Amado:2008hw,
    author = "Amado, Irene and Hoyos-Badajoz, Carlos",
    title = "{AdS black holes as reflecting cavities}",
    eprint = "0807.2337",
    archivePrefix = "arXiv",
    primaryClass = "hep-th",
    reportNumber = "IFT-UAM-CSIC-08-45",
    doi = "10.1088/1126-6708/2008/09/118",
    journal = "JHEP",
    volume = "09",
    pages = "118",
    year = "2008"
}

@article{Iliesiu:2018fao,
    author = "Iliesiu, Luca and Kolo{\u{g}}lu, Murat and Mahajan, Raghu and Perlmutter, Eric and Simmons-Duffin, David",
    title = "{The Conformal Bootstrap at Finite Temperature}",
    eprint = "1802.10266",
    archivePrefix = "arXiv",
    primaryClass = "hep-th",
    reportNumber = "CALT-TH-2018-013, PUPT-2550",
    doi = "10.1007/JHEP10(2018)070",
    journal = "JHEP",
    volume = "10",
    pages = "070",
    year = "2018"
}

@article{Dodelson:2023vrw,
    author = "Dodelson, Matthew and Iossa, Cristoforo and Karlsson, Robin and Zhiboedov, Alexander",
    title = "{A thermal product formula}",
    eprint = "2304.12339",
    archivePrefix = "arXiv",
    primaryClass = "hep-th",
    reportNumber = "CERN-TH-2023-062",
    doi = "10.1007/JHEP01(2024)036",
    journal = "JHEP",
    volume = "01",
    pages = "036",
    year = "2024"
}

@article{Ceplak:2024bja,
    author = "{\v{C}}eplak, Nejc and Liu, Hong and Parnachev, Andrei and Valach, Samuel",
    title = "{Black hole singularity from OPE}",
    eprint = "2404.17286",
    archivePrefix = "arXiv",
    primaryClass = "hep-th",
    doi = "10.1007/JHEP10(2024)105",
    journal = "JHEP",
    volume = "10",
    pages = "105",
    year = "2024"
}

@article{Ceplak:2025dds,
    author = "{\v{C}}eplak, Nejc and Liu, Hong and Parnachev, Andrei and Valach, Samuel",
    title = "{Fooling the censor: going beyond inner horizons with the OPE}",
    eprint = "2511.09638",
    archivePrefix = "arXiv",
    primaryClass = "hep-th",
    doi = "10.1007/JHEP05(2026)001",
    journal = "JHEP",
    volume = "05",
    pages = "001",
    year = "2026"
}

@article{Dodelson:2025jff,
    author = "Dodelson, Matthew and Iossa, Cristoforo and Karlsson, Robin",
    title = "{Bouncing off a stringy singularity}",
    eprint = "2511.09616",
    archivePrefix = "arXiv",
    primaryClass = "hep-th",
    doi = "10.1007/JHEP07(2026)161",
    journal = "JHEP",
    volume = "07",
    pages = "161",
    year = "2026"
}

@article{AliAhmad:2026wem,
    author = "Ali Ahmad, Shadi and Almheiri, Ahmed and Lin, Simon",
    title = "{Continuing past the inner horizon using WKB}",
    eprint = "2601.02354",
    archivePrefix = "arXiv",
    primaryClass = "hep-th",
    month = "1",
    year = "2026"
}

@article{Simpson:1973ua,
    author = "Simpson, M. and Penrose, R.",
    title = "{Internal instability in a Reissner-Nordstrom black hole}",
    journal = "Int. J. Theor. Phys.",
    volume = "7",
    pages = "183--197",
    year = "1973",
    doi = "10.1007/BF00792069"
}

@article{Poisson:1989zz,
    author = "Poisson, Eric and Israel, W.",
    title = "{Inner-horizon instability and mass inflation in black holes}",
    journal = "Phys. Rev. Lett.",
    volume = "63",
    pages = "1663--1666",
    year = "1989",
    doi = "10.1103/PhysRevLett.63.1663"
}

@article{Poisson:1990eh,
    author = "Poisson, Eric and Israel, W.",
    title = "{Internal structure of black holes}",
    journal = "Phys. Rev. D",
    volume = "41",
    pages = "1796--1809",
    year = "1990",
    doi = "10.1103/PhysRevD.41.1796"
}

@article{Ori:1991zz,
    author = "Ori, Amos",
    title = "{Inner structure of a charged black hole: An exact mass-inflation solution}",
    journal = "Phys. Rev. Lett.",
    volume = "67",
    pages = "789--792",
    year = "1991",
    doi = "10.1103/PhysRevLett.67.789"
}

@article{Dafermos:2003wr,
    author = "Dafermos, Mihalis",
    title = "{Stability and instability of the Cauchy horizon for the spherically symmetric Einstein-Maxwell-scalar field equations}",
    journal = "Annals Math.",
    volume = "158",
    pages = "875--928",
    year = "2003",
    doi = "10.4007/annals.2003.158.875"
}

@article{Dafermos:2003yw,
    author = "Dafermos, Mihalis",
    title = "{The interior of charged black holes and the problem of uniqueness in general relativity}",
    eprint = "gr-qc/0307013",
    archivePrefix = "arXiv",
    journal = "Commun. Pure Appl. Math.",
    volume = "58",
    number = "4",
    pages = "445--504",
    year = "2005",
    doi = "10.1002/cpa.20071"
}

@article{Papadodimas:2019ery,
    author = "Papadodimas, Kyriakos and Raju, Suvrat and Shrivastava, Pushkal",
    title = "{A simple quantum test for smooth horizons}",
    eprint = "1910.02992",
    archivePrefix = "arXiv",
    primaryClass = "hep-th",
    journal = "JHEP",
    volume = "12",
    pages = "003",
    year = "2020",
    doi = "10.1007/JHEP12(2020)003"
}

@article{Balasubramanian:2019qwk,
    author = "Balasubramanian, Vijay and Kar, Arjun and S{\'a}rosi, G{\'a}bor",
    title = "{Holographic Probes of Inner Horizons}",
    eprint = "1911.12413",
    archivePrefix = "arXiv",
    primaryClass = "hep-th",
    reportNumber = "CERN-TH-2019-207",
    doi = "10.1007/JHEP06(2020)054",
    journal = "JHEP",
    volume = "06",
    pages = "054",
    year = "2020"
}

@article{Hartnoll:2026vhu,
    author = "Hartnoll, Sean A. and Zhiboedov, Alexander",
    title = "{Can one hear the shape of a black hole singularity?}",
    eprint = "2609.07514",
    archivePrefix = "arXiv",
    primaryClass = "hep-th",
    reportNumber = "CERN-TH-2026-217",
    month = "9",
    year = "2026"
}

@article{Alday:2020vdk,
    author = "Alday, Luis F. and Kologlu, Murat and Zhiboedov, Alexander",
    title = "{Holographic correlators at finite temperature}",
    eprint = "2009.10062",
    archivePrefix = "arXiv",
    primaryClass = "hep-th",
    journal = "JHEP",
    volume = "06",
    pages = "082",
    year = "2021",
    doi = "10.1007/JHEP06(2021)082"
}

@article{Karlsson:2022mzg,
    author = "Karlsson, Robin and Parnachev, Andrei and Prilepina, Valentina and Valach, Samuel",
    title = "{Thermal stress tensor correlators, OPE and holography}",
    eprint = "2206.05544",
    archivePrefix = "arXiv",
    primaryClass = "hep-th",
    journal = "JHEP",
    volume = "09",
    pages = "234",
    year = "2022",
    doi = "10.1007/JHEP09(2022)234"
}

@article{Esper:2023fiv,
    author = "Esper, Chantelle and Huang, Kuo-Wei and Karlsson, Robin and Parnachev, Andrei and Valach, Samuel",
    title = "{Thermal stress tensor correlators near lightcone and holography}",
    eprint = "2306.00787",
    archivePrefix = "arXiv",
    primaryClass = "hep-th",
    journal = "JHEP",
    volume = "11",
    pages = "107",
    year = "2023",
    doi = "10.1007/JHEP11(2023)107"
}

@article{Louko:2000tp,
    author = "Louko, Jorma and Marolf, Donald and Ross, Simon F.",
    title = "{On geodesic propagators and black hole holography}",
    eprint = "hep-th/0002111",
    archivePrefix = "arXiv",
    journal = "Phys. Rev. D",
    volume = "62",
    pages = "044041",
    year = "2000",
    doi = "10.1103/PhysRevD.62.044041"
}

@article{Kraus:2002iv,
    author = "Kraus, Per and Ooguri, Hirosi and Shenker, Stephen",
    title = "{Inside the horizon with AdS/CFT}",
    eprint = "hep-th/0212277",
    archivePrefix = "arXiv",
    journal = "Phys. Rev. D",
    volume = "67",
    pages = "124022",
    year = "2003",
    doi = "10.1103/PhysRevD.67.124022"
}

@article{Maldacena:2001kr,
    author = "Maldacena, Juan Martin",
    title = "{Eternal black holes in Anti-de Sitter}",
    eprint = "hep-th/0106112",
    archivePrefix = "arXiv",
    journal = "JHEP",
    volume = "04",
    pages = "021",
    year = "2003",
    doi = "10.1088/1126-6708/2003/04/021"
}

@article{Romatschke:2015gic,
    author = "Romatschke, Paul",
    title = "{Retarded correlators in kinetic theory: branch cuts, poles and hydrodynamic onset transitions}",
    eprint = "1512.02641",
    archivePrefix = "arXiv",
    primaryClass = "hep-th",
    doi = "10.1140/epjc/s10052-016-4169-7",
    journal = "Eur. Phys. J. C",
    volume = "76",
    number = "6",
    pages = "352",
    year = "2016"
}

@article{Xiao:2026pir,
    author = "Xiao, Zi-Qing and Nian, Jun and Li, Li",
    title = "{Channel-Resolved High-Overtone Quasinormal Modes Decode Kasner Scaling inside Black Holes}",
    eprint = "2609.17426",
    archivePrefix = "arXiv",
    primaryClass = "hep-th",
    month = "9",
    year = "2026"
}

@article{Buric:2026cmn,
    author = "Buri{\'c}, Ilija and Gusev, Ivan and Parnachev, Andrei",
    title = "{Spin-resolved double-trace thermal coefficients in holography}",
    eprint = "2607.24499",
    archivePrefix = "arXiv",
    primaryClass = "hep-th",
    month = "7",
    year = "2026"
}

@article{Arnaudo:2026axe,
    author = "Arnaudo, Paolo and Iossa, Cristoforo and Karlsson, Robin and Withers, Benjamin",
    title = "{OPE = QNM}",
    eprint = "2607.24909",
    archivePrefix = "arXiv",
    primaryClass = "hep-th",
    month = "7",
    year = "2026"
}

@article{Coviello:2026hdc,
    author = "Coviello, Chiara and Gupta, Ansh and Hennigar, Robie A. and Shi, Kai and Svesko, Andrew",
    title = "{Quasinormal modes and quantum black hole interiors}",
    eprint = "2609.26724",
    archivePrefix = "arXiv",
    primaryClass = "hep-th",
    month = "9",
    year = "2026"
}

@article{Coviello:2026tol,
    author = "Coviello, Chiara and Gupta, Ansh and Hennigar, Robie A. and Shi, Kai and Svesko, Andrew",
    title = "{Looking inside a quantum black hole}",
    eprint = "2609.20788",
    archivePrefix = "arXiv",
    primaryClass = "hep-th",
    month = "9",
    year = "2026"
}

@article{Saha:2026ows,
    author = "Saha, Arnab Priya and Shivanna, Prajwal",
    title = "{Asymptotic expansion of cavity Quasi-Normal Modes in Schwarzschild de-Sitter black hole}",
    eprint = "2609.29243",
    archivePrefix = "arXiv",
    primaryClass = "gr-qc",
    month = "9",
    year = "2026"
}

@article{Buric:2026qsp,
    author = "Buri{\'c}, Ilija and Chang, Chi-Ming and Gusev, Ivan and Helfenberger, Elizabeth and Parnachev, Andrei and Rangamani, Mukund",
    title = "{Thermal two-point functions in SYK and complex-time singularities}",
    eprint = "2607.05258",
    archivePrefix = "arXiv",
    primaryClass = "hep-th",
    month = "7",
    year = "2026"
}

@article{Dodelson:2026gak,
    author = "Dodelson, Matthew and Gupta, Om and Mezei, M{\'a}rk and Wang, Diandian",
    title = "{On the temperature dependence of quasinormal modes in SYK and holography}",
    eprint = "2606.22679",
    archivePrefix = "arXiv",
    primaryClass = "hep-th",
    month = "6",
    year = "2026"
}

@article{Aniceto:2026zmc,
    author = "Aniceto, In{\^e}s and Arnaudo, Paolo and Ratcliffe, Alex and Spali{\'n}ski, Micha{\l}",
    title = "{Analytic approaches to perturbations of strongly coupled Yang-Mills plasma}",
    eprint = "2606.12529",
    archivePrefix = "arXiv",
    primaryClass = "hep-th",
    month = "6",
    year = "2026"
}

@article{Grozdanov:2026ktq,
    author = "Grozdanov, Sa{\v{s}}o and Movrin, Vita and Valach, Samuel",
    title = "{Bouncing Geodesics, Singularities, and the Cavity Thermal Product Formula in Asymptotically Flat and de Sitter Black Holes}",
    eprint = "2606.11297",
    archivePrefix = "arXiv",
    primaryClass = "hep-th",
    month = "6",
    year = "2026"
}

@article{Alkalaev:2026sha,
    author = "Alkalaev, K. B. and Mandrygin, Semyon and Samsonov, Vladimir",
    title = "{Thermal conformal partial waves from flat-space and defect CFT}",
    eprint = "2605.27256",
    archivePrefix = "arXiv",
    primaryClass = "hep-th",
    month = "5",
    year = "2026"
}

@article{Chakravarty:2025ncy,
    author = "Chakravarty, Joydeep",
    title = "{Imprint of black hole interior on thermal four-point correlators}",
    eprint = "2512.10912",
    archivePrefix = "arXiv",
    primaryClass = "hep-th",
    doi = "10.1007/JHEP09(2026)034",
    journal = "JHEP",
    volume = "09",
    pages = "034",
    year = "2026"
}

@article{Arnaudo:2026tcy,
    author = "Arnaudo, Paolo and Withers, Benjamin",
    title = "{Bouncing singularities in Schwarzschild: a geometric origin of the QNM convergence region}",
    eprint = "2605.16489",
    archivePrefix = "arXiv",
    primaryClass = "gr-qc",
    month = "5",
    year = "2026"
}

@article{Aalsma:2026kqj,
    author = "Aalsma, Lars and Faruk, Mir Mehedi",
    title = "{Complex geodesics in the Nariai geometry}",
    eprint = "2604.26662",
    archivePrefix = "arXiv",
    primaryClass = "hep-th",
    doi = "10.1007/JHEP07(2026)140",
    journal = "JHEP",
    volume = "07",
    pages = "140",
    year = "2026"
}

@article{Jia:2026ryl,
    author = "Jia, Hewei Frederic and Rangamani, Mukund",
    title = "{Exact holographic thermal spectral functions: OPE, non-perturbative corrections, and black hole singularity}",
    eprint = "2604.10803",
    archivePrefix = "arXiv",
    primaryClass = "hep-th",
    month = "4",
    year = "2026"
}

@article{Grozdanov:2026cut,
    author = "Grozdanov, Sa{\v{s}}o and Valach, Samuel and Vrbica, Mile",
    title = "{Bouncing geodesics, black hole singularities, and singularities of thermal correlators}",
    eprint = "2603.15598",
    archivePrefix = "arXiv",
    primaryClass = "hep-th",
    doi = "10.1007/JHEP08(2026)022",
    journal = "JHEP",
    volume = "08",
    pages = "022",
    year = "2026"
}

@article{Arnaudo:2026der,
    author = "Arnaudo, Paolo and Withers, Benjamin",
    title = "{Analytic structure of holographic thermal correlators from Fourier series}",
    eprint = "2603.13469",
    archivePrefix = "arXiv",
    primaryClass = "hep-th",
    doi = "10.1007/JHEP06(2026)205",
    journal = "JHEP",
    volume = "06",
    pages = "205",
    year = "2026"
}

@article{Giombi:2026kdz,
    author = "Giombi, Simone and Li, Yue-Zhou and Shan, Jieru",
    title = "{Bouncing singularities and thermal correlators on line defects}",
    eprint = "2603.11012",
    archivePrefix = "arXiv",
    primaryClass = "hep-th",
    doi = "10.1007/JHEP07(2026)170",
    journal = "JHEP",
    volume = "07",
    pages = "170",
    year = "2026"
}

@article{Araya:2026shz,
    author = "Araya, Ignacio J. and Esper, Chantelle and Jia, Yueke and Kulaxizi, Manuela and Parnachev, Andrei",
    title = "{Bulkcone Singularities and Complex Geodesics}",
    eprint = "2602.12893",
    archivePrefix = "arXiv",
    primaryClass = "hep-th",
    month = "2",
    year = "2026"
}

@article{Jia:2026pmv,
    author = "Jia, Yueke and Kulaxizi, Manuela",
    title = "{Bulk Phase Shift and Singularity}",
    eprint = "2602.06558",
    archivePrefix = "arXiv",
    primaryClass = "hep-th",
    month = "2",
    year = "2026"
}

@article{Chen:2025jbf,
    author = "Chen, Heng-Yu and Hikida, Yasuaki and Koga, Yasutaka",
    title = "{Bulk-cone singularities and echoes from AdS exotic compact objects}",
    eprint = "2512.21535",
    archivePrefix = "arXiv",
    primaryClass = "hep-th",
    doi = "10.1007/JHEP04(2026)210",
    journal = "JHEP",
    volume = "04",
    pages = "210",
    year = "2026"
}

@article{Barrat:2025twb,
    author = "Barrat, Julien and Bozkurt, Deniz N. and Marchetto, Enrico and Miscioscia, Alessio and Pomoni, Elli",
    title = "{Analytic thermal bootstrap meets holography}",
    eprint = "2510.20894",
    archivePrefix = "arXiv",
    primaryClass = "hep-th",
    reportNumber = "DESY-25-139 , YITP-SB-2025-16",
    doi = "10.1007/JHEP05(2026)180",
    journal = "JHEP",
    volume = "05",
    pages = "180",
    year = "2026"
}

@article{Son:2002sd,
    author = "Son, Dam T. and Starinets, Andrei O.",
    title = "{Minkowski space correlators in AdS / CFT correspondence: Recipe and applications}",
    eprint = "hep-th/0205051",
    archivePrefix = "arXiv",
    reportNumber = "INT-PUB-02-34",
    doi = "10.1088/1126-6708/2002/09/042",
    journal = "JHEP",
    volume = "09",
    pages = "042",
    year = "2002"
}

@article{Policastro:2002se,
    author = "Policastro, Giuseppe and Son, Dam T. and Starinets, Andrei O.",
    title = "{From AdS / CFT correspondence to hydrodynamics}",
    eprint = "hep-th/0205052",
    archivePrefix = "arXiv",
    reportNumber = "INT-PUB-02-32",
    doi = "10.1088/1126-6708/2002/09/043",
    journal = "JHEP",
    volume = "09",
    pages = "043",
    year = "2002"
}

@article{Policastro:2001yc,
    author = "Policastro, G. and Son, Dan T. and Starinets, Andrei O.",
    title = "{The Shear viscosity of strongly coupled N=4 supersymmetric Yang-Mills plasma}",
    eprint = "hep-th/0104066",
    archivePrefix = "arXiv",
    reportNumber = "NYU-TH-01-04-02, SNS-PH-01-05",
    doi = "10.1103/PhysRevLett.87.081601",
    journal = "Phys. Rev. Lett.",
    volume = "87",
    pages = "081601",
    year = "2001"
}

@book{AS,
  title     = {Handbook of Mathematical Functions with Formulas, Graphs, and Mathematical Tables},
  author    = {Abramowitz, Milton and Stegun, Irene A.},
  year      = {1965},
  publisher = {Dover Publications},
  address   = {New York},
  edition   = {Reprint},
  isbn      = {978-0-486-61272-0}
}

@article{Afkhami-Jeddi:2025wra,
  author = {Afkhami-Jeddi, Nima and Caron-Huot, Simon and Chakravarty, Joydeep and Maloney, Alexander},
  title = {{Imprint of the black hole singularity on thermal two-point functions}},
  eprint = {2510.21673},
  archiveprefix = {arXiv},
  primaryclass = {hep-th},
  month = {10},
  year = {2025},
}

@article{Caron-Huot:2025hmk,
  author = {Caron-Huot, Simon and Chakravarty, Joydeep and Namjou, Keivan},
  title = {{Looking at bulk points in general geometries}},
  eprint = {2502.14963},
  archiveprefix = {arXiv},
  primaryclass = {hep-th},
  doi = {10.1007/JHEP06(2025)197},
  journal = {JHEP},
  volume = {06},
  pages = {197},
  year = {2025},
}

@article{Caron-Huot:2025she,
  author = {Caron-Huot, Simon and Chakravarty, Joydeep and Namjou, Keivan},
  title = {{Boundary imprint of bulk causality}},
  eprint = {2501.13182},
  archiveprefix = {arXiv},
  primaryclass = {hep-th},
  doi = {10.1007/JHEP07(2025)076},
  journal = {JHEP},
  volume = {07},
  pages = {076},
  year = {2025},
}

@article{Zigdon:2024ljl,
  author = {Zigdon, Yoav},
  title = {{Stringy forces in the black hole interior}},
  eprint = {2407.12903},
  archiveprefix = {arXiv},
  primaryclass = {hep-th},
  doi = {10.1007/JHEP11(2024)063},
  journal = {JHEP},
  volume = {11},
  pages = {063},
  year = {2024},
}

@article{Batra:2026gbq,
    author = "Batra, Gauri and Levine, Adam and Shenker, Stephen H.",
    title = "{A backreacted bouncing geodesic in 2D}",
    eprint = "2609.32071",
    archivePrefix = "arXiv",
    primaryClass = "hep-th",
    month = "9",
    year = "2026"
}

@article{Gao:2026ieh,
    author = "Gao, Ling-Long and Liu, Yan and Lyu, Hong-Da",
    title = "{Critical Kasner scaling from Cauchy-horizon collapse near holographic phase transitions}",
    eprint = "2609.24227",
    archivePrefix = "arXiv",
    primaryClass = "hep-th",
    month = "9",
    year = "2026"
}

@article{Myers:2007we,
    author = "Myers, Robert C. and Starinets, Andrei O. and Thomson, Rowan M.",
    title = "{Holographic spectral functions and diffusion constants for fundamental matter}",
    eprint = "0706.0162",
    archivePrefix = "arXiv",
    primaryClass = "hep-th",
    doi = "10.1088/1126-6708/2007/11/091",
    journal = "JHEP",
    volume = "11",
    pages = "091",
    year = "2007"
}

@article{Ceplak:2026ard,
    author = "{\v{C}}eplak, Nejc and Valach, Samuel",
    title = "{Where's Kasner? OPE Signature of Kasner Exponents}",
    eprint = "2609.23776",
    archivePrefix = "arXiv",
    primaryClass = "hep-th",
    month = "9",
    year = "2026"
}

@article{Martinec:1994xj,
  author = {Martinec, Emil J.},
  title = {{Space - like singularities and string theory}},
  eprint = {hep-th/9412074},
  archiveprefix = {arXiv},
  reportnumber = {EFI-94-62},
  doi = {10.1088/0264-9381/12/4/005},
  journal = {Class. Quant. Grav.},
  volume = {12},
  pages = {941--950},
  year = {1995},
}

@article{Sharma:2025hbk,
    author = "Sharma, Shivam K.",
    title = "{Holographic fluctuation-dissipation relations in finite density systems}",
    eprint = "2501.17852",
    archivePrefix = "arXiv",
    primaryClass = "hep-th",
    doi = "10.1007/JHEP01(2026)093",
    journal = "JHEP",
    volume = "01",
    pages = "093",
    year = "2026"
}

@article{Jia:2025jbi,
    author = "Jia, Hewei Frederic and Rangamani, Mukund",
    title = "{Thermal spectral function asymptotics and black hole singularity in holography}",
    eprint = "2512.15114",
    archivePrefix = "arXiv",
    primaryClass = "hep-th",
    doi = "10.1103/tzm2-jjvg",
    journal = "Phys. Rev. D",
    volume = "114",
    number = "4",
    pages = "046025",
    year = "2026"
}

@article{Buric:2025fye,
  author = {Buri{\'c}, Ilija and Gusev, Ivan and Parnachev, Andrei},
  title = {{Holographic Correlators from Thermal Bootstrap}},
  eprint = {2508.08373},
  archiveprefix = {arXiv},
  primaryclass = {hep-th},
  month = {8},
  year = {2025},
}

@article{Teaney:2006nc,
  author = {Teaney, Derek},
  title = {{Finite temperature spectral densities of momentum and R-charge correlators in N=4 Yang Mills theory}},
  eprint = {hep-ph/0602044},
  archiveprefix = {arXiv},
  doi = {10.1103/PhysRevD.74.045025},
  journal = {Phys. Rev. D},
  volume = {74},
  pages = {045025},
  year = {2006},
}

@article{Chakrabarty:2019aeu,
  author = {Chakrabarty, Bidisha and Chakravarty, Joydeep and Chaudhuri, Soumyadeep and Jana, Chandan and Loganayagam, R. and Sivakumar, Akhil},
  title = {{Nonlinear Langevin dynamics via holography}},
  eprint = {1906.07762},
  archiveprefix = {arXiv},
  primaryclass = {hep-th},
  doi = {10.1007/JHEP01(2020)165},
  journal = {JHEP},
  volume = {01},
  pages = {165},
  year = {2020},
}

@article{Komargodski:2026oob,
  author = {Komargodski, Zohar},
  title = {{Fate of Bouncing Singularities at Finite $G_N$}},
  eprint = {2609.31309},
  archiveprefix = {arXiv},
  primaryclass = {hep-th},
  month = {9},
  year = {2026},
}

@article{Shi:2026cfq,
    author = "Shi, Yu and Ding, Chikun and Zhou, Yuebing and Liu, Changqing and Deng, Weike and Long, Sheng",
    title = "{Metric and dilaton information from high-overtone quasinormal modes of analytic Einstein-Maxwell-dilaton black holes}",
    eprint = "2610.16973",
    archivePrefix = "arXiv",
    primaryClass = "hep-th",
    month = "10",
    year = "2026"
}

@article{Batra:2026otu,
    author = "Batra, Gauri",
    title = "{A non-prompt null ray cures its own singularity}",
    eprint = "2610.22478",
    archivePrefix = "arXiv",
    primaryClass = "hep-th",
    month = "10",
    year = "2026"
}

@article{Simon_unpub,
  author = {Caron-Huot, Simon and Simmons-Duffin, David and Stanford, Douglas},
  title = {{Private communication}},
}

@article{Caron-Huot:2022lff,
  author = {Caron-Huot, Simon},
  title = {{Holographic cameras: an eye for the bulk}},
  eprint = {2211.11791},
  archiveprefix = {arXiv},
  primaryclass = {hep-th},
  doi = {10.1007/JHEP03(2023)047},
  journal = {JHEP},
  volume = {03},
  pages = {047},
  year = {2023},
}

@article{Anegawa:2024kdj,
    author = "Anegawa, Takanori and Tamaoka, Kotaro",
    title = "{Black hole singularity and timelike entanglement}",
    eprint = "2406.10968",
    archivePrefix = "arXiv",
    primaryClass = "hep-th",
    doi = "10.1007/JHEP10(2024)182",
    journal = "JHEP",
    volume = "10",
    pages = "182",
    year = "2024"
}

@article{Doi:2023zaf,
    author = "Doi, Kazuki and Harper, Jonathan and Mollabashi, Ali and Takayanagi, Tadashi and Taki, Yusuke",
    title = "{Timelike entanglement entropy}",
    eprint = "2302.11695",
    archivePrefix = "arXiv",
    primaryClass = "hep-th",
    reportNumber = "YITP-23-22",
    doi = "10.1007/JHEP05(2023)052",
    journal = "JHEP",
    volume = "05",
    pages = "052",
    year = "2023"
}

@article{Bernamonti:2026pxo,
    author = "Bernamonti, Alice and Galli, Federico and Heller, Michal P. and Ori, Fabio and Serantes, Alexandre",
    title = "{Temporal Entanglement from Twist Correlators in 2d Conformal Field Theory and Holography}",
    eprint = "2607.14012",
    archivePrefix = "arXiv",
    primaryClass = "hep-th",
    month = "7",
    year = "2026"
}

@article{Sorokhaibam:2025nur,
    author = "Sorokhaibam, Nilakash",
    title = "{ETH-monotonicity and the black hole singularity}",
    eprint = "2508.02895",
    archivePrefix = "arXiv",
    primaryClass = "hep-th",
    month = "8",
    year = "2025"
}

@article{Dey:2026ckh,
    author = "Dey, Parijat and Goldar, Arundhati and Kajuri, Nirmalya",
    title = "{Geodesics, one-point functions, and black hole perturbations}",
    eprint = "2601.09397",
    archivePrefix = "arXiv",
    primaryClass = "hep-th",
    doi = "10.1103/ktch-p9h8",
    journal = "Phys. Rev. D",
    volume = "114",
    number = "2",
    pages = "026037",
    year = "2026"
}

@article{Niarchos:2026vow,
    author = "Niarchos, Vasilis and Papageorgakis, Constantinos",
    title = "{Neural Networks, Dispersion Relations and the Thermal Bootstrap}",
    eprint = "2605.13183",
    archivePrefix = "arXiv",
    primaryClass = "hep-th",
    reportNumber = "ITCP-2026-11, CCTP-2026-11, QMUL-PH-26-18",
    month = "5",
    year = "2026"
}

@article{Niarchos:2025cdg,
    author = "Niarchos, V. and Papageorgakis, C. and Stratoudakis, A. and Woolley, M.",
    title = "{Deep finite temperature bootstrap}",
    eprint = "2508.08560",
    archivePrefix = "arXiv",
    primaryClass = "hep-th",
    reportNumber = "CCTP-2025-8, QMUL-PH-25-16, ITCP-2025/8",
    doi = "10.1103/qrjx-w4md",
    journal = "Phys. Rev. D",
    volume = "112",
    number = "12",
    pages = "126012",
    year = "2025"
}

@article{Buric:2026sit,
    author = "Buri{\'c}, Ilija and Gusev, Ivan and Parnachev, Andrei",
    title = "{Stringy resolution of bouncing singularities}",
    eprint = "2610.12311",
    archivePrefix = "arXiv",
    primaryClass = "hep-th",
    month = "10",
    year = "2026"
}

@article{David:2022nfn,
  author = {David, Justin R. and Kumar, Srijan},
  title = {{Thermal one point functions, large d and interior geometry of black holes}},
  eprint = {2212.07758},
  archiveprefix = {arXiv},
  primaryclass = {hep-th},
  doi = {10.1007/JHEP03(2023)256},
  journal = {JHEP},
  volume = {03},
  pages = {256},
  year = {2023},
}

@article{Singhi:2024sdr,
  author = {Singhi, Kaustubh},
  title = {{Proper time to singularity and thermal correlators}},
  eprint = {2406.08553},
  archiveprefix = {arXiv},
  primaryclass = {hep-th},
  month = {6},
  year = {2024},
}

@article{Rodriguez-Gomez:2021pfh,
  author = {Rodriguez-Gomez, D. and Russo, J. G.},
  title = {{Correlation functions in finite temperature CFT and black hole singularities}},
  eprint = {2102.11891},
  archiveprefix = {arXiv},
  primaryclass = {hep-th},
  reportnumber = {ICCUB-21-002},
  doi = {10.1007/JHEP06(2021)048},
  journal = {JHEP},
  volume = {06},
  pages = {048},
  year = {2021},
}

\end{document}